\documentclass[10pt,preprint]{aastex}
\usepackage{amssymb,amsmath,graphicx,lscape,rotating}
\usepackage[T1]{fontenc}
\usepackage{enumitem}
\setlist{itemsep=0pt,topsep=0pt,parsep=0pt,partopsep=0pt,itemindent=0pt}

\begin{document}

\title{The CO-to-H$_{2}$ Conversion Factor across the Perseus Molecular Cloud}

\author{Min-Young Lee\altaffilmark{1}, Sne\v{z}ana Stanimirovi\'c\altaffilmark{1}, Mark G. Wolfire\altaffilmark{2}, 
Rahul Shetty\altaffilmark{3}, Simon C. O. Glover\altaffilmark{3}, \\
Faviola Z. Molina\altaffilmark{3}, Ralf S. Klessen\altaffilmark{3}}

\altaffiltext{1}{Department of Astronomy, University of Wisconsin, Madison, WI 53706, USA; lee@astro.wisc.edu}
\altaffiltext{2}{Department of Astronomy, University of Maryland, College Park, MD 20742, USA}
\altaffiltext{3}{Zentrum f\"ur Astronomie der Universit\"at Heidelberg, 
Institut f\"ur Theoretische Astrophysik, Albert-Ueberle-Str. 2, 69120 Heidelberg, Germany} 
 
\begin{abstract} 
\noindent We derive the CO-to-H$_{2}$ conversion factor, $X_{\rm CO}$ = $N$(H$_{2}$)/$I_{\rm CO}$, 
across the Perseus molecular cloud on sub-parsec scales 
by combining the dust-based $N$(H$_{2}$) data with the $I_{\rm CO}$ data from the COMPLETE Survey.
We estimate an average $X_{\rm CO}$ $\sim$ 3 $\times$ 10$^{19}$ cm$^{-2}$ K$^{-1}$ km$^{-1}$ s 
and find a factor of $\sim$3 variations in $X_{\rm CO}$ between the five sub-regions in Perseus. 
Within the individual regions, $X_{\rm CO}$ varies by a factor of $\sim$100, 
suggesting that $X_{\rm CO}$ strongly depends on local conditions in the interstellar medium. 
We find that $X_{\rm CO}$ sharply decreases at $A_{V}$ $\lesssim$ 3 mag but gradually increases at $A_{V}$ $\gtrsim$ 3 mag,
with the transition occuring at $A_{V}$ where $I_{\rm CO}$ becomes optically thick.
We compare the $N$(HI), $N$(H$_{2}$), $I_{\rm CO}$, and $X_{\rm CO}$ distributions with two models of the formation of molecular gas, 
a one-dimensional photodissociation region (PDR) model and a three-dimensional magnetohydrodynamic (MHD) model 
tracking both the dynamical and chemical evolution of gas. 
The PDR model based on the steady state and equilibrium chemistry reproduces our data very well 
but requires a diffuse halo to match the observed $N$(HI) and $I_{\rm CO}$ distributions. 
The MHD model matches our data reasonably well, suggesting that time-dependent effects on H$_{2}$ 
and CO formation are insignificant for an evolved molecular cloud like Perseus. 
However, we find interesting discrepancies, including a broader range of $N$(HI), likely underestimated $I_{\rm CO}$,
and a large scatter of $I_{\rm CO}$ at small $A_{V}$. 
These discrepancies most likely result from strong compressions/rarefactions and density fluctuations in the MHD model.

\end{abstract}

\section{Introduction}

Stars form exclusively in molecular clouds, 
although the question whether molecular gas is a prerequisite or a byproduct of star formation   
is currently under debate (e.g., Glover \& Clark 2012; Kennicutt \& Evans 2012; Krumholz 2012).
In either case, accurate measuruments of the physical properties of molecular clouds
are critical to constrain the initial conditions for star and molecular gas formation.
However, obtaining such measurements is hampered by the fact that 
molecular hydrogen (H$_{2}$), the most abundant molecular species in the interstellar medium (ISM), 
is not directly observed under the typical conditions in molecular clouds. 
As a homonuclear diatomic molecule, H$_{2}$ does not have a permanent electric dipole moment 
and its ro-vibrational states change only via weak quadrupole transitions. 
Therefore, alternative tracers have been employed to infer the abundance and distribution of H$_{2}$.

Carbon monoxide (CO) is one of the most commonly used tracers of H$_{2}$ 
due to its large abundance and low rotational transitions that
are readily excited in molecular clouds through collisions with H$_{2}$. 
In particular, the $^{12}$CO($J = 1 \rightarrow 0$) integrated intensity\footnote[4]{Hereafter 
$^{12}$CO($J = 1 \rightarrow 0$) is quoted as CO.}, $I_{\rm CO}$, 
is often used to estimate the H$_{2}$ column density, $N$(H$_{2}$), 
via the so-called ``$X$-factor''\footnote[5]{Hereafter $X_{\rm CO}$ is quoted without its units.},
which is defined by 
\begin{equation}
X_{\rm CO} = \frac{N(\rm H_{2})}{I_{\rm CO}} \quad{\rm cm^{-2}~K^{-1}~km^{-1}~s}. 
\end{equation}
Accurate knowledge of $X_{\rm CO}$ is crucial to address some of the fundamental questions in astrophysics.
For example, one of the most intriguing properties of galaxies is a strong power-law relation
between the surface density of star formation rate, 
$\Sigma_{\rm SFR}$, and the surface density of H$_{2}$, $\Sigma_{\rm H2}$, 
generally known as the ``Kennicutt-Schmidt relation'' (e.g., Schmidt 1959; Kennicutt 1989;
Bigiel et al. 2008; Schruba et al. 2011; Rahman et al. 2012; Shetty et al. 2013).
While this empirical relation provides important insights into the physical process of star formation  
(e.g., a close connection between the chemical or thermal state of the ISM and star formation),
its precise form has been a subject of debate and strongly depends on $X_{\rm CO}$.  

From an observational perspective, $X_{\rm CO}$ is usually adopted as a conversion factor.
Its estimate relies on the derivation of $N$(H$_{2}$) using observational methods 
independent of CO (Bolatto et al. 2013 for a review). 
One of the methods to derive $N$(H$_{2}$) utilizes dust as a tracer of total gas column density.   
Dust has been observed to be well mixed with gas (e.g., Boulanger et al. 1996)
and can be mapped through its emission at far-infrared (FIR) wavelengths 
or its absorption at near-infrared (NIR) wavelengths.
The procedure is to estimate the dust column density or the $V$-band extinction, $A_{V}$, 
from the FIR emission or the NIR absorption (e.g., Cardelli et al. 1989) 
and to assume or to estimate a dust-to-gas ratio (DGR) that 
linearly relates $A_{V}$ to the total gas column density $N$(H) = $N$(HI) + 2$N$(H$_{2}$).
The atomic gas column density, $N$(HI), is then measured from the 21-cm emission  
and is removed from $N$(H) for an estimate of $N$(H$_{2}$) (e.g., Israel 1997; Dame et al. 2001; 
Leroy et al. 2007, 2011; Lee et al. 2012; Sandstrom et al. 2013).
The derived $N$(H$_{2}$) is finally combined with $I_{\rm CO}$ to estimate $X_{\rm CO}$.

This procedure has been applied to the Milky Way and a number of nearby galaxies. 
For the Milky Way, Dame et al. (2001) showed that 
$X_{\rm CO}$ does not change significantly with Galactic latitude (for $|b|$ $\sim$ 5$^{\circ}$--30$^{\circ}$) 
from the mean value of (1.8 $\pm$ 0.3) $\times$ 10$^{20}$ when molecular clouds are averaged over $\sim$kpc scales. 
Several studies of individual molecular clouds at 3$'$--9$'$ angular resolution have estimated similar average $X_{\rm CO}$ values 
(e.g., Frerking et al. 1982 for Ophiuchus; Lombardi et al. 2006 for Pipe; Pineda et al. 2008 for Perseus;  
Pineda et al. 2010 for Taurus; Paradis et al. 2012 for Aquila-Ophiuchus, Cepheus-Polaris, Taurus, and Orion).
At the same time, $X_{\rm CO}$ values different from the Galactic mean value have been occasionally found,  
e.g., $X_{\rm CO}$ $\sim$ 0.5 $\times$ 10$^{20}$ for infrared cirrus clouds in Ursa Major (de Vries et al. 1987) 
and $X_{\rm CO}$ $\sim$ 6.1 $\times$ 10$^{20}$ for high-latitude clouds (Magnani et al. 1988), 
suggesting cloud-to-cloud variations in $X_{\rm CO}$. 
Rare studies of $X_{\rm CO}$ in spatially resolved molecular clouds have shown some variations as well,
e.g., $X_{\rm CO}$ $\sim$ (1.6--12) $\times$ 10$^{20}$ for Taurus (Pineda et al. 2010) 
and $X_{\rm CO}$ $\sim$ (0.9--1.8) $\times$ 10$^{20}$ for Perseus (Pineda et al. 2008). 
In studies of nearby galaxies on $\sim$kpc scales,
$X_{\rm CO}$ values are similar with the Galactic mean value and are relatively constant within individual galaxies. 
However, systematically smaller and larger $X_{\rm CO}$ values have been found from the central regions of star-forming galaxies 
(down to $\sim$0.1 $\times$ 10$^{20}$; e.g., Smith et al. 1991; Sandstrom et al. 2013) 
and low-metallicity dwarf irregular galaxies 
(up to $\sim$130 $\times$ 10$^{20}$; e.g., Israel 1997; Leroy et al. 2007; Gratier et al. 2010; Leroy et al. 2011), 
indicating the dependence of the average $X_{\rm CO}$ on interstellar environments.

From a theoretical perspective, $X_{\rm CO}$ has been primarily studied using photodissociation region (PDR) models 
because the majority of the CO emission originates from the outskirts of molecular clouds, 
where the interstellar radiation field (ISRF) illuminates the cloud
(e.g., Taylor et al. 1993; Le Bourlot et al. 1993; Wolfire et al. 1993; Kaufman et al. 1999; Bell et al. 2006; Wolfire et al. 2010).  
For example, Bell et al. (2006) used the \textsc{ucl}\textunderscore\textsc{pdr} code (Papadopoulos et al. 2002) 
to calculate chemical abundances and emission strengths
and showed that $X_{\rm CO}$ changes by more than an order of magnitude with varying depths within molecular clouds. 
In addition, they found significant variations in $X_{\rm CO}$ between molecular clouds with a wide range of physical parameters, 
e.g., density, metallicity, and cloud age. 
While the PDR models are limited to simple geometries and density distributions, 
three-dimensional magnetohydrodynamic (MHD) simulations have been recently performed to investigate 
$X_{\rm CO}$ in turbulent molecular clouds (e.g., Glover \& Mac Low 2011; Shetty et al. 2011a,b). 
These simulations model chemistry for simple molecules such as H$_{2}$ and CO as a function of time
and show that $X_{\rm CO}$ is not constant within individual clouds. 
Moreover, $X_{\rm CO}$ in simulations varies over four orders of magnitude 
between clouds with low densities, low metallicities, and strong radiation fields. 
Such variability of $X_{\rm CO}$ within resolved clouds and between clouds with different properties predicted by the PDR and MHD models 
has been rarely found in observations, largely due to the lack of high-resolution observations.

In this paper, we derive $X_{\rm CO}$ for the Perseus molecular cloud on sub-pc scales 
and test two theoretical models of the formation of molecular gas,
in an attempt to understand the origins of the variations in $X_{\rm CO}$ and the physical processes of H$_{2}$ and CO formation. 
One model is the one-dimensional PDR model originally developed by Tielens \& Hollenbach (1985) 
and updated by Kaufman et al. (2006), Wolfire et al. (2010), and Hollenbach et al. (2012).
Here we use a further modification of this model, which allows for a two-sided illumination 
and either a constant density or a simple formulation of the density distribution (hereafter the modified W10 model).  
The other model is the three-dimensional MHD model by Shetty et al. (2011a)
that is based on the modified \textsc{zeus--mp} code described in Glover et al. (2010) (hereafter the S11 model). 
There are two primary differences between these two models.  
First, the S11 model simulates H$_{2}$ and CO formation in turbulent molecular clouds 
by coupling the chemical and dynamical evolution of gas,
while the modified W10 model takes into account the impact of turbulence
only via a constant supersonic linewidth for spectral line formation and cooling. 
Second, the S11 model follows the time-dependent evolution of a number of chemical species, including H$_{2}$ and CO, 
while the modified W10 model uses a detailed time-independent chemical network 
that explicitly assumes chemical equilibrium for every atomic and molecular species.
Therefore, we consider the modified W10 model and the S11 model 
as representative ``microturbulent time-independent model'' and ``macroturbulent time-dependent model''. 
Our study is one of the first attempts to test the MHD model tracking both the chemical and dynamical evolution of the ISM 
and compare it with a more traditional view of the formation of molecular gas (PDR model).
In addition, considering that small-scale ISM models are starting to be implemented  
in large-scale simulations of galaxy formation and evolution 
(e.g., Feldmann et al. 2012a,b; Lagos et al. 2012; Narayanan et al. 2012),
our study will serve as a ``zero point test'' for the models of gas contents in galaxies.

\begin{figure}
\begin{center}
\includegraphics[scale=0.25]{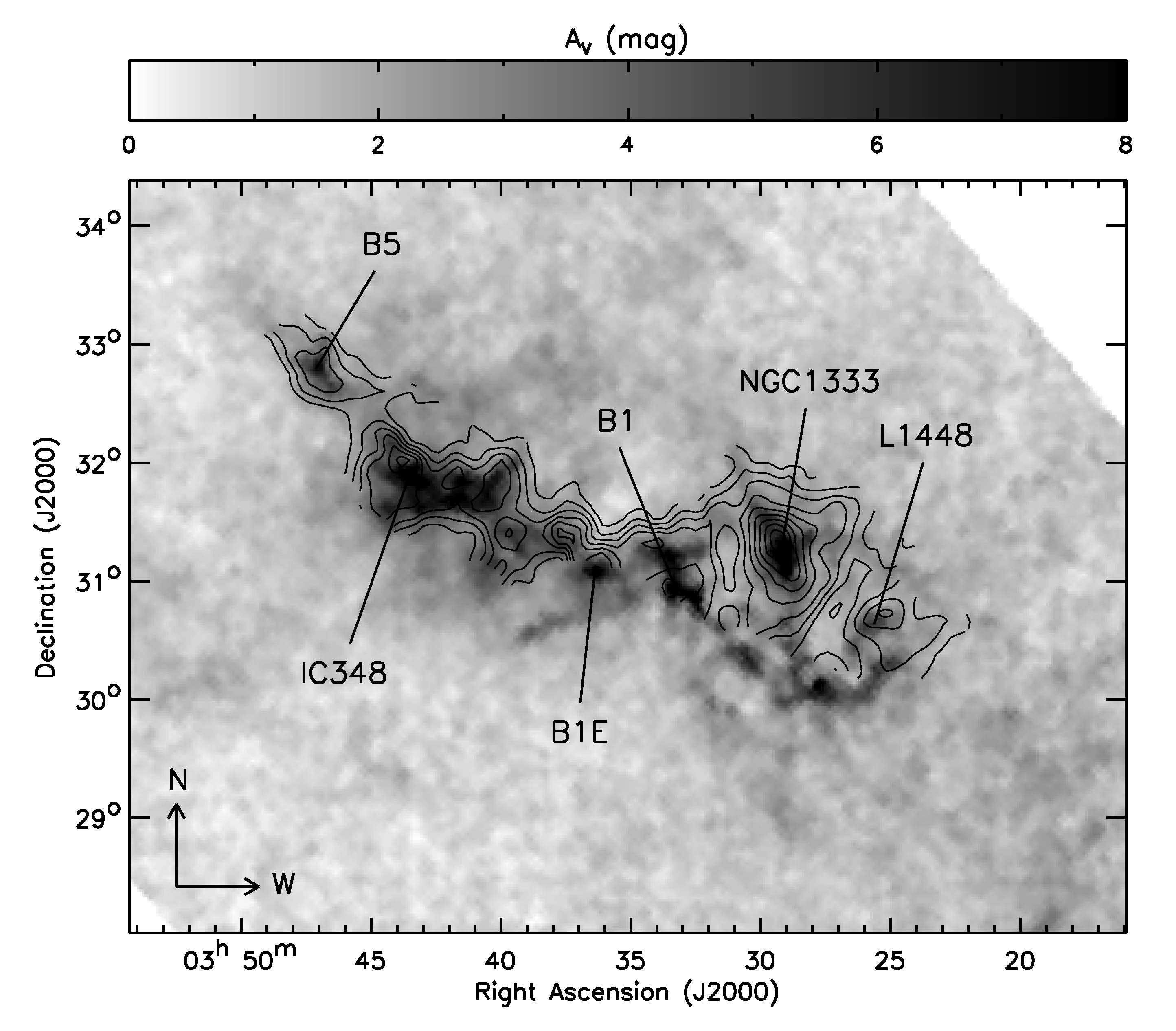}
\caption{\label{f:Av_intro} 
COMPLETE $A_{V}$ image of Perseus overlaid with the COMPLETE $I_{\rm CO}$ contours (Ridge et al. 2006). 
The contour levels range from 10\% to 90\% of the peak (80 K km s$^{-1}$) with 10\% steps. 
The angular resolution of the $A_{V}$ and $I_{\rm CO}$ images here is 5$'$ and 4.3$'$ respectively. 
A number of dark (B5, B1E, B1, and L1448) and star-forming regions (IC348 and NGC1333) are labelled.}
\end{center}
\end{figure}

We focus on the Perseus molecular cloud because of its proximity and a wealth of multi-wavelength observations.  
Located at a distance of $\sim$200--350 pc (Herbig \& Jones 1983; \v{C}ernis 1990), 
Perseus has a projected angular size of $\sim$6$^{\circ}$ $\times$ 3$^{\circ}$ on the sky. 
In this paper, we adopt the distance to Perseus of 300 pc.
With a mass of $\sim$2 $\times$ 10$^{4}$ M$_{\odot}$ (Sancisi et al. 1974; Lada et al. 2010), 
Perseus is considered as a low-mass molecular cloud with an intermediate level of star formation (Bally et al. 2008). 
The cloud contains a number of dark (B5, B1E, B1, and L1448) and star-forming regions (IC348 and NGC1333) shown in Figure \ref{f:Av_intro}.

This paper is organized as follows.
In Section 2, we summarize the results from previous studies highly relevant to our investigation  
and provide constraints on important physical parameters of Perseus.  
In Section 3, we describe the multi-wavelength observations used in our study. 
In Section 4, we divide Perseus into a number of individual regions and select data points for each region. 
We then derive the $X_{\rm CO}$ image (Section 5) 
and investigate the large-scale spatial variations of $X_{\rm CO}$ 
and their correlations with physical parameters such as the strength of the radiation field and the CO velocity dispersion (Section 6). 
In addition, we examine how $I_{\rm CO}$ and $X_{\rm CO}$ change with $A_{V}$ in Perseus.   
In Section 7, we summarize the details of the modified W10 model and the S11 model 
and compare our observational data with model predictions.  
Finally, we discuss and summarize our conclusions (Sections 8 and 9).

\section{Background} 

\subsection{Relevant Previous Studies of Perseus}

Pineda et al. (2008) used the $I_{\rm CO}$ and $A_{V}$ data from the COMPLETE Survey of Star Forming Regions (COMPLETE; Ridge et al. 2006) 
to investigate $X_{\rm CO}$ in Perseus.  
They fitted a linear function to $I_{\rm CO}$ vs $A_{V}$ to estimate $X_{\rm CO}$ 
and found $X_{\rm CO}$ $\sim$ 1.4 $\times$ 10$^{20}$ for the whole cloud 
and a range of $X_{\rm CO}$ $\sim$ (0.9--1.8) $\times$ 10$^{20}$ for six sub-regions,
suggesting a factor of $\sim2$ spatial variations of $X_{\rm CO}$ caused by different ISM conditions. 
In the process of performing a linear fit, 
they noticed that $X_{\rm CO}$ is heavily affected by the saturation of $I_{\rm CO}$ at $A_{V}$ $\gtrsim$ 4 mag
and re-estimated $X_{\rm CO}$ $\sim$ 0.7 $\times$ 10$^{20}$ from the linear fit only to the unsaturated $I_{\rm CO}$.
In addition, Pineda et al. (2008) compared the observed CO and $^{13}$CO($J = 1 \rightarrow 0$) integrated intensities 
with predictions from the Meudon PDR code (Le Petit et al. 2006) 
and found that the PDR models reproduce the CO and $^{13}$CO observations reasonably well 
and the variations among the six sub-regions can be explained by 
variations in physical parameters, in particular density and non-thermal gas motion.

In our recent study, we derived the $\Sigma_{\rm HI}$ and $\Sigma_{\rm H2}$ images of Perseus on $\sim$0.4 pc scales (Section 3.1) 
and investigated how the H$_{2}$-to-HI ratio, 
$R_{\rm H2}$ = $\Sigma_{\rm H2}$/$\Sigma_{\rm HI}$ = 2$N(\rm H_{2})$/$N$(HI), changes across the cloud (Lee et al. 2012).
We found that $\Sigma_{\rm HI}$ is relatively uniform with $\sim$6--8 M$_{\odot}$ pc$^{-2}$, 
while $\Sigma_{\rm H2}$ significantly varies from $\sim$0 M$_{\odot}$ pc$^{-2}$ to $\sim$73 M$_{\odot}$ pc$^{-2}$,
resulting in $R_{\rm H2}$ $\sim$ 0--10 with a mean of $\sim$0.7.
Due to the nearly constant $\Sigma_{\rm HI}$, a strong linear relation 
between $R_{\rm H2}$ and $\Sigma_{\rm HI} + \Sigma_{\rm H2}$ was found.  
Interestingly, these results are consistent with the time-independent H$_{2}$ formation model 
by Krumholz et al. (2009; hereafter the K09 model).  
In the K09 model, a spherical cloud is embedded in a uniform and isotropic radiation field 
and the H$_{2}$ abundance is estimated based on the balance between H$_{2}$ formation on dust grains and 
H$_{2}$ photodissociation by Lyman-Werner (LW) photons.
The most important prediction of the K09 model is the minimum $\Sigma_{\rm HI}$ 
required to shield H$_{2}$ against photodissociation.
This minimum $\Sigma_{\rm HI}$ for H$_{2}$ formation depends on metallicity 
(e.g., $\Sigma_{\rm HI}$ $\sim$ 10 M$_{\odot}$ pc$^{-2}$ for solar metallicity) 
but only weakly on the strength of the radiation field. 
Once the minimum $\Sigma_{\rm HI}$ is achieved, 
additional $\Sigma_{\rm HI}$ is fully converted into $\Sigma_{\rm H2}$,
resulting in the uniform $\Sigma_{\rm HI}$ distribution and 
the linear increase of $R_{\rm H2}$ with $\Sigma_{\rm HI}$+$\Sigma_{\rm H2}$.

\subsection{Constraints on Physical Parameters}

We summarize estimates of several important physical parameters of Perseus obtained from previous studies.
We will use these parameters in later sections of this paper.

\textbf{Density} $\boldsymbol{n}$ $\boldsymbol{\sim}$ $\boldsymbol{10^{3-4}}$ $\boldsymbol{\rm cm^{-3}}$\textbf{:}
Young et al. (1982) estimated $n$ $\sim$ (1.7--5) $\times$ 10$^{3}$ cm$^{-3}$ for B5
based on the large velocity gradient (LVG) model applied to CO and CO($J = 2 \rightarrow 1$) observations.
Bensch (2006) derived larger $n$ $\sim$ (3--30) $\times$ 10$^{3}$ cm$^{-3}$ for the same cloud
by comparing PDR models with CO, $^{13}$CO, and [CI] observations.
Similarly, Pineda et al. (2008) found that PDR models with $n$ $\sim$ a few $\times$ 10$^{3-4}$ cm$^{-3}$
can reproduce the CO and $^{13}$CO($J = 1 \rightarrow 0$) observations of Perseus.
In summary, gas traced by the CO emission in Perseus is likely to have $n\sim10^{3-4}$ cm$^{-3}$.

\textbf{ISRF} $\boldsymbol{G}$ $\boldsymbol{\sim}$ \textbf{0.4} $\boldsymbol{G_{0}'}$\textbf{:}
Lee et al. (2012) investigated the dust temperature, $T_{\rm dust}$, across Perseus and potential heating sources  
and concluded that the cloud is embedded in the uniform Galactic ISRF heating dust grains to $\sim$17 K, 
except for the central parts of IC348 and NGC1333, 
where the radiation from internal B-type stars likely dominates.   
Under the assumption that dust grains are in thermal equilibrium, 
we can use $T_{\rm dust}$ $\sim$ 17 K to estimate the strength of the radiation field by 
\begin{equation}
G = 4.6 \times 10^{-11} \left(\frac{a}{0.1~\rm \mu m}\right) 
T_{\rm dust}^{6}~\rm{erg~cm^{-2}~s^{-1}}, 
\end{equation}

\noindent where $G$ is the flux at ultraviolet (UV) wavelengths and $a$ is the size of dust grains (Lequeux 2005).
Equation (2) assumes the absorption efficiency $Q_{\rm a}$ = 1 and the dust emissivity index $\beta$ = 2.
For dust grains with $a$ $\sim$ 0.1 $\mu$m, whose size is comparable to UV wavelengths therefore $Q_{\rm a}$ $\simeq$ 1,
we estimate $G$ $\sim$ 1.1 $\times$ 10$^{-3}$ erg cm$^{-2}$ s$^{-1}$ $\sim$ 0.4 $G_{0}'$ 
($G_{0}'$ = the local field measured in the solar neighborhood by Draine 1978 $\sim$ 2.7 $\times$ 10$^{-3}$ erg cm$^{-2}$ s$^{-1}$)
for the uniform ISRF incident upon Perseus. 
The exceptions are the central regions of IC348 and NGC1333, 
where the radiation from the B-type stars is dominant.

\textbf{Cosmic-ray ionization rate} $\boldsymbol{\zeta}$
$\boldsymbol{\gtrsim}$ $\boldsymbol{10^{-17}}$ $\boldsymbol{\rm s^{-1}}$\textbf{:}
There is emerging evidence that $\zeta$ likely lies between $\sim$10$^{-17}$ s$^{-1}$ to $\sim$10$^{-15}$ s$^{-1}$
with lower values in dense molecular clouds
and $\sim$10$^{-16}$ s$^{-1}$ to $\sim$10$^{-15}$ s$^{-1}$ in the diffuse ISM
(e.g., Dalgarno 2006; Indriolo \& McCall 2012; Hollenbach et al. 2012).
This suggests that $\zeta$ could be larger than the canonical $\zeta$ $\sim$ 10$^{-17}$ s$^{-1}$
by a factor of $\sim$10--100 in the regions where the CO emission arises.

\textbf{Metallicity} $\boldsymbol{Z}$ $\boldsymbol{\sim}$ $\boldsymbol{1}$ $\boldsymbol{\rm Z_{\odot}}$\textbf{:}
Gonz\'alez Hern\'andez et al. (2009) performed a chemical abundance analysis
for \u Cernis 52, a member of IC348 whose spectral type is A3 V,
and derived [Fe/H] = $-$0.01 $\pm$ 0.15 (corresponding to $Z$ $\sim$ 0.7--1.4 $\rm Z_{\odot}$).
In addition, Lee et al. (2012) compared the intensity at 100 $\mu$m, $I_{\rm 100}$, with $N$(HI) for Perseus 
and found an overall linear relation.
As $I_{\rm 100}$/$N$(HI) is an approximation of DGR,
the fact that a single $I_{\rm 100}$/$N$(HI) fits most of the diffuse regions suggests
no significant variation of DGR or $Z$ across the cloud.
Therefore, $Z$ $\sim$ 1 Z$_{\odot}$ would be a reasonable estimate for Perseus.
Note that Lee et al. (2012) derived DGR = $A_{V}$/$N$(H) $\sim$ 1.1 $\times$ 10$^{-21}$ mag cm$^{2}$ for Perseus,
which is $\sim$2 times larger than the typical Galactic DGR $\sim$ 5.3 $\times$ 10$^{-22}$ mag cm$^{2}$ (Bohlin et al. 1978).

\textbf{Turbulent linewidth} $\boldsymbol{v_{\rm turb}}$ $\boldsymbol{\lesssim}$
$\boldsymbol{2}$$\boldsymbol{-}$$\boldsymbol{5}$ $\boldsymbol{\rm km~s^{-1}}$\textbf{:}
Pineda et al. (2008) compared the CO excitation temperature, $T_{\rm ex}$, with $A_{V}$ 
and found that $T_{\rm ex}$ increases from $\sim$5 K at $A_{V}$ $\sim$ 2 mag to $\sim$20 K at $A_{V}$ $\gtrsim$ 4 mag.
If $n$ > $n_{\rm crit}$ $\sim$ 10$^{3}$ cm$^{-3}$ where $n_{\rm crit}$ is the critical density for the CO emission, 
the case likely for the regions with $A_{V}$ $\gtrsim$ 4 mag,
we expect that the CO emission is in local thermodynamic equilibrium (LTE)
and $T_{\rm ex}$ $\sim$ $T_{\rm k}$ where $T_{\rm k}$ is the kinetic temperature.
When we assume $T_{\rm ex}$ $\sim$ $T_{\rm k}$ $\sim$ 20 K for Perseus,
the mean thermal velocity of CO-emitting gas would be 
$\langle$$v_{\rm th}$$\rangle$ = $\sqrt{2k_{\rm B}T_{\rm k}/\mu m_{\rm H}}$ $\sim$ 0.1 km s$^{-1}$
($k_{\rm B}$ = the Boltzmann constant, $\mu$ = the mass of a molecule in amu = 28 for CO,
$m_{\rm H}$ = the mass of a hydrogen atom).
This $\langle$$v_{\rm th}$$\rangle$ $\sim$ 0.1 km s$^{-1}$ is an order of magnitude smaller than
the CO velocity dispersion $\sigma_{\rm CO}$ $\sim$ 0.9--2 km s$^{-1}$ 
(corresponding to FWHM = $(\rm 8 ln 2)^{1/2}$ $\sigma_{\rm CO}$ $\sim$ 2.1--4.7 km s$^{-1}$) 
measured across Perseus (Pineda et al. 2008).
This suggests that there are most likely contributions from other processes, 
e.g., interstellar turbulence, systematic motions such as inflow, outflow, rotation, etc.,  
and/or multiple components along a line of sight. 
For example, B1 and NGC1333 contain a large number of Herbig-Haro objects 
that are known to trace currently active shocks in outflows (e.g., Bally et al. 2008). 
Therefore, not all the observed $\sigma_{\rm CO}$ should be attributed to interstellar turbulence alone.
As a result, we expect $v_{\rm turb}$ to be smaller than the measured FWHM of $\sim$2--5 km s$^{-1}$.

\textbf{Cloud age} $\boldsymbol{t_{\rm age}}$ $\boldsymbol{\sim}$ $\boldsymbol{10}$ $\boldsymbol{\rm Myr}$\textbf{:}
For IC348, Muench et al. (2003) derived a mean age of $\sim$2 Myr 
with a spread of $\sim$3 Myr using published spectroscopic observations.
However, there are some indications for the existence of older stars in IC348.
For example, Herbig (1998) found that H$\alpha$ emission line stars in IC348
have an age spread from $\sim$0.7 Myr to $\sim$12 Myr.
A similar spread in stellar age, from $\sim$0.5 Myr to $\sim$10 Myr,
has been found by Luhman et al. (1998) from their infrared and optical spectroscopic observations.
Considering this duration of star formation in IC348, 
$t_{\rm age}$ $\sim$ 10 Myr would be a reasonable age estimate for Perseus.
 
\section{Data}

\subsection{Derived H$\boldsymbol{_{2}}$ Distribution}

We use the $N$(H$_{2}$) image derived in our recent study, Lee et al. (2012).
We used the 60 $\mu$m and 100 $\mu$m data from the Improved Reprocessing of the \textit{IRAS} Survey 
(IRIS; Miville-Desch\^enes \& Lagache 2005) to derive the dust optical depth at 100 $\mu$m, $\tau_{100}$.
Dust grains were assumed to be in thermal equilibrium 
and the contribution from very small grains (VSGs) to the intensity at 60 $\mu$m was accounted for 
by calibrating the derived $T_{\rm dust}$ image with the $T_{\rm dust}$ data from Schlegel et al. (1998).  
The $\tau_{100}$ image was then converted into the $A_{V}$ image 
by finding the conversion factor $X$ for $A_{V} = X\tau_{100}$ 
that results in the best agreement between the derived $A_{V}$ and COMPLETE $A_{V}$.   
This calibration of $\tau_{100}$ to COMPLETE $A_{V}$ was motivated by Goodman et al. (2009), 
who found that dust extinction at NIR wavelengths is the best probe of total gas column density.
Finally, Lee et al. (2012) estimated a local DGR for Perseus 
and derived the $N$(H$_{2}$) image in combination with the HI data 
from the Galactic Arecibo L-band Feed Array HI Survey (GALFA-HI; Peek et al. 2011).
The HI emission was integrated from $-$5 km s$^{-1}$ to $+$15 km s$^{-1}$, 
the range that maximizes the spatial correlation between the HI integrated intensity and the dust column density, 
and $N$(HI) was calculated under the assumption of optically thin HI. 
The derived $N$(H$_{2}$) has a mean of $\sim$1.3 $\times$ 10$^{20}$ cm$^{-2}$
and peaks at $\sim$4.5 $\times$ 10$^{21}$ cm$^{-2}$. 
Its mean 1$\sigma$ uncertainty is $\sim$3.6 $\times$ 10$^{19}$ cm$^{-2}$.
See Section 4 of Lee et al. (2012) for details on the derivation of the $N$(H$_{2}$) image and its 1$\sigma$ uncertainty.

The $T_{\rm dust}$, $A_{V}$, $N$(HI), and $N$(H$_{2}$) images derived by Lee et al. (2012) 
are all at 4.3$'$ angular resolution, corresponding to $\sim$0.4 pc at the distance of 300 pc.
We present the $N$(H$_{2}$) image at 4.3$'$ angular resolution in Figure \ref{f:H2_cden_display}. 
The blank data points correspond to point sources and regions with possible contamination  
(the Taurus molecular cloud and a background HII region). 
See Sections 4.2 and 4.3 of Lee et al. (2012) for details. 

 
\subsection{Observed CO Distribution} 

We use the COMPLETE CO data cube obtained with the 14-m FCRAO telescope (Ridge et al. 2006).  
This cube covers the main body of Perseus 
with a spatial area of $\sim$6$^{\circ}$ $\times$ 3$^{\circ}$ at 46$''$ angular resolution.
We correct the CO data for the main-beam efficiency of 0.45, 
following Ridge et al. (2006) and Pineda et al. (2008). 
The rms noise per channel\footnote[6]{In this paper, 
all temperatures are in main-beam brightness units 
and all velocities are quoted in the local standard of rest (LSR) frame.}
ranges from $\sim$0.3 K to $\sim$3.5 K with a mean of $\sim$0.8 K. 
We show the average CO spectrum for Perseus in Figure \ref{f:CO_spectrum_Perseus}.
To produce this spectrum, we average the spectra of all data points 
where the ratio of the peak main-beam brightness temperature to the rms noise is greater than 3.
The CO emission is clearly contained between $-5$ km s$^{-1}$ and $+$15 km s$^{-1}$ 
and shows multiple velocity components.

To derive $I_{\rm CO}$, we integrate
the CO emission from $-5$ km s$^{-1}$ to $+$15 km s$^{-1}$, 
the range where Lee et al. (2012) found the HI emission associated with Perseus, 
with a spectral resolution of 0.064 km s$^{-1}$. 
At 46$''$ angular resolution, the derived $I_{\rm CO}$ ranges from $-$19.9 K km s$^{-1}$ to 116.6 K km s$^{-1}$. 
Its mean 1$\sigma$ uncertainty is $\sim$0.9 K km s$^{-1}$. 
We note that some data points in the CO cube are affected by an artificial absorption feature at $v$ $\sim$ 7.5 km s$^{-1}$.
This artifact is due to the contaminated off-position\footnote[7]{See http://www.cfa.harvard.edu/COMPLETE/data\_html\_pages/PerA\_12coFCRAO\_F.html.} 
and is responsible for a number of blanked data points in Figure \ref{f:CO_vel_dispersion} 
that do not correspond to point sources and regions with possible contamination. 
We find that this artifact does not affect our estimate of $I_{\rm CO}$. 

\section{Region Division}

As pointed out by Pineda et al. (2008) and Lee et al. (2012), 
there are considerable region-to-region variations in physical parameters across Perseus.
We therefore divide the cloud into five regions and perform analyses mainly on the individual regions. 
To define the individual regions, we draw the COMPLETE $I_{\rm CO}$ contours 
from 4 K km s$^{-1}$ (5\% of the peak) to 72 K km s$^{-1}$ (90\% of the peak) with 4 K km s$^{-1}$ intervals 
and use the contours to determine the boundaries of each region.
Note that the minimum $I_{\rm CO}$ of 4 K km s$^{-1}$ for the regional boundaries does not mean that 
there is no data point with $I_{\rm CO}$ < 4 K km s$^{-1}$. 
In addition, we select data points that have 
(1) $-5$ km s$^{-1}$ < CO velocity centroid < $+$15 km s$^{-1}$, 
(2) $I_{\rm CO}$ > 0 K km s$^{-1}$, and (3) $N$(H$_{2}$) > 0 cm$^{-2}$.
These criteria are to select data points that are reliable and kinematically associated with Perseus.
Applying these criteria results in 1160 independent data points 
and all except three data points have S/N > 1 for both $I_{\rm CO}$ and $N$(H$_{2}$).  
We show the selected data points for each region (B5, B1E/B1, L1448 as dark regions and IC348, NGC1333 as star-forming regions) 
with a different color in Figure \ref{f:COMPLETE_five_regions}.
The individual regions have an average size of $\sim$5--7 pc at the distance of 300 pc (Table 1).  

 
\section{Deriving $\boldsymbol{X_{\rm CO}}$}

We derive the $X_{\rm CO}$ image at 4.3$'$ angular resolution by applying Equation (1) to the $N$(H$_{2}$) image 
and the COMPLETE $I_{\rm CO}$ image (smoothed to match the angular resolution of the $N$(H$_2$) image)
on a pixel-by-pixel basis (Figure \ref{f:COMPLETE_X_CO}).  
For the five regions defined in Section 4, 
$X_{\rm CO}$ ranges from $\sim$5.7 $\times$ 10$^{15}$ to $\sim$4.4 $\times$ 10$^{21}$.
While $X_{\rm CO}$ shows a substantial range, 
most data points ($\sim$80\%) have 10$^{19}$ < $X_{\rm CO}$ < 10$^{20}$. 
Summing both $N$(H$_{2}$) and $I_{\rm CO}$ over all five regions results in 
an average $\langle X_{\rm CO} \rangle$ = $\Sigma$$N$(H$_{2}$)/$\Sigma$$I_{\rm CO}$ $\sim$ 3 $\times$ 10$^{19}$.
Applying a single criterion of $N$(H$_{2}$) > 0 cm$^{-2}$ to the whole cloud
to include the regions with H$_{2}$ but without CO detection 
results in the same average $\langle X_{\rm CO} \rangle$ $\sim$ 3 $\times$ 10$^{19}$.
The 1$\sigma$ uncertainty of $X_{\rm CO}$ is derived based on the propagation of errors 
(Bevington \& Robinson 2003) and its mean value is $\sim$1.6 $\times$ 10$^{19}$.

\section{Results}

\subsection{Large-scale Spatial Variations of $\boldsymbol{X_{\rm CO}}$}

Figure \ref{f:COMPLETE_X_CO} shows interesting spatial variations of $X_{\rm CO}$ across Perseus.
To quantify these variations, we estimate the $\langle X_{\rm CO} \rangle$ values for the dark and star-forming regions 
by summing $N$(H$_{2}$) and $I_{\rm CO}$ over each region (Table 1). 
We find a factor of $\sim$3 decrease in $X_{\rm CO}$ from the northeastern regions (B5 and IC348) 
where $\langle X_{\rm CO} \rangle$ $\sim$ 6 $\times$ 10$^{19}$ 
to the southwestern regions (B1E/B1, NGC1333, and L1448) where $\langle X_{\rm CO} \rangle$ $\sim$ 2 $\times$ 10$^{19}$.
Our result is consistent with Pineda et al. (2008) 
in that both studies found regional variations of $X_{\rm CO}$ across Perseus. 
However, while they estimated a single $X_{\rm CO}$ for each sub-region, 
we derived the spatial distribution of $X_{\rm CO}$.  
Based on this distribution, we investigate large-scale trends in several physical parameters 
and their possible connections with the variations of $X_{\rm CO}$.

We first derive the $\sigma_{\rm CO}$ image using the COMPLETE CO data cube. 
Figure \ref{f:CO_vel_dispersion} shows that the southwestern part 
has systematically larger $\sigma_{\rm CO}$ than the northeastern part.
For example, $\sim$70\% of the data points in the southwestern part have $\sigma_{\rm CO}$ > 1.5 km s$^{-1}$, 
while $\sim$40\% of the data points in the northeastern part have $\sigma_{\rm CO}$ > 1.5 km s$^{-1}$. 
In particular, B1E/B1 and NGC1333 have the largest median $\sigma_{\rm CO}$ $\sim$ 2 km s$^{-1}$
compared to other regions where the median $\sigma_{\rm CO}$ is $\sim$1.3 km s$^{-1}$ (Table 1). 
The large $\sigma_{\rm CO}$ in the southwestern part could be caused by more complex velocity structure
and/or multiple components along a line of sight.
In addition, outflows from embedded protostars could contribute to broaden the CO spectra.
For example, B1 and NGC1333 have many Herbig-Haro objects
identified from the surveys of H$\alpha$ and [SII] emission,
which trace currently active shocks in outflows (e.g., Bally et al. 2008). 

The $T_{\rm dust}$ image derived by Lee et al. (2012) also shows systematic variations across Perseus. 
Specifically, $T_{\rm dust}$ slightly decreases toward the southwestern part.
This is consistent with Pineda et al. (2008), who found $T_{\rm dust}$ $\sim$ 17 K for B5/IC348
and $T_{\rm dust}$ $\sim$ 16 K for B1E/B1/NGC1333.
To investigate the variations of ISRF,
We evaluate $G$ using Equation (2) and assess its distributions for B5/IC348 (\textit{East}) and B1E/B1/NGC1333/L1448 (\textit{West}).
Figure \ref{f:ISRF_histo} shows the median $G$ of \textit{East} ($\sim$10$^{-2.86}$ erg cm$^{-2}$ s$^{-1}$) as a dashed line.
We find that $\sim$50\% and $\sim$2\% of the data points have $G$ > 10$^{-2.86}$ erg cm$^{-2}$ s$^{-1}$
for \textit{East} and \textit{West} respectively.
This suggests that $G$ systematically decreases toward the southwestern part of Perseus.
However, the variation of $G$ is very mild: 
the median $G$ decreases from \textit{East} to \textit{West} by only a factor of $\sim$1.4. 
This result does not change even when we examine the median $G$ for each dark and star-forming region.
 
Finally, Lee et al. (2012) noticed a considerable difference between the northeastern and southwestern parts of Perseus 
regarding the relative spatial distribution of H$_{2}$ and CO.
They estimated the fraction of ``CO-dark'' H$_{2}$,
which refers to interstellar gas in the form of H$_{2}$ along with CI and CII but little or no CO, 
and found a factor of $\sim$3 decrease in the fraction toward the southwestern part.  
In other words, ``CO-free'' H$_{2}$ envelopes exist in the northeastern part, 
while CO traces H$_{2}$ reasonably well in the southwestern part (e.g., Figure 14 of Lee et al. 2012).
This suggests that H$_{2}$ takes up a larger volume than CO in the northeastern region, which could result in larger $X_{\rm CO}$.


Many theoretical studies have shown that $X_{\rm CO}$ can vary over several orders of magnitude
with changes in density, metallicity, turbulent linewidth, ISRF, etc. 
(e.g., Maloney \& Black 1988; Le Bourlot et al. 1993; Wolfire et al. 1993;
Sakamoto 1996, 1999; Kaufman et al. 1999; Bell et al. 2006; Glover \& Mac Low 2011; Shetty et al. 2011a,b), 
suggesting that various physical parameters play a role in determining $X_{\rm CO}$. 
This likely applies to Perseus as well. 
While $\sigma_{\rm CO}$ and $G$ show some interesting variations across the cloud, 
their correlations with $X_{\rm CO}$ are not strong 
(Spearman's rank correlation coefficient $r_{\rm s}$ = $-0.2$ and 0.6 respectively;
the null hypothesis is rejected at the 99\% two-tailed confidence level).  
In addition, as we will show in comparison with the modified W10 model (Section 7.1), 
changes in density appear to contribute to the observed variations in $X_{\rm CO}$ as well.  
It is most likely, therefore, that combinations of changes in 
density, turbulent linewidth, ISRF, and possibly other parameters we do not test in our study  
result in the variations in $X_{\rm CO}$ across the cloud. 
This conclusion is consistent with Pineda et al. (2008), 
who suggested that local variations in density, non-thermal gas motion, and ISRF 
can explain the observed scatter of $X_{\rm CO}$ among the sub-regions in Perseus. 

Because $X_{\rm CO}$ depends on many properties of the ISM, 
constraining physical conditions by matching models to the observed value of $X_{\rm CO}$ 
requires a search through a large parameter space. 
Nevertheless, from a theoretical standpoint, $X_{\rm CO}$ has an interesting characteristic dependance on $A_{V}$ 
(e.g., Taylor et al. 1993, Bell et al. 2006; Glover \& Mac Low 2011; Shetty et al. 2011a; Feldmann et al. 2012a). 
We focus on investigating this characteristic dependance over a broad range of $A_{V}$ 
by comparing our observations with two different theoretical models, 
with an aim of understanding the important physical processes of H$_{2}$ and CO formation.
To do so, we use the models with a simple set of input parameters reasonable for Perseus
and focus mainly on the general trends of $N$(HI), $N$(H$_{2}$), $I_{\rm CO}$, and $X_{\rm CO}$ with $A_{V}$.

\subsection{$\boldsymbol{I_{\rm CO}}$ versus $\boldsymbol{A_{V}}$}

\subsubsection{Global Properties} 

To understand how $X_{\rm CO}$ varies with $A_{V}$,
we begin by plotting $I_{\rm CO}$ as a function of $A_{V}$ in Figure \ref{f:CO_int_Av_plot_Perseus} 
for all five regions defined in Section 4. 
We use the $A_{V}$ image at 4.3$'$ angular resolution derived by Lee et al. (2012).
Even though there is a large amount of scatter, several important features are noticeable. 

First, there appears to be some threshold $A_{V,\rm th}$ $\sim$ 1 mag below which no CO emission is detected. 
The sharp increase of $I_{\rm CO}$ with $A_{V}$ found from the individual regions (Section 6.2.2)
strongly supports the existence of such threshold. 
This may suggest that CO becomes shielded against photodissociation at $A_{V}$ $\sim$ 1 mag in Perseus.
Previous observations of molecular clouds have found the similar $A_{V,\rm th}$ $\sim$ 1 mag  
(e.g., Lombardi et al. 2006; Pineda et al. 2008; Leroy et al. 2009).
Note that a lack of CO detection at $A_{V}$ $\lesssim$ 1 mag is not due to our sensitivity, 
considering that our mean 1$\sigma$ uncertainty of $A_{V}$ is $\sim$0.2 mag.
In addition, the threshold is not the result of the limited spatial coverage of the COMPLETE $I_{\rm CO}$ image. 
We made a comparison between our $A_{V}$ image
and the $I_{\rm CO}$ image with a large spatial area of $\sim$10$^{\circ}$ $\times$ 7$^{\circ}$ 
from the Center for Astrophysics CO Survey (CfA; Dame et al. 2001) at the common angular resolution of 8.4$'$
and found essentially the same threshold. 

Second, $I_{\rm CO}$ significantly increases from $\sim$0.1 K km s$^{-1}$ 
to $\sim$70 K km s$^{-1}$ for a narrow range of $A_{V}$ $\sim$ 1--3 mag.
This steep increase of $I_{\rm CO}$ may suggest that the transition from CII/CI to CO is sharp  
once shielding becomes sufficiently strong to prevent photodissociation (e.g., Taylor et al. 1993; Bell et al. 2006). 

Third, $I_{\rm CO}$ gradually increases and saturates to $\sim$50--80 K km s$^{-1}$ at $A_{V, \rm sat}$ $\gtrsim$ 3 mag.
This is consistent with Pineda et al. (2008), who found $A_{V,\rm sat}$ $\sim$ 4 mag for Perseus.
Similarly, Lombardi et al. (2006) found the saturation of $I_{\rm CO}$ $\sim$ 30 K km s$^{-1}$ 
for the Pipe nebula at $A_{V, \rm sat}$ $\sim$ 6 mag (with their adopted relation $A_{V}$ = $A_{K}$/0.112).  
The saturation of $I_{\rm CO}$ is expected based on the relation between $I_{\rm CO}$ and $\tau$,
$I_{\rm CO}$ $\propto$ $1 - e^{-\tau}$, where $\tau$ $\propto$ $A_{V}$.
Therefore, $I_{\rm CO}$ does not faithfully trace $A_{V}$ once it becomes optically thick.
The presence of optically thick CO emission in Perseus was hinted by Pineda et al. (2008),  
who performed the curve of growth analysis for the CO and $^{13}$CO($J$ = 1 $\rightarrow$ 0) observations.  
 
\subsubsection{Individual Regions} 

In agreement with Pineda et al. (2008),
we find that the relation between $I_{\rm CO}$ and $A_{V}$ has significant region-to-region variations across Perseus,  
contributing to the large scatter in Figure \ref{f:CO_int_Av_plot_Perseus}. 
We therefore show $I_{\rm CO}$ vs $A_{V}$ for each dark and star-forming region in Figure \ref{f:CO_int_Av_plot}. 
To emphasize the steep increase and saturation of $I_{\rm CO}$,
we plot $I_{\rm CO}$ as a function of $A_{V}$ on a log-log scale. 

Among the five regions, B5 and L1448 have the narrowest range of $A_{V}$ $\sim$ 1--3 mag, 
simply reflecting their smaller $N$(H) range on average. 
On the other hand, IC348 has the largest ranges of $A_{V}$ $\sim$ 1--11 mag and $I_{\rm CO}$ $\sim$ 0.2--50 K km s$^{-1}$. 
$I_{\rm CO}$ steeply increases from $\sim$0.2 K km s$^{-1}$ to $\sim$35 K km s$^{-1}$ at $A_{V}$ $\sim$ 1--3 mag 
and then saturates to $\sim$50 K km s$^{-1}$ at $A_{V}$ $\gtrsim$ 3 mag. 
In the case of B1E/B1, two components are apparent. 
The first component corresponds to the relatively steep increase of $I_{\rm CO}$ 
from $\sim$1 K km s$^{-1}$ to $\sim$20 K km s$^{-1}$ at $A_{V}$ $\sim$ 1.5--3 mag. 
The second component corresponds to the gradual increase of $I_{\rm CO}$ 
from $\sim$20 K km s$^{-1}$ to $\sim$60 K km s$^{-1}$ at $A_{V}$ $\sim$ 1.5--5 mag. 
Considering the two components together, $I_{\rm CO}$ saturates to $\sim$60 K km s$^{-1}$ at $A_{V}$ $\gtrsim$ 3 mag. 
Lastly, NGC1333 has the majority of the data points ($\sim$90\%) at $A_{V}$ $\lesssim$ 3 mag.
$I_{\rm CO}$ increases from $\sim$0.5 K km s$^{-1}$ to $\sim$70 K km s$^{-1}$ at $A_{V}$ $\sim$ 1--3 mag
and then shows a hint of the saturation to $\sim$80 K km s$^{-1}$ at $A_{V}$ $\gtrsim$ 3 mag.
Note that NGC1333 is the region where $I_{\rm CO}$ saturates to the largest value in Perseus.

In summary, the most important properties we find from the individual regions are 
the abrupt increase of $I_{\rm CO}$ at $A_{V}$ $\lesssim$ 3 mag 
and the saturation of $I_{\rm CO}$ at $A_{V}$ $\gtrsim$ 3 mag.
However, $I_{\rm CO}$ saturates to different values,  
from $\sim$50 K km s$^{-1}$ for IC348 to $\sim$80 K km s$^{-1}$ for NGC1333.

\subsection{$\boldsymbol{X_{\rm CO}}$ versus $\boldsymbol{A_{V}}$}


Our derived spatial distribution of $X_{\rm CO}$ allows us to test interesting theoretical predictions 
such as the dependence of $X_{\rm CO}$ on $A_{V}$.
In Figure \ref{f:X_CO_Av}, we plot $X_{\rm CO}$ as a function of $A_{V}$ for each dark and star-forming region.  
While B5 and L1448 do not show a clear relation between $X_{\rm CO}$ and $A_{V}$ due to their narrow range of $A_{V}$,
IC348 has a distinct trend of $X_{\rm CO}$ decreasing at small $A_{V}$ and increasing at large $A_{V}$.
$X_{\rm CO}$ decreases by a factor of $\sim$70 at $A_{V}$ $\sim$ 1--2.5 mag 
and increases by only a factor of $\sim$4 at $A_{V}$ $\sim$ 2.5--11 mag. 
In the case of B1E/B1, there appears to be two components.
The majority of the data points show a linear increase of $X_{\rm CO}$
from $\sim$7 $\times$ 10$^{18}$ to $\sim$5 $\times$ 10$^{19}$ for $A_{V}$ $\sim$ 1.5--5 mag.
The additional group of the data points is located at $A_{V}$ $\sim$ 2--3 mag
and $X_{\rm CO}$ $\sim$ 10$^{20}$ with some scatter.
Finally, NGC1333 has the majority of the data points ($\sim$83\%) at $A_{V}$ $\lesssim$ 3 mag 
and $X_{\rm CO}$ $\lesssim$ 5 $\times$ 10$^{19}$ with a large degree of scatter (a factor of $\sim$10).
At $A_{V}$ $\sim$ 3--10 mag, $X_{\rm CO}$ increases by only a factor of $\sim$4.
Overall, we find a factor of up to $\sim$100 variations in $X_{\rm CO}$ for IC348, B1E/B1, and NGC1333 
with a size of $\sim$6--7 pc. 

We notice that the shape of the $X_{\rm CO}$ vs $A_{V}$ profiles is primarily driven by how $I_{\rm CO}$ changes with $A_{V}$.
Specifically, decreasing $X_{\rm CO}$ with $A_{V}$ results from the steep increase of $I_{\rm CO}$ at small $A_{V}$,
while increasing $X_{\rm CO}$ with $A_{V}$ is due to the saturation of $I_{\rm CO}$ at large $A_{V}$.
$X_{\rm CO}$ decreases because $I_{\rm CO}$ increases more steeply than $N$(H$_{2}$)
likely due to the sharp transition from CII/CI to CO.
On the other hand, $X_{\rm CO}$ increases because $I_{\rm CO}$ increases gradually compared to $N$(H$_{2}$) 
likely due to the saturation of $I_{\rm CO}$ resulted from the large optical depth. 
Therefore, the transition from decreasing to increasing $X_{\rm CO}$ occurs in the $X_{\rm CO}$ vs $A_{V}$ profile
where the CO emission becomes optically thick.
This is particulary prominent for IC348, where this transition occurs at $A_{V}$ $\sim$ 3 mag.
B1E/B1 is relatively similar with IC348, while we do not observe a clear indication of this transition for NGC1333.  
Several theoretical studies have predicted the similar shape for the $X_{\rm CO}$ vs $A_{V}$ profile
(e.g., Taylor et al. 1993; Bell et al. 2006; Glover \& Mac Low 2011; Shetty et al. 2011a; Feldmann et al. 2012a).
In the next sections, we compare our $X_{\rm CO}$ data with predictions from two models.  

\section{$\boldsymbol{X_{\rm CO}}$: Comparison between Observations and Theory} 

\subsection{Microturbulent Time-independent Model} 

\subsubsection{Summary of the Modified W10 Model}

We use a modified form of the model in Wolfire et al. (2010) to calculate H$_{2}$ and CO abundances and CO line emission. 
The model in Wolfire et al. (2010) uses a plane-parallel PDR code with one-sided illumination 
to estimate the distributions of atomic and molecular species as a function of $A_{V}$ into a cloud. 
The density distribution is taken to be the median density as expected from turbulence 
and the distribution is converted into a spherical geometry. 
In our modified W10 model, a plane-parallel slab of gas is illuminated by UV photons on two sides 
and has either a uniform density distribution or a distribution described with a simple step function. 
The gas temperature and the abundances of atomic and molecular species are calculated as a function of $A_{V}$ 
under the assumptions of thermal balance and chemical equilibrium.
For details on the chemical and thermal processes, 
we refer the reader to Tielens \& Hollenbach (1985), 
Kaufman et al. (2006), Wolfire et al. (2010), and Hollenbach et al. (2012). 

The input parameters for the modified W10 model are $n$, $G$, $\zeta$, $v_{\rm turb}$, $Z$, and DGR. 
Considering the constraints on the physical parameters of Perseus (Section 2.2),
we use a set of the modified W10 models with the following inputs: 
$G$ = 0.5 $G_{0}'$, $\zeta$ = 10$^{-16}$ s$^{-1}$, $v_{\rm turb}$ = 4 km s$^{-1}$, 
$Z$ = 1 Z$_{\odot}$, and DGR = 1 $\times$ 10$^{-21}$ mag cm$^{2}$.
For the density distribution, we use both a uniform density distribution with $n$ = 10$^{3}$, 5 $\times$ 10$^{3}$, and 10$^{4}$ cm$^{-3}$ 
and a ``core-halo'' density distribution. 
The ``halo'' consists of HI with a fixed density $n_{\rm halo}$ = 40 cm$^{-3}$, 
comparable to diffuse cold neutral medium (CNM) clouds (Wolfire et al. 2003), 
and has $N$(HI) = 4.5 $\times$ 10$^{20}$ cm$^{-2}$ on each side of the slab.
In the ``core'', on the other hand, $n$ abruptly increases to a large density  
$n_{\rm core}$ = 10$^{3}$, 5 $\times$ 10$^{3}$, or 10$^{4}$ cm$^{-3}$.
This ``core-halo'' structure is motivated by observations of molecular clouds 
that have found HI envelopes with $N$(HI) $\sim$ 10$^{21}$ cm$^{-2}$ (e.g., Imara \& Blitz 2011; Imara, Bigiel, \& Blitz 2011; Lee et al. 2012).
As the minimum density of the densest regions for both the uniform and ``core-halo'' density distributions 
($\sim$10$^{3}$ cm$^{-3}$) has already been constrained by previous comparisons between CO observations and LVG/PDR models (Section 2.2), 
we expect that the modified W10 model with density much smaller 
than 10$^{3}$ cm$^{-3}$ would not reproduce the observed $I_{\rm CO}$ in Perseus 
and therefore do not demonstrate the effect of $n$, $n_{\rm core}$ < 10$^{3}$ cm$^{-3}$ in this paper. 
In addition, we note that the modified W10 model is not sensitive to the exact value of $n_{\rm halo}$, 
as long as this is small enough to contain a small amount of H$_{2}$ and CO in the halo (Section 7.1.2 for details). 

We run the model for $A_{V}$ = 0.6, 0.8, 1, 1.5, 2, 2.8, 4.8, 7.2, 10 mag (uniform density)
and $A_{V}$ = 1.25, 1.3, 1.5, 1.7, 2, 2.8, 4.8, 7.2, 10 mag (``core-halo'') 
and the output quantities are $N$(HI), $N$(H$_{2}$), and $I_{\rm CO}$ for a given $A_{V}$.
We summarize the ranges of the output quantities in Tables 2 (``core-halo'') and 3 (uniform density).
Note that for both the uniform and ``core-halo'' density distributions 
an increase in $A_{V}$ can be thought of as an increase in size of the dense region.
For example, 
$A_{V}$ = DGR $\times$ $N$(H) = DGR$(n_{\rm core}$$L_{\rm core}$ + $n_{\rm halo}$$L_{\rm halo})$ = 
3.1 $\times$ 10$^{-3}$ $L_{\rm core}$$n_{\rm core}$ + 0.9 mag,  
with $L_{\rm core}$ in units of pc and $n_{\rm core}$ in units of cm$^{-3}$ for the ``core-halo'' density distribution.
The ``core'' has a typical size $L_{\rm core}$ $\lesssim$ 1 pc, 
while the ``halo'' is significantly more extended with $L_{\rm halo}$ $\sim$ 7 pc.
For the uniform density distribution, the size of the slab is generally $L_{\rm uniform}$ $\lesssim$ 1 pc. 
We note that in the most extreme case the size of the dense region is much smaller than our spatial resolution 
($L_{\rm core}$ $\sim$ 0.01 pc), 
implying a considerably small filling factor of the ``core'' relative to the ``halo'', 
but comparable to the size of small-scale clumps observed in the CO emission (e.g., Heithausen et al. 1998; Kramer et al. 1998). 


\subsubsection{Comparison with Observations: ``Core-halo'' Density Distribution} 

We compare $X_{\rm CO}$ vs $A_{V}$ with predictions from the modified W10 model (``core-halo'') in Figure \ref{f:X_CO_Av_W10}(a).
While B5 and L1448 probe too narrow ranges of $A_{V}$ for significant comparisons,
the model curves with $n_{\rm core}$ = 10$^{3-4}$ cm$^{-3}$ follow the observed trends for IC348 and B1E/B1.  
The situation is more complicated for NGC1333,  
where the model matches the observed $X_{\rm CO}$ only for a partial range of $A_{V}$ 
and has difficulties in reproducing the observations at $A_{V}$ $\lesssim$ 3 mag and $X_{\rm CO}$ $\lesssim$ 10$^{19}$.
In addition, NGC1333 lacks the decreasing portion of the $X_{\rm CO}$ vs $A_{V}$ profile  
because of the missing data points with small $I_{\rm CO}$ $\lesssim$ 10 K km s$^{-1}$.  
Here we provide a description of the detailed comparison between our data of IC348, B1E/B1, and NGC1333 and the modified W10 model.  

\begin{enumerate}

\item[(1)] For IC348, the model with $n_{\rm core}$ = 10$^{3}$ cm$^{-3}$ reproduces well the observed shape of the $X_{\rm CO}$ vs $A_{V}$ profile 
(decreasing $X_{\rm CO}$ at $A_{V}$ $\lesssim$ 3 mag and increasing $X_{\rm CO}$ at $A_{V}$ $\gtrsim$ 3 mag).

\item[(2)] For B1E/B1, the model with $n_{\rm core}$ varying from 10$^{3}$ cm$^{-3}$ to 10$^{4}$ cm$^{-3}$   
can reproduce the observed shape of the $X_{\rm CO}$ and $A_{V}$ profile. 

\item[(3)] For NGC1333, the observed scatter at small $A_{V}$ calls for a range of $n_{\rm core}$ $\sim$ 10$^{3-4}$ cm$^{-3}$. 
Considering that the models with $n_{\rm core}$ = 5 $\times$ 10$^{3}$ cm$^{-3}$ and 10$^{4}$ cm$^{-3}$ are essentially identical, however,  
the data points at $A_{V}$ $\lesssim$ 3 mag with $X_{\rm CO}$ $\lesssim$ 10$^{19}$ would not be reproduced  
by the model with $n_{\rm core}$ > 10$^{4}$ cm$^{-3}$. 
In addition, our observational data lack the decreasing portion of the $X_{\rm CO}$ vs $A_{V}$ profile. 
We suspect that this is due to the limited spatial coverage of the COMPLETE $I_{\rm CO}$ image, 
which does not adequately sample low column density regions for NGC1333 
(only $\sim$10\% of the data points have $I_{\rm CO}$ < 10 K km s$^{-1}$).

\end{enumerate}

In Figure \ref{f:X_CO_Av_W10}(b), 
we compare the observed $X_{\rm CO}$ vs $N$(H$_{2}$) profiles with the model and find similar results. 
In summary, the modified W10 model with the ``core-halo'' structure and the input parameters appropriate for Perseus 
predicts the ranges of $I_{\rm CO}$ and $N$(H$_{2}$) in good agreement with our data. 
IC348 and B1E/B1 are the best cases where the shape of the $X_{\rm CO}$ vs $A_{V}$ profiles 
and the location of the minimum $X_{\rm CO}$ are well described by the model. 
We note that there are some discrepancies at low column densities in NGC1333, 
where the data points with $X_{\rm CO}$ $\lesssim$ 10$^{19}$ are not reproduced by the model 
and at the same time the observed data with $X_{\rm CO}$ $\gtrsim$ 10$^{20}$ are missing 
due to the limited observational coverage.  



Next, we plot $N$(HI) as a function of $N$(H) in Figure \ref{f:X_CO_Av_W10}(c) 
and compare the profiles with the modified W10 model. 
As summarized in Section 2.1, Lee et al. (2012) found a relatively uniform $N$(HI) distribution across Perseus 
with $\sim$(8--10) $\times$ 10$^{20}$ cm$^{-2}$. 
Here we use the same $N$(HI) image as in Lee et al. (2012) 
and apply the same boundaries for the five regions as in Section 4. 
We find that the mean $N$(HI) varies from $\sim$7.4 $\times$ 10$^{20}$ cm$^{-2}$ (B5) 
to $\sim$9.6 $\times$ 10$^{20}$ cm$^{-2}$ (NGC1333 and L1448). 
The model predicts $N$(HI) $\sim$ (9--9.6) $\times$ 10$^{20}$ cm$^{-2}$, 
with essentially no difference between $n_{\rm core}=10^3$ cm$^{-3}$ and 10$^4$ cm$^{-3}$ models. 
The predicted $N$(HI) distribution with $\sim$9 $\times$ 10$^{20}$ cm$^{-2}$ and its uniformity 
are consistent with what we observe in Perseus. 
This agreement will persist even if the $N$(HI) distribution is corrected for high optical depth HI.
Our preliminary work on the effect of high optical depth HI 
that is missing in the HI emission image of Perseus shows that 
$N$(HI) increases by a factor of $\sim$1.5 at most due to the optical depth correction 
(the corrected $N$(HI) $\sim$ (8--18) $\times$ 10$^{20}$ cm$^{-2}$; Stanimirovi\'c et al. in prep). 
The ranges of the predicted $N$(HI) and $N$(H$_{2}$) distributions are comparable to what we find in Perseus.
In Figure \ref{f:X_CO_Av_W10}(d),
we plot $R_{\rm H2}$ against $N$(H) and indeed find that the model matches well our observations. 
In particular, the linearly increasing $R_{\rm H2}$ with $N$(H) is reproduced well by the model, 
mainly driven by the uniform $N$(HI) distribution.

\subsubsection{Comparison with Observations: Uniform Density Distribution} 

So far we made comparisons between the observations of Perseus and the modified W10 model with the ``core-halo'' structure. 
To investigate the role of the diffuse halo in determining H$_{2}$ and CO distributions, 
we show our data for IC348 and predictions from the modified W10 model both with the ``core-halo'' structure 
and the uniform density distribution in Figure \ref{f:W_compare}.
The uniform density distribution simply assumes a dense core 
with $n$ = 10$^{3}$, 5 $\times$ 10$^{3}$, or 10$^{4}$ cm$^{-3}$.
Clearly, the ``core-halo'' model describes our data better. 
For example, the uniform density model underestimates the $N$(HI) distribution 
compared to the observed one across the cloud.  
In addition, it predicts the decreasing portion of the $X_{\rm CO}$ vs $A_{V}$ profile shallower than our data, 
while reproducing the observed range of $X_{\rm CO}$ reasonably well.  
We compare the ``core-halo'' model with the uniform density model in detail as follows. 

$N$(HI) vs $N$(H): 
The uniform density model predicts $N$(HI) significantly smaller than what we measure across Perseus, 
$N$(HI) $\sim$ 9 $\times$ 10$^{20}$ cm$^{-2}$.
The discrepancy ranges from a factor of $\sim$10--20 for $n$ = 10$^{3}$ cm$^{-3}$ to a factor of $\sim$70--160 for $n$ = 10$^{4}$ cm$^{-3}$.
This large discrepancy results from the fact that H$_{2}$ self-shielding is so strong that almost all hydrogen is converted into H$_{2}$. 
On the other hand, the density of the halo is small enough that dust shielding is more important than H$_{2}$ self-shielding.
To provide the sufficient dust shielding for H$_{2}$ formation, 
the entire halo remains atomic with its initial $N$(HI) $\sim$ 9 $\times$ 10$^{20}$ cm$^{-2}$, 
resulting in the uniform $N$(HI) distribution.
We expect that if the density of the halo is significantly larger than the current $n_{\rm halo}$ = 40 cm$^{-3}$, 
the halo will no longer be purely atomic due to the increased H$_{2}$ self-shielding. 

$N$(H$_{2}$) vs $N$(H):  
All models predict the $N$(H$_{2}$) vs $N$(H) profile in good agreement with our data, 
even though the uniform density model slightly overestimates $N$(H$_{2}$) at small $N$(H). 
For example, the uniform density model with $n$ = 10$^{4}$ cm$^{-3}$ predicts 
$N$(H$_{2}$) = 9.96 $\times$ 10$^{20}$ cm$^{-2}$ at $N$(H) = 2 $\times$ 10$^{21}$ cm$^{-2}$,  
larger than our data by less than a factor of 2.
However, this discrepancy is significant at such small $N$(H)  
and results in the small amount of $N$(HI) $\lesssim$ 10$^{19}$ cm$^{-2}$. 
In addition, models with different densities predict essentially the same $N$(H$_{2}$) for a given $N$(H). 
All these results imply that neither density nor its distribution is critical for the H$_{2}$ abundance.   
Instead, $N$(H) primarily determines $N$(H$_{2}$). 

$R_{\rm H2}$ vs $N$(H): 
While the ``core-halo'' model reproduces both the range of $R_{\rm H2}$ and the linear increase of $R_{\rm H2}$ with $N$(H), 
the uniform density model overestimates $R_{\rm H2}$ for a given $N$(H) by up to a factor of $\sim$300.
This discrepancy mainly results from the significantly underestimated $N$(HI) in the uniform density model.

$I_{\rm CO}$ vs $N$(H$_{2}$): 
All models reproduce the observed $I_{\rm CO}$ vs $N$(H$_{2}$) profile reasonably well. 
In particular, both the ``core-halo'' and uniform density models with the smallest density 
show an excellent agreement with our data for IC348. 
While the models with $n$ $\gtrsim$ 5 $\times$ 10$^{3}$ cm$^{-3}$ and $n_{\rm core}$ $\gtrsim$ 5 $\times$ 10$^{3}$ cm$^{-3}$ predict 
larger $I_{\rm CO}$ at small $N$(H$_{2}$) (up to a factor of $\sim$10), 
the difference between the models with different densities becomes negligible at $N$(H$_{2}$) $\gtrsim$ 1 $\times$ 10$^{21}$ cm$^{-2}$, 
where $I_{\rm CO}$ saturates to $\sim$45--60 K km s$^{-1}$ for the uniform density model  
and $\sim$30--40 K km s$^{-1}$ for the ``core-halo'' model. 
All these results suggest that $I_{\rm CO}$ depends on density but only at small $N$(H$_{2}$) 
and changes in physical parameters other than density (e.g., $v_{\rm turb}$) will be required 
to produce larger $I_{\rm CO}$ values once $I_{\rm CO}$ becomes optically thick.

$I_{\rm CO}$ vs $A_{V}$: 
While the ``core-halo'' model reproduces the sharp increase of $I_{\rm CO}$ observed at $A_{V}$ $\gtrsim$ 1 mag, 
the uniform density model predicts the increase of $I_{\rm CO}$ at $A_{V}$ $\gtrsim$ 0.6 mag much more gradually than our data. 
This difference comes from the fact that the uniform density model has larger density than the ``core-halo'' model, 
resulting in the larger $I_{\rm CO}$ for a given $A_{V}$ $\lesssim$ 3 mag
($n$ $\gtrsim$ 10$^{3}$ cm$^{-3}$ for the uniform density model vs
$\langle n \rangle$ $\sim$ 55--125 cm$^{-3}$ for the ``core-halo'' model; Table 2).  
On the other hand, all models predict the saturation of $I_{\rm CO}$ to similar values at $A_{V}$ $\gtrsim$ 3 mag, 
suggesting that the larger density in the uniform density model 
no longer has a significant impact on $I_{\rm CO}$ due to the large optical depth of $I_{\rm CO}$
($n$ $\gtrsim$ 10$^{3}$ cm$^{-3}$ for the uniform density model vs
$\langle n \rangle$ $\sim$ 180--430 cm$^{-3}$ for the ``core-halo'' model; Table 2).  

$X_{\rm CO}$ vs $A_{V}$: 
All models reproduce the observed increase of $X_{\rm CO}$ at $A_{V}$ $\gtrsim$ 3 mag,  
because they predict both the range of $N$(H$_{2}$) and the saturation of $I_{\rm CO}$ comparable to our data.  
On the other hand, the uniform density model shows the decrease of $X_{\rm CO}$ at $A_{V}$ $\lesssim$ 3 mag much shallower than our data. 
This discrepancy mainly results from the less steep increase of $I_{\rm CO}$ predicted by the 
model at $A_{V}$ $\lesssim$ 3 mag.

Summary: 
While we do not perform a full parameter space search, 
our comparison between the ``core-halo'' and uniform density models is illustrative
and demonstrates that the diffuse halo is essential for reproducing the following observed properties:  
the uniform $N$(HI) distribution, the H$_{2}$-to-HI ratio for a given $N$(H), 
the sharp increase of $I_{\rm CO}$ and decrease of $X_{\rm CO}$ at 1 mag $\lesssim$ $A_{V}$ $\lesssim$ 3 mag.
Considering that the uniform density model predicts the $I_{\rm CO}$ distribution extended toward smaller $A_{V}$, 
while producing the $N$(H$_{2}$) distribution in reasonably good agreement with 
our data (Figures \ref{f:W_compare}d and j), 
we expect that the neglect of the diffuse halo will result in the underestimation of the size of ``CO-free'' H$_{2}$ envelope.

\subsection{Macroturbulent Time-dependent Model} 

\subsubsection{Summary of the S11 Model}

The S11 model is essentially comprised of two parts. 
The first part is a modified version of the \textsc{zeus--mp} MHD code (Stone \& Norman 1992; Norman 2000). 
Gas in a periodic box is set to have a uniform density distribution 
and is driven by a turbulent velocity field with uniform power 1 $\leq$ $k$ $\leq$ 2 where $k$ is the wavenumber.
In addition, the magnetic field has initially orientation parallel to the $z$-axis, with a strength of 1.95 $\mu$G.
To model the chemical evolution of the gas, Glover \& Mac Low (2007a,b), Glover et al. (2010), and Glover \& Clark (2012) 
updated the \textsc{zeus-mp} MHD code with chemical reactions of several atomic and molecular species.
The photodissociation of molecules by a radiation field is treated by  
the ``six-ray approximation'' method developed by Glover \& Mac Low (2007a).
The effect of self-gravity is not included.  
We refer to Glover \& Mac Low (2007a,b), Glover et al (2010), and Glover \& Clark (2012) 
for details on MHD, thermodynamics, and chemistry included in the S11 model.
The second part is a three-dimensional radiative transfer code \textsc{radmc--3d} (Dullemond et al. in prep)\footnote[8]{See 
http://www.ita.uni-heidelberg.de/$\sim$dullemond/software/radmc-3d/.}.
Once the simulated molecular cloud reaches a statistically steady state, 
\textsc{radmc--3d} is executed to model molecular line emission (e.g., CO).  
To solve the population levels of atomic and molecular species, 
\textsc{radmc--3d} implements the LVG method (Sobolev 1957), 
which has been shown to be a good approximation for molecular clouds (e.g., Ossenkopf 1997).   
We refer to Shetty et al. (2011a) for details on \textsc{radmc--3d}. 

The MHD simulation follows the evolution of an initially atomic gas in a (20 pc)$^{3}$ box with a numerical resolution of 512$^3$. 
In this paper, we use the S11 model with the following input parameters:
initial $n$ = 100 cm$^{-3}$, 
$G$ = 1 $G_{0}'$, $\zeta$ = 10$^{-17}$ s$^{-1}$, $Z$ = 1 Z$_{\odot}$, and DGR = 5.3 $\times$ 10$^{-22}$ mag cm$^{2}$.  
This simulation is essentially the same as the ``n100 model'' in S11 
but has a higher numerical resolution and a simpler CO formation model based on Nelson \& Langer (1999).
We choose this particular simulation because it has a mass of $\sim$2 $\times$ 10$^{4}$ M$_{\odot}$, consistent with that of Perseus. 
The input parameters for the S11 model are reasonably close to what we expect for Perseus
but not exactly the same as what we used for the modified W10 model.
As it has been shown in S11 and Glover \& Mac Low (2007b) that the simulated H$_2$ and CO column densities 
do not depend on small changes in $G$ and $\zeta$, 
this simulation would be appropriate for the comparison with our observations (Section 8.4.1 for details). 

Compared to the modified W10 model, the final density distribution in the S11 model 
has a majority of the data points ($\sim$99\%) with $n$ < 10$^{3}$ cm$^{-3}$,
resulting in the small median density of $\sim$30 cm$^{-3}$. 
Another important difference between the modified W10 model and the S11 model is that 
H$_{2}$ formation in the S11 model does not achieve chemical equilibrium until the end of the simulation. 
For example, Glover et al. (2010) found from their MHD simulations 
that the H$_{2}$ abundance primarily depends on the time available for H$_{2}$ formation   
and shows no indication of chemical equilibrium up to $t$ $\sim$ 20 Myr.
The gas will eventually become fully H$_{2}$ unless the molecular cloud is destroyed by stellar feedback 
such as photoevaporation by HII regions and protostellar outflows.
On the other hand, the CO abundance is controlled by photodissociation and reaches chemical equilibrium within $t$ $\sim$ 2 Myr.

The final products of the S11 model include the $N$(HI), $N$(H$_{2}$), and $I_{\rm CO}$ images obtained at $t$ $\sim$ 5.7 Myr. 
We smooth and regrid the simulated $N$(HI), $N$(H$_{2}$), and $I_{\rm CO}$ images 
so that they have both a spatial resolution of 0.4 pc and a pixel size of 0.4 pc.
Recently, Beaumont et al. (2013) compared the COMPLETE data of Perseus with the S11 model 
and found that the S11 model systematically overestimates $N$(H$_{2}$) (e.g., Figure 5 of Beaumont et al. 2013). 
One of the possible explanations for this discrepancy is the different size between the simulation box and the individual regions in Perseus. 
Because the simulation box is larger than the individual regions in Perseus (20 pc vs $\sim$5--7 pc),
the integrated quantities $N$(HI), $N$(H$_{2}$), and $I_{\rm CO}$ would need to be scaled. 
In the case of $N$(HI) and $N$(H$_{2}$), 
the scaling is straightforward under the assumption of isotropic density distribution, 
which is appropriate for the S11 model\footnote[9]{We found that the assumption of isotropic density distribution is reasonable.
For the optimal line of sight depth of 7 pc that minimizes the difference between our data and the S11 model, 
we derived three different versions of $N$(H) image by integrating the simulated number density cube for 7 pc but with three different intervals. 
These images were then compared with the image we derived by multiplying the original $N$(H) image from the S11 model by 7/20. 
The histograms of all four $N$(H) images were very similar with each other.}, 
and we simply need to account for the difference between the box and region sizes.
However, estimating a proper scaling for $I_{\rm CO}$ is much more complicated because of the following reasons. 
First, the $I_{\rm CO}$ image was produced from the S11 model by integrating the CO brightness temperature, 
which was estimated by three-dimensional radiative transfer calculations, along a full radial velocity range.
Second, the CO emission is optically thick in some parts of the simulation ($\sim$10\% of the volume). 
Re-running the simulation with a smaller box does not solve the problem
as molecular cores/clouds form out of initially larger-scale diffuse ISM.
We therefore take an approach of determining the optimal line of sight (LoS) depth 
that minimizes the difference between our observations and the S11 model  
by investigating the $N$(HI) and $N$(H$_{2}$) images simultaneously. 
For the simulated $I_{\rm CO}$ image, on the other hand, we do not apply any scaling. 

To do this, we estimate the difference between the observed mean and the simulated mean 
for each of $N$(HI) and $N$(H$_{2}$) with varying LoS depths.
For example, we divide the simulated $N$(HI) and $N$(H$_{2}$) images by two 
to calculate the mean $N$(HI) and $N$(H$_{2}$) for the simulation with the LoS depth of 10 pc.
We then normalize the difference by the observed mean of each quantity 
and calculate the sum of the two normalized differences in quadrature.
The results are shown in Figure \ref{f:box_size} 
and we find that the LoS depth that minimizes the difference between our data and the simulation products is 7 pc (Figure \ref{f:box_size}c). 
While the final quantity in Figure \ref{f:box_size}(c) has a broad minimum,
it is encouraging that the estimated scale length is comparable to both the characteristic size of the five regions in Perseus  
and the total size of the slab for the ``core-halo'' model (Tables 1 and 2).
As a double check that this scale length is appropriate, 
we use Larson's law established for turbulent molecular clouds from both observations and MHD simulations:
$\sigma_{\rm CO}=(0.96 \pm 0.17) L_{\rm pc}^{0.59 \pm 0.07}$ km s$^{-1}$ (Heyer \& Brunt 2004).
For a region size of 20 pc we expect $\sigma_{\rm CO}$ $\sim$ 6 km s$^{-1}$, 
while for a region size of 7 pc we expect $\sigma_{\rm CO}$ $\sim$ 3 km s$^{-1}$. 
This level of CO velocity dispersion is in agreement with what is shown in Figure \ref{f:CO_vel_dispersion},
confirming that scaling the simulation products to the LoS depth of 7 pc is reasonable.
In summary, when we compare our observations with the S11 model,
we scale
the simulated $N$(HI), $N$(H$_{2}$), and $N$(H) images by multiplying them by 7/20 
(Figures \ref{f:histo_Perseus_s11}a, b, c, and Figures \ref{f:HI_tot_S11}a, b). 
On the other hand, because of the uncertainty in $I_{\rm CO}$ scaling, 
we use the original $I_{\rm CO}$ image produced by the S11 model 
(Figure \ref{f:histo_Perseus_s11}d and Figures \ref{f:HI_tot_S11}c, d).   

Finally, we apply the following thresholds to the simulated data 
to mimic the sensitivity limits of our observational data: 
$N$(H$_{2}$) > 3.3 $\times$ 10$^{19}$ cm$^{-2}$ and $I_{\rm CO}$ > 0.09 K km s$^{-1}$
(our mean 1$\sigma$ uncertainties calculated for the data points with $N$(H$_{2}$) > 0 cm$^{-2}$ and $I_{\rm CO}$ > 0 K km s$^{-1}$). 
This application of the thresholds to the S11 model is reasonable, 
considering the minimum $N$(H$_{2}$) $\sim$ 3.8 $\times$ 10$^{19}$ cm$^{-2}$ and $I_{\rm CO}$ $\sim$ 0.2 K km s$^{-1}$ 
for the five regions in Perseus. 



\subsubsection{Comparison with Observations: Global Properties} 

\begin{figure}
\begin{center}
\includegraphics[scale=0.22]{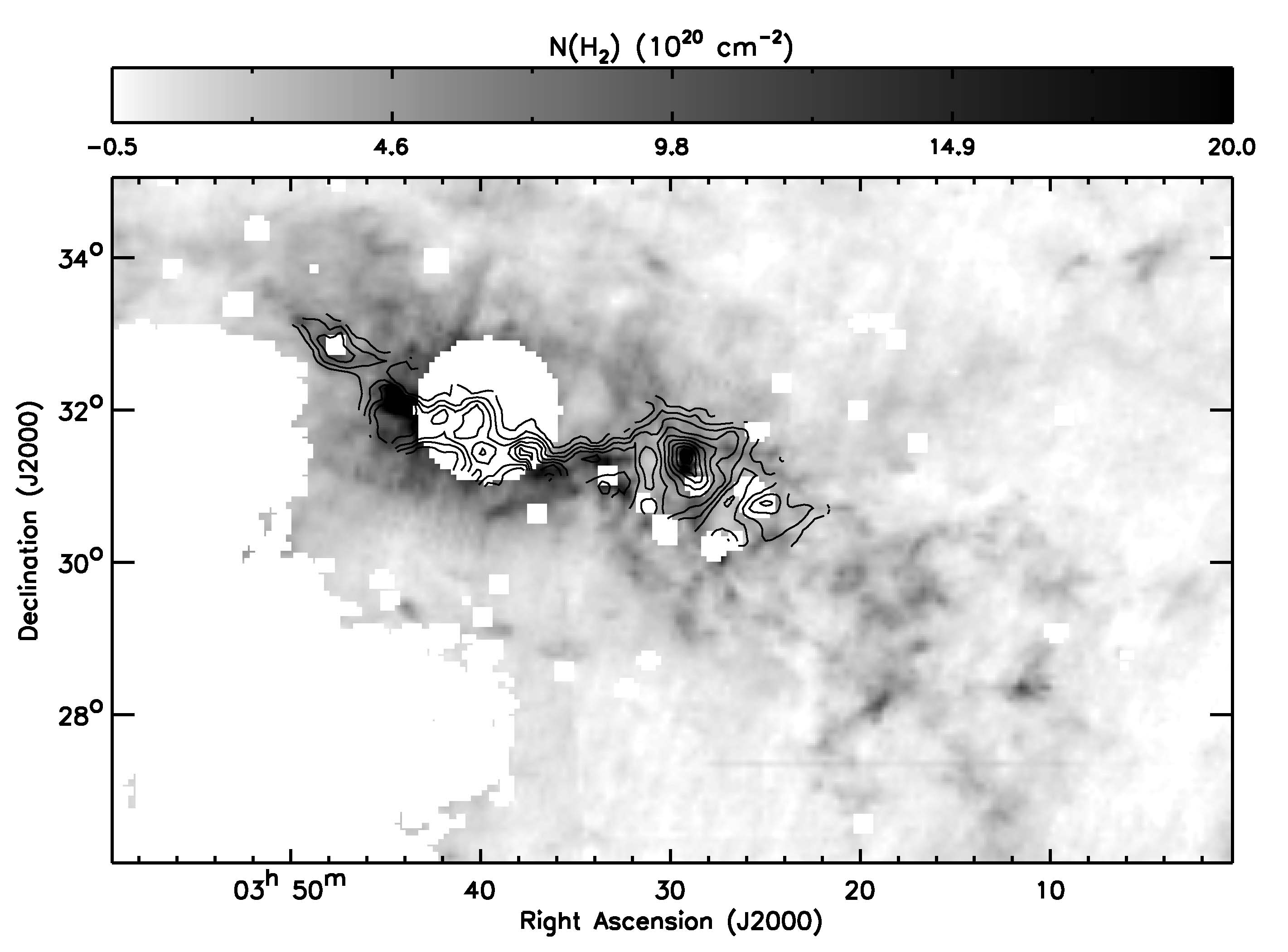}
\caption{\label{f:H2_cden_display} $N$(H$_{2}$) image at 4.3$'$ angular resolution derived by Lee et al. (2012).
The COMPLETE $I_{\rm CO}$ contours are overlaid in black
and their levels range from 10\% to 90\% of the peak (80 K km s$^{-1}$) with 10\% steps. 
The blank data points correspond to point sources and regions with possible contamination
(the Taurus molecular cloud and a background HII region).
See Sections 4.2 and 4.3 of Lee et al. (2012) for details.}
\end{center}
\end{figure}
\begin{figure} 
\begin{center} 
\includegraphics[scale=0.4]{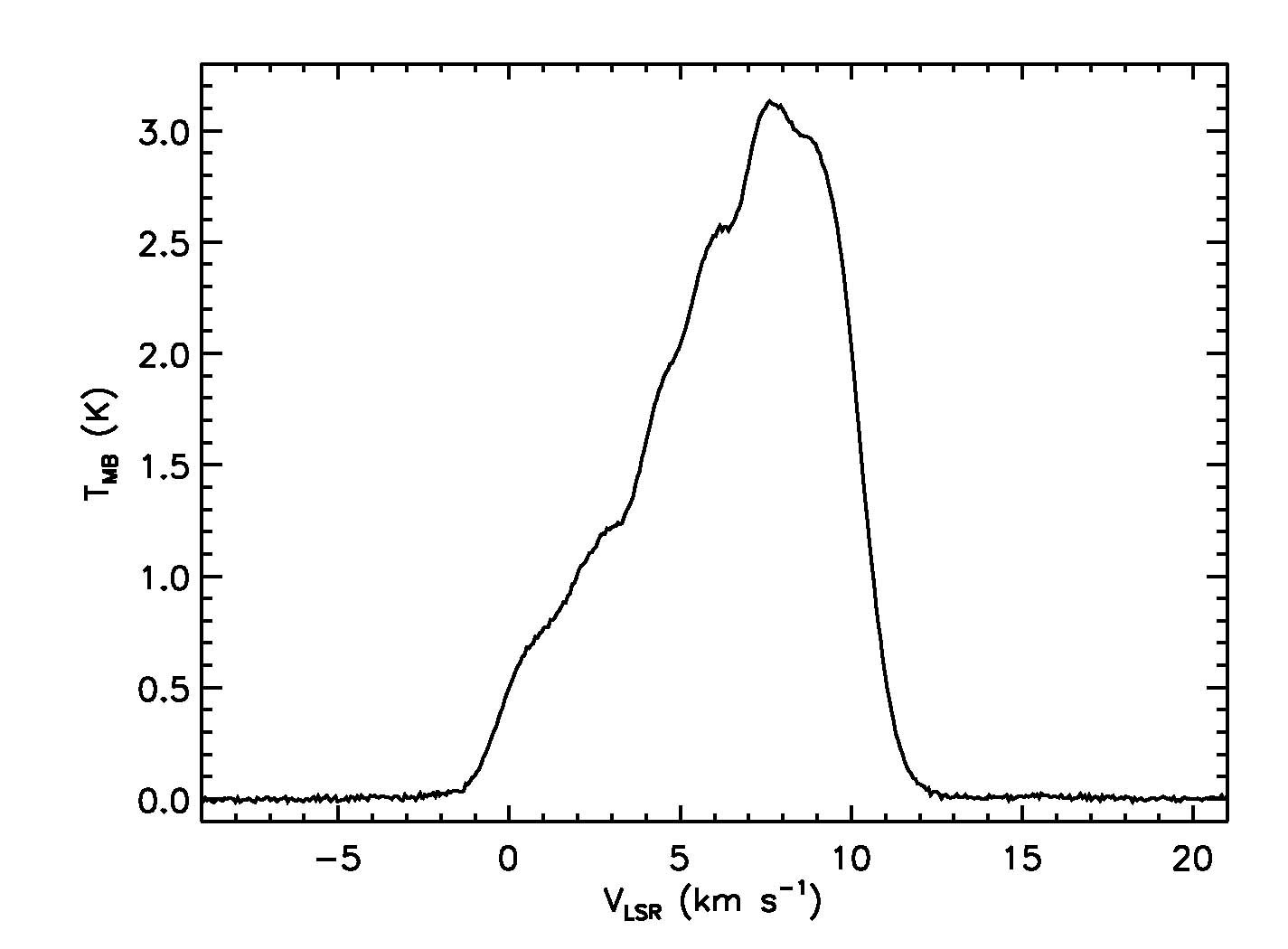}
\caption{\label{f:CO_spectrum_Perseus} CO spectrum obtained by averaging the COMPLETE CO spectra of all data points 
where the ratio of the peak main-beam brightness temperature to the rms noise is greater than 3. 
Note that the CO emission shows multiple velocity components.} 
\end{center}
\end{figure}
\begin{figure} 
\begin{center} 
\includegraphics[scale=0.45]{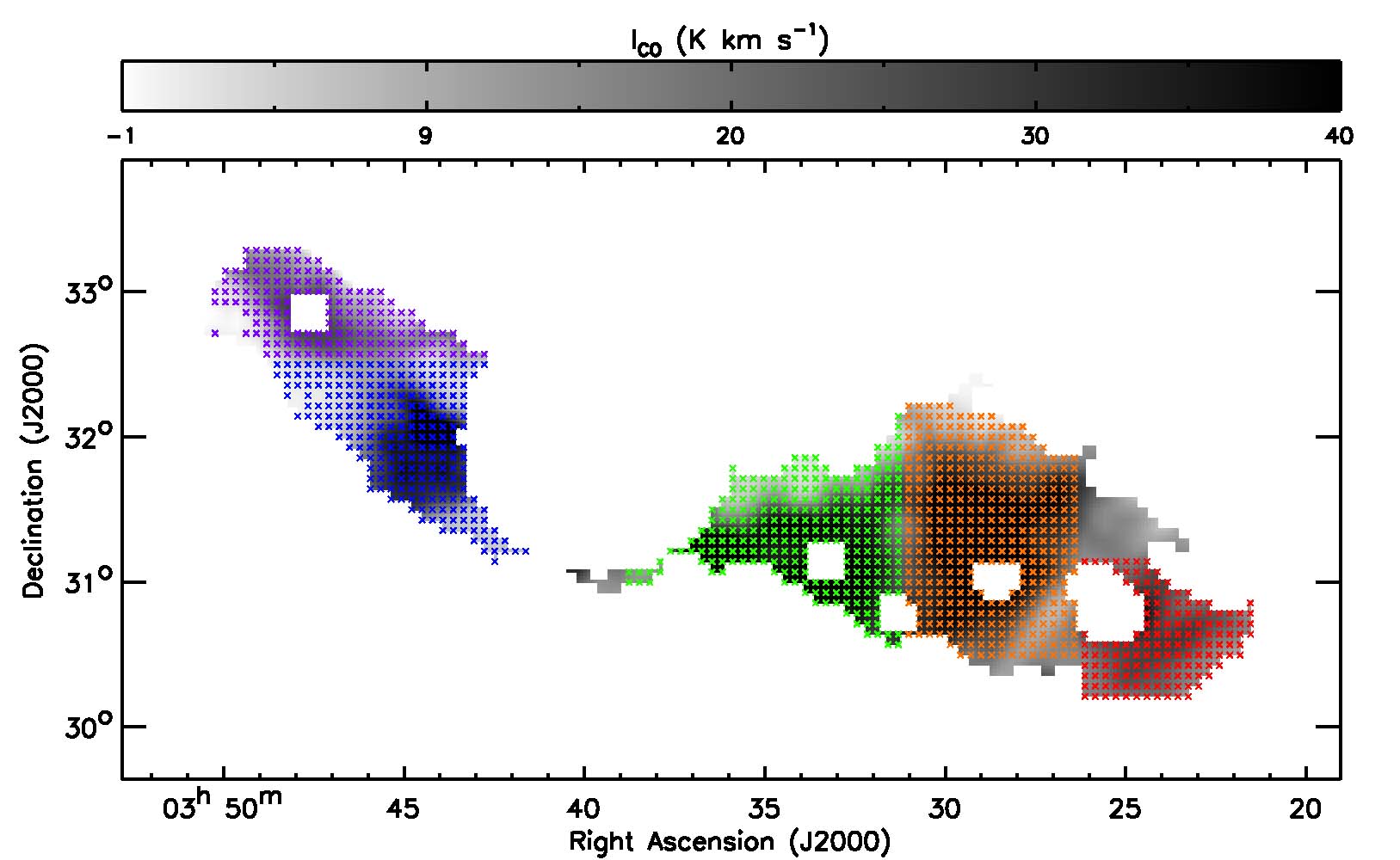}
\caption{\label{f:COMPLETE_five_regions} Each of the five regions is overlaid on 
the COMPLETE $I_{\rm CO}$ image in different color. 
B5 is purple, IC348 is blue, B1E/B1 is green, NGC1333 is orange, and L1448 is red. 
See Section 4 for details on how we determined these regions.}
\end{center}
\end{figure}
\begin{figure}
\begin{center}
\includegraphics[scale=0.48]{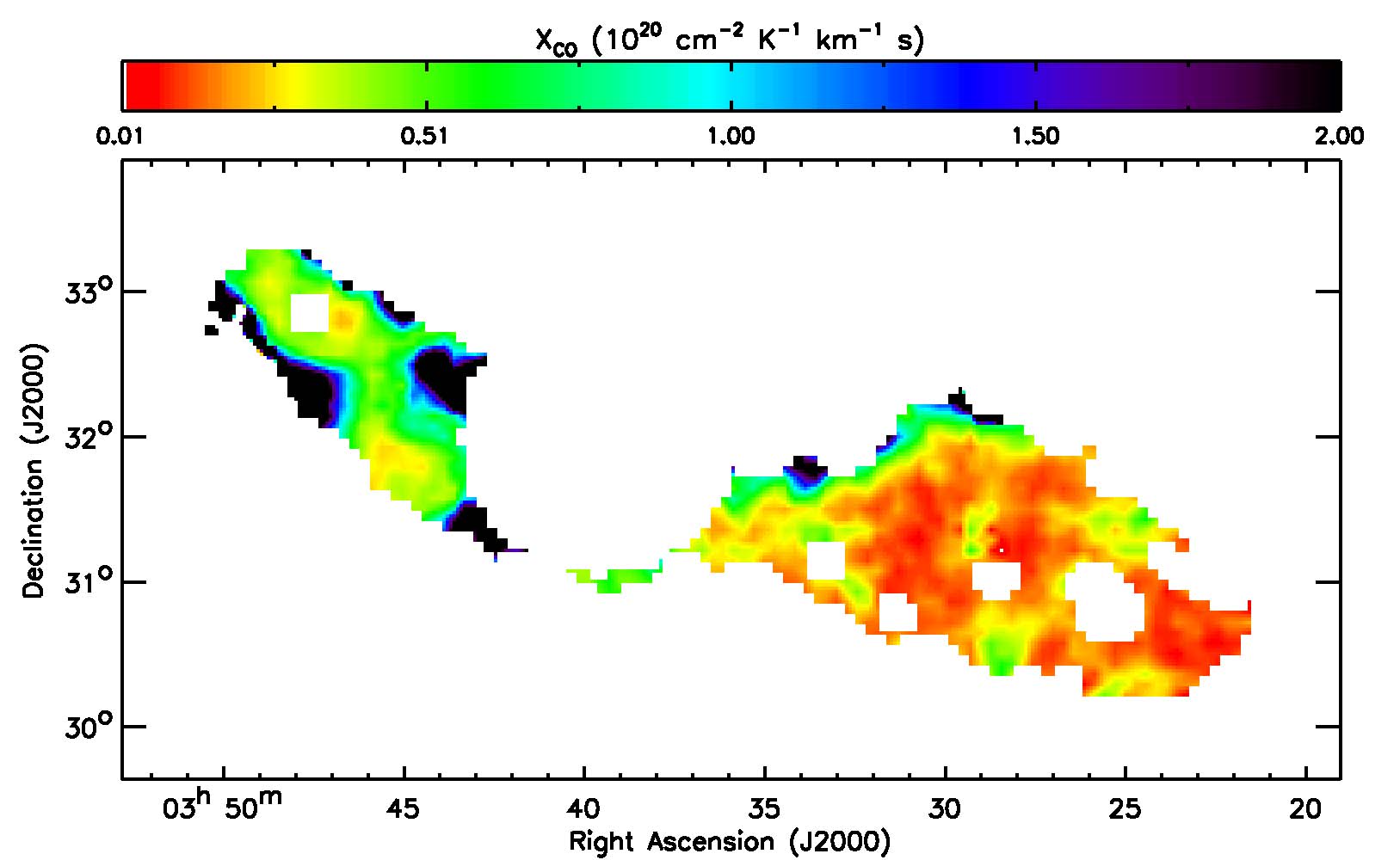}
\caption{\label{f:COMPLETE_X_CO} $X_{\rm CO}$ image at 4.3$'$ angular resolution.}
\end{center}
\end{figure}
\begin{figure}
\begin{center}
\includegraphics[scale=0.48]{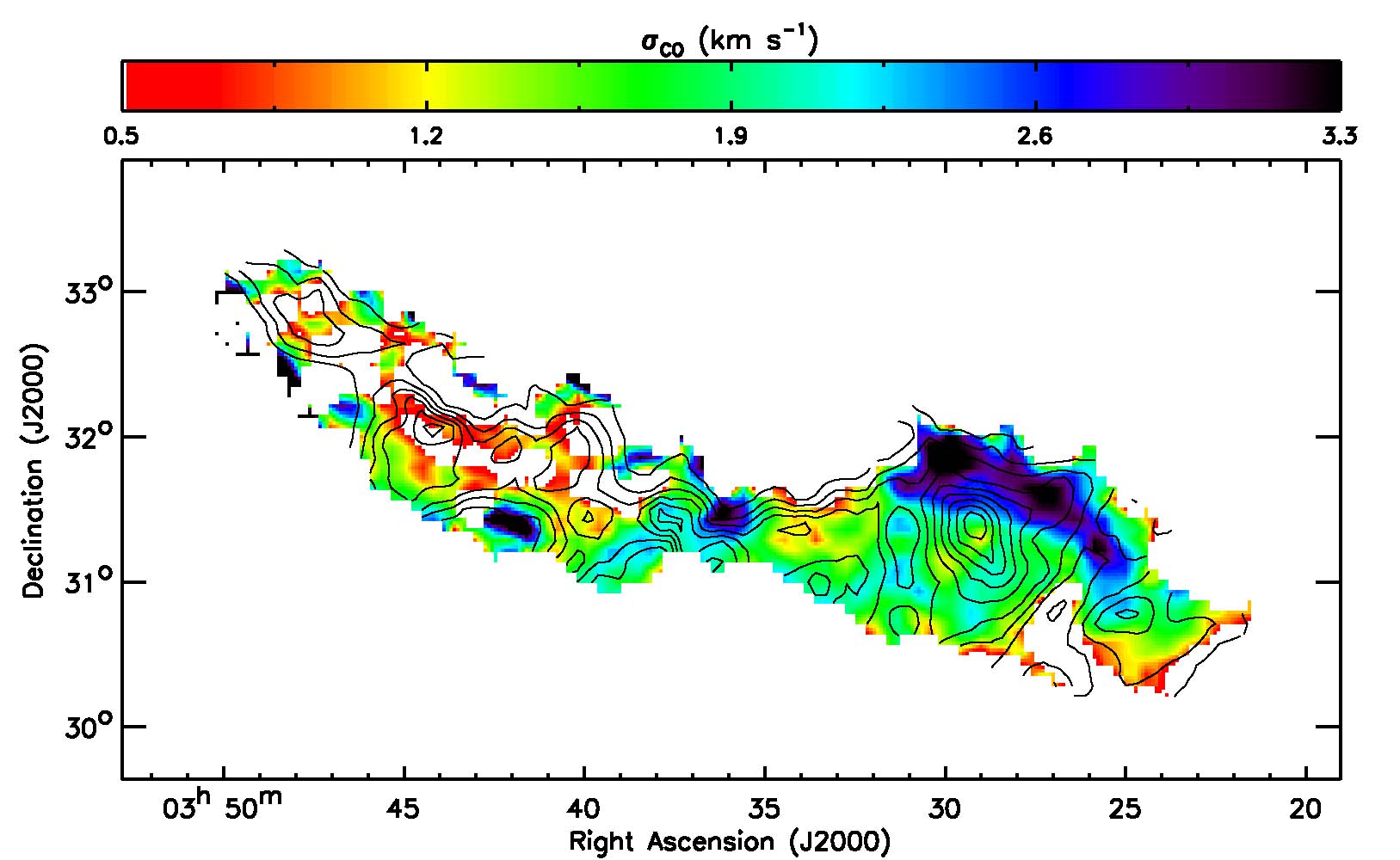}
\caption{\label{f:CO_vel_dispersion} $\sigma_{\rm CO}$ image overlaid with the COMPLETE $I_{\rm CO}$ contours.
The contour levels range from 10\% to 90\% of the peak (80 K km s$^{-1}$) with 10\% steps.}
\end{center}
\end{figure}
\begin{figure}
\begin{center}
\includegraphics[scale=0.5]{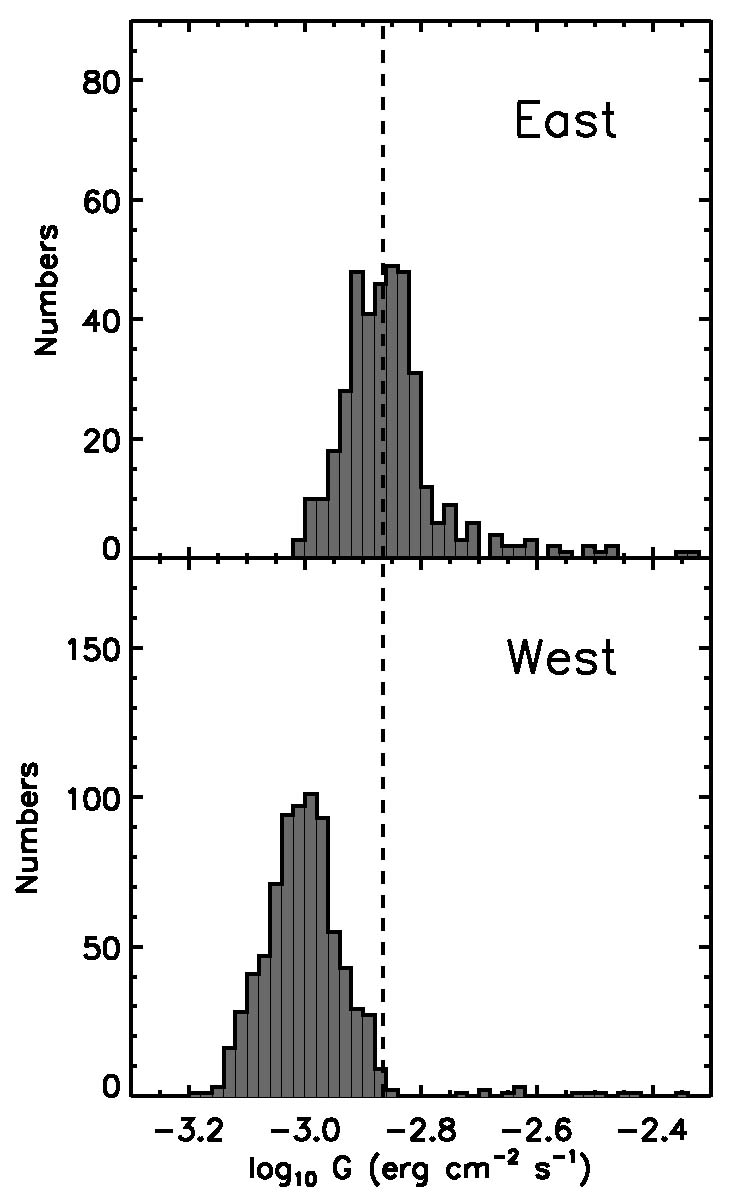}
\caption{\label{f:ISRF_histo} Histograms of $G$.
B5/IC348 and B1E/B1/NGC1333/L1448 are combined to produce the histograms of \textit{East} and \textit{West}. 
The median $G$ of \textit{East} ($\sim$10$^{-2.86}$ erg cm$^{-2}$ s$^{-1}$) is shown as a dashed line.}
\end{center}
\end{figure}
\begin{figure}
\begin{center}
\includegraphics[scale=0.45]{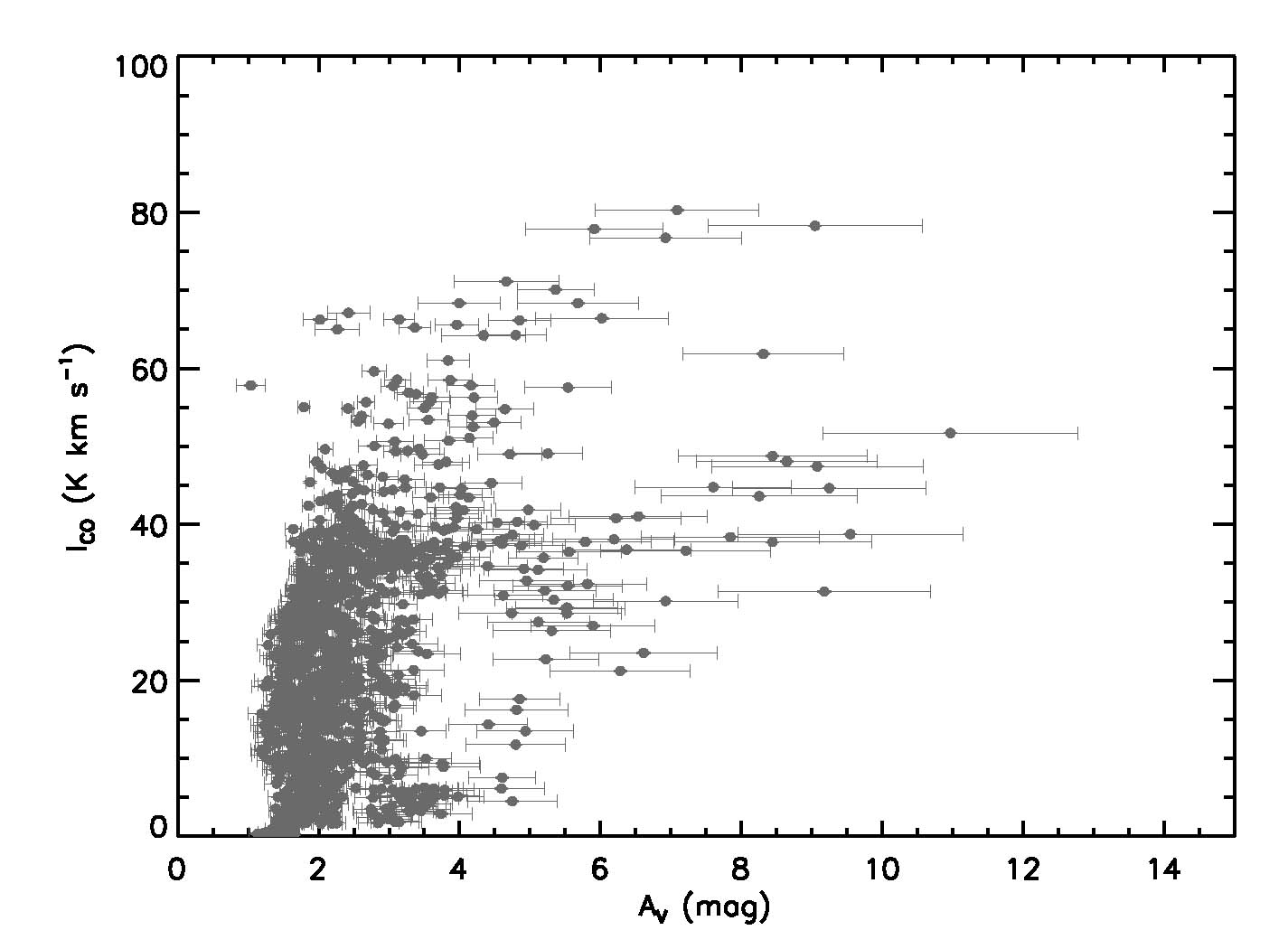}
\caption{\label{f:CO_int_Av_plot_Perseus} COMPLETE $I_{\rm CO}$ as a function of $A_{V}$ for all five regions defined in Section 4.} 
\end{center}
\end{figure}
\begin{figure}
\begin{center}
\includegraphics[scale=0.43]{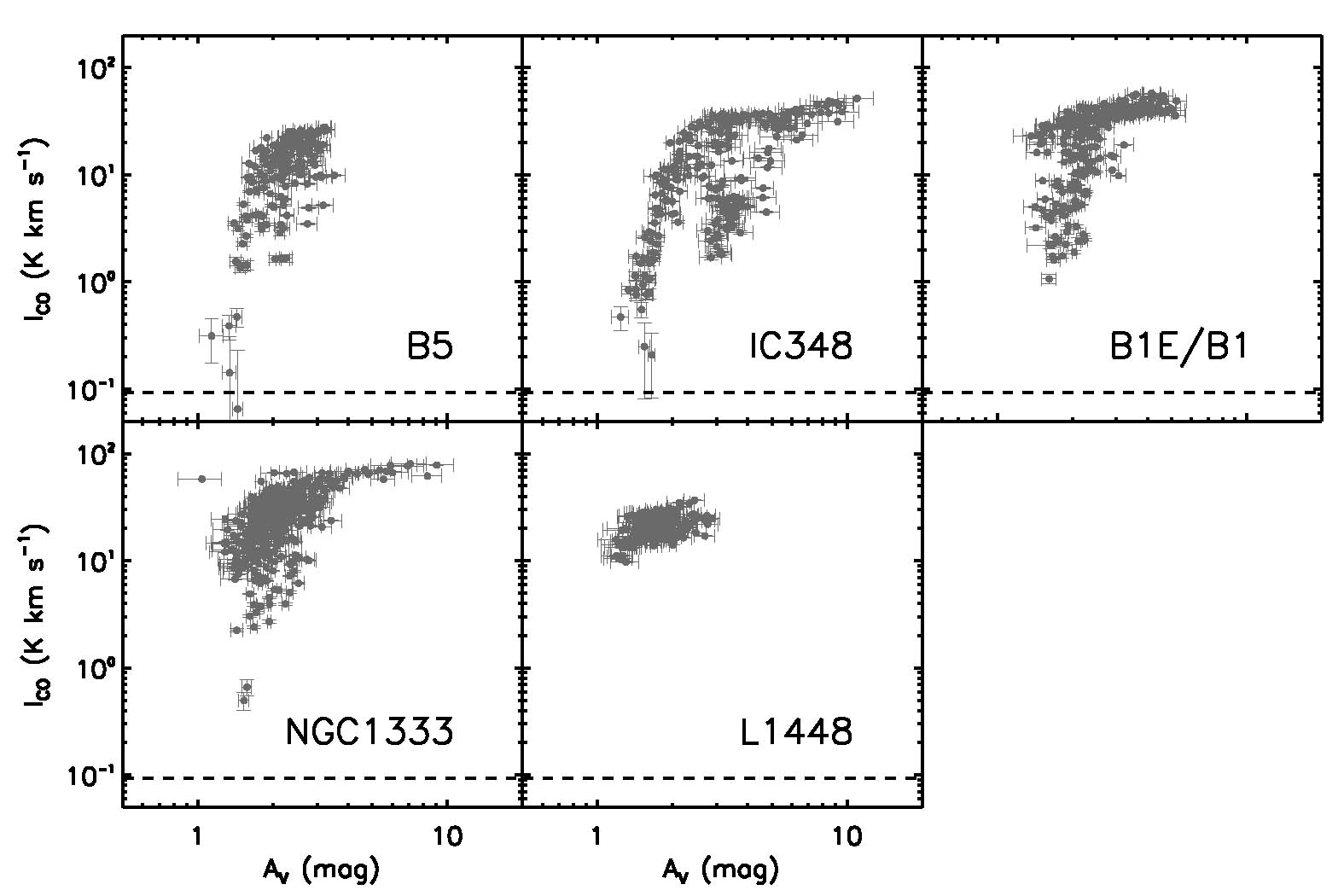}
\caption{\label{f:CO_int_Av_plot} COMPLETE $I_{\rm CO}$ as a function of $A_{V}$ for each dark and star-forming region.
The mean 1$\sigma$ uncertainty of $I_{\rm CO}$ ($\sim$0.09 K km s$^{-1}$) is shown as a dashed line, 
while that of $A_{V}$ ($\sim$0.2 mag) is too small to be shown.}
\end{center}
\end{figure}
\begin{figure} 
\begin{center} 
\includegraphics[scale=0.43]{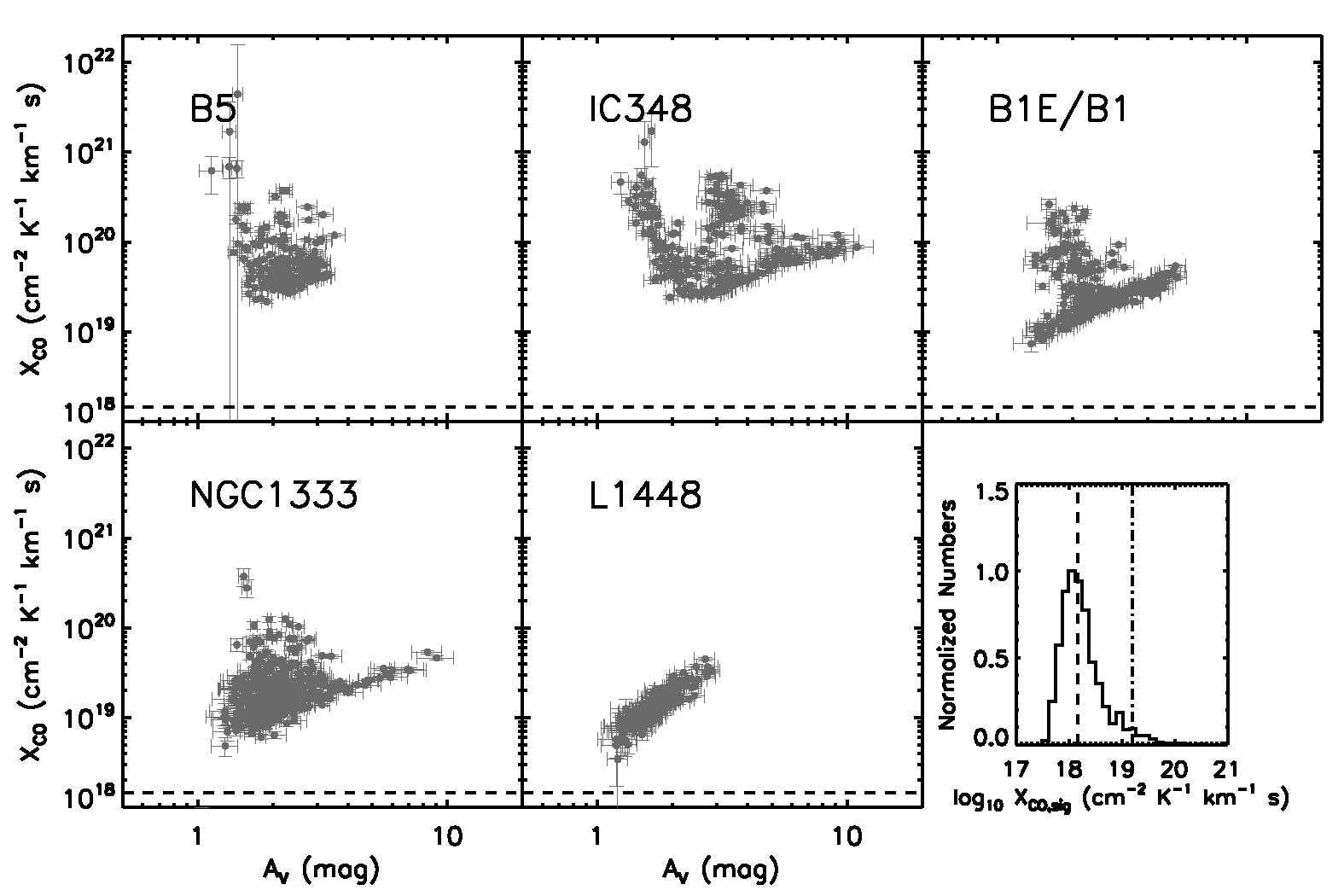}
\caption{\label{f:X_CO_Av} $X_{\rm CO}$ as a function of $A_{V}$ for each dark and star-forming region.
The median 1$\sigma$ uncertainty of $X_{\rm CO}$ ($\sim$1.5 $\times$ 10$^{18}$) is shown as a dashed line, 
while the mean 1$\sigma$ uncertainty of $A_{V}$ ($\sim$0.2 mag) is too small to be shown. 
Note that we show the median 1$\sigma$ instead of the mean 1$\sigma$,  
because it is a better representative of the uncertainty in $X_{\rm CO}$. 
The right lowermost panel shows a normalized histogram of the 1$\sigma$ uncertainty in $X_{\rm CO}$ 
and it is clear that the mean 1$\sigma$ shown as a dashed-dot line corresponds to the high end of the distribution, 
affected by a small fraction of the data points with large uncertainties.}
\end{center}
\end{figure}
\begin{sidewaysfigure}
\begin{center}
\includegraphics[scale=0.44]{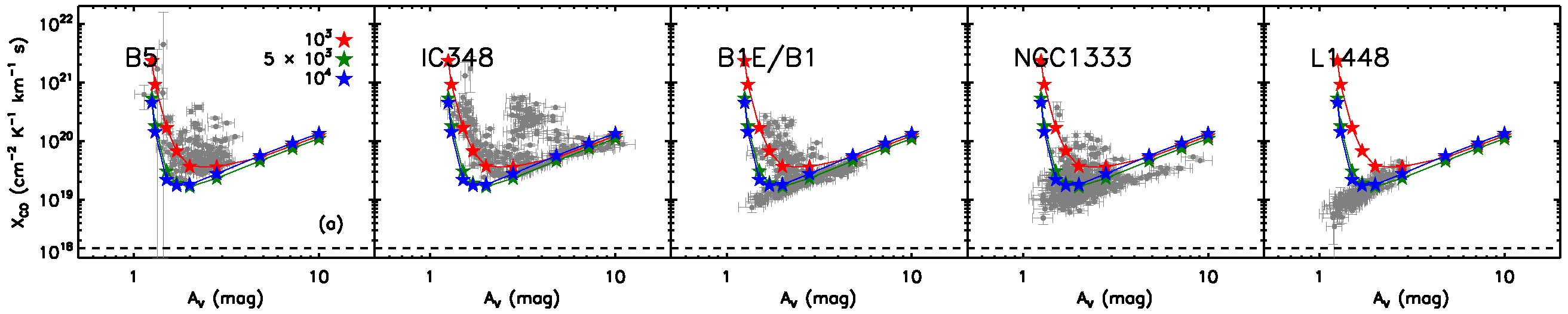}
\includegraphics[scale=0.44]{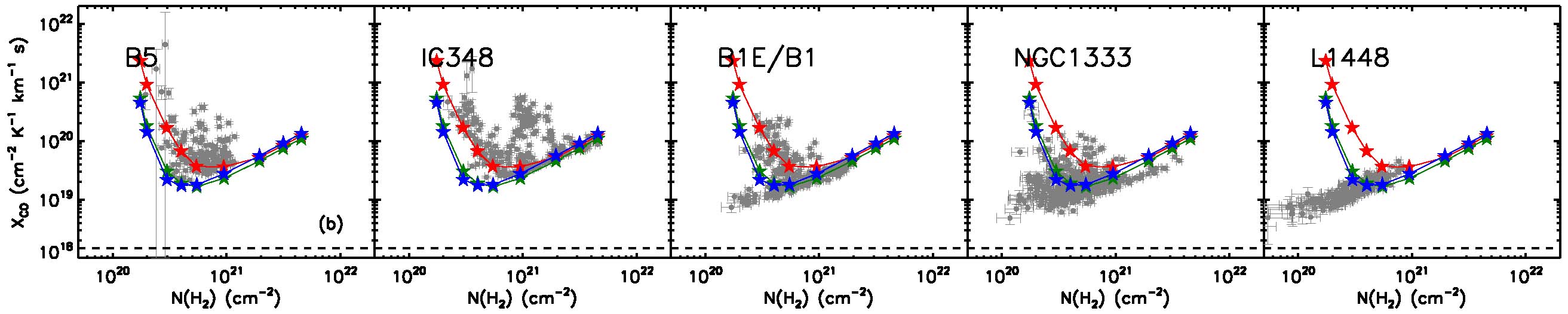}
\includegraphics[scale=0.44]{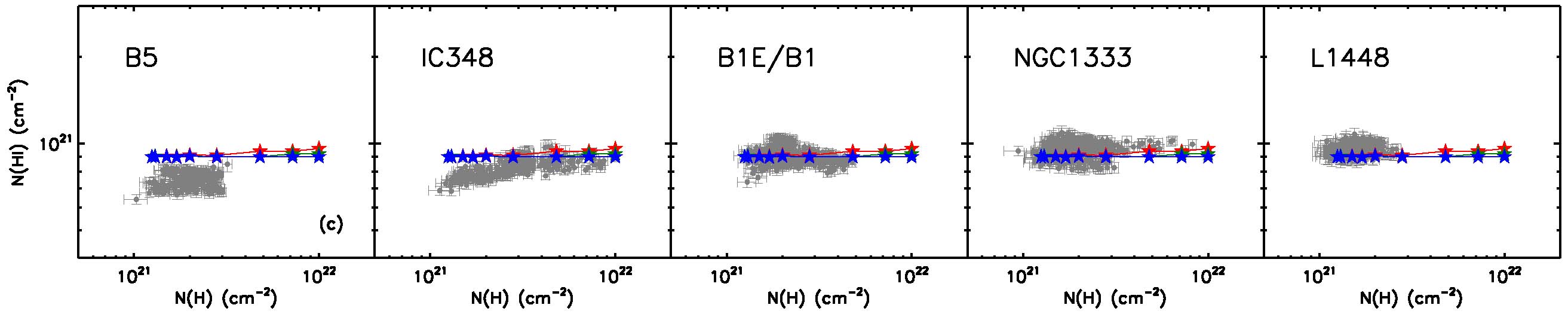}
\includegraphics[scale=0.44]{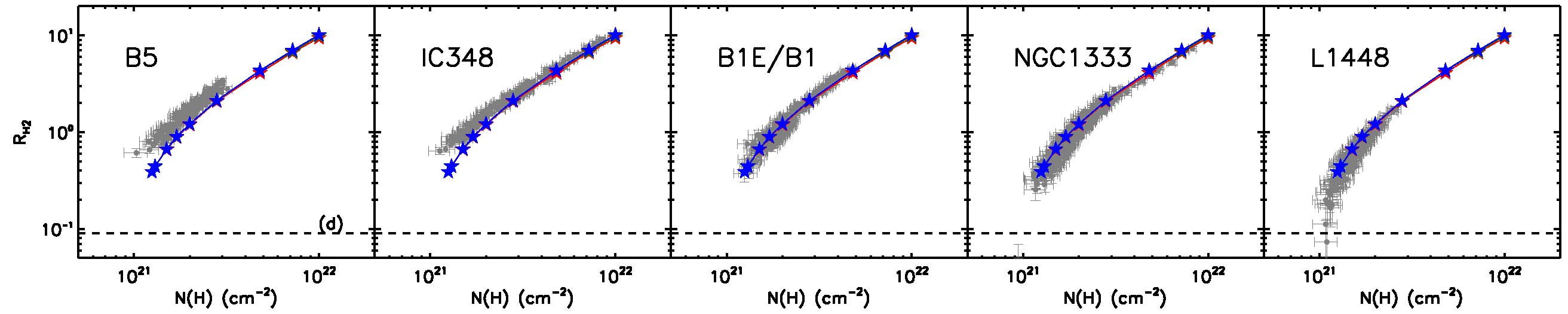}
\caption{\label{f:X_CO_Av_W10} 
Comparison with the modified W10 model. 
Model predictions for $n_{\rm core}$ = 10$^{3}$, 5 $\times$ 10$^{3}$, and 10$^{4}$ cm$^{-3}$ are shown 
as red, green, and blue stars respectively.
The mean 1$\sigma$ uncertainty of $R_{\rm H2}$ ($\sim$0.09) is shown as a dashed line, 
while those of $A_{V}$ ($\sim$0.2 mag), $N$(H$_{2}$) ($\sim$3.3 $\times$ 10$^{19}$ cm$^{-2}$), $N$(HI) ($\sim$3.5 $\times$ 10$^{19}$ cm$^{-2}$),
and $N$(H) ($\sim$1.6 $\times$ 10$^{20}$ cm$^{-2}$) are too small to be shown. 
The median 1$\sigma$ uncertainty of $X_{\rm CO}$ ($\sim$1.5 $\times$ 10$^{18}$) is shown as a dashed line as well. 
(a) $X_{\rm CO}$ vs $A_{V}$. 
(b) $X_{\rm CO}$ vs $N$(H$_{2}$). 
(c) $N$(HI) vs $N$(H). 
(d) $R_{\rm H2}$ vs $N$(H).}
\end{center}
\end{sidewaysfigure}
\begin{figure} 
\begin{center} 
\includegraphics[scale=0.45]{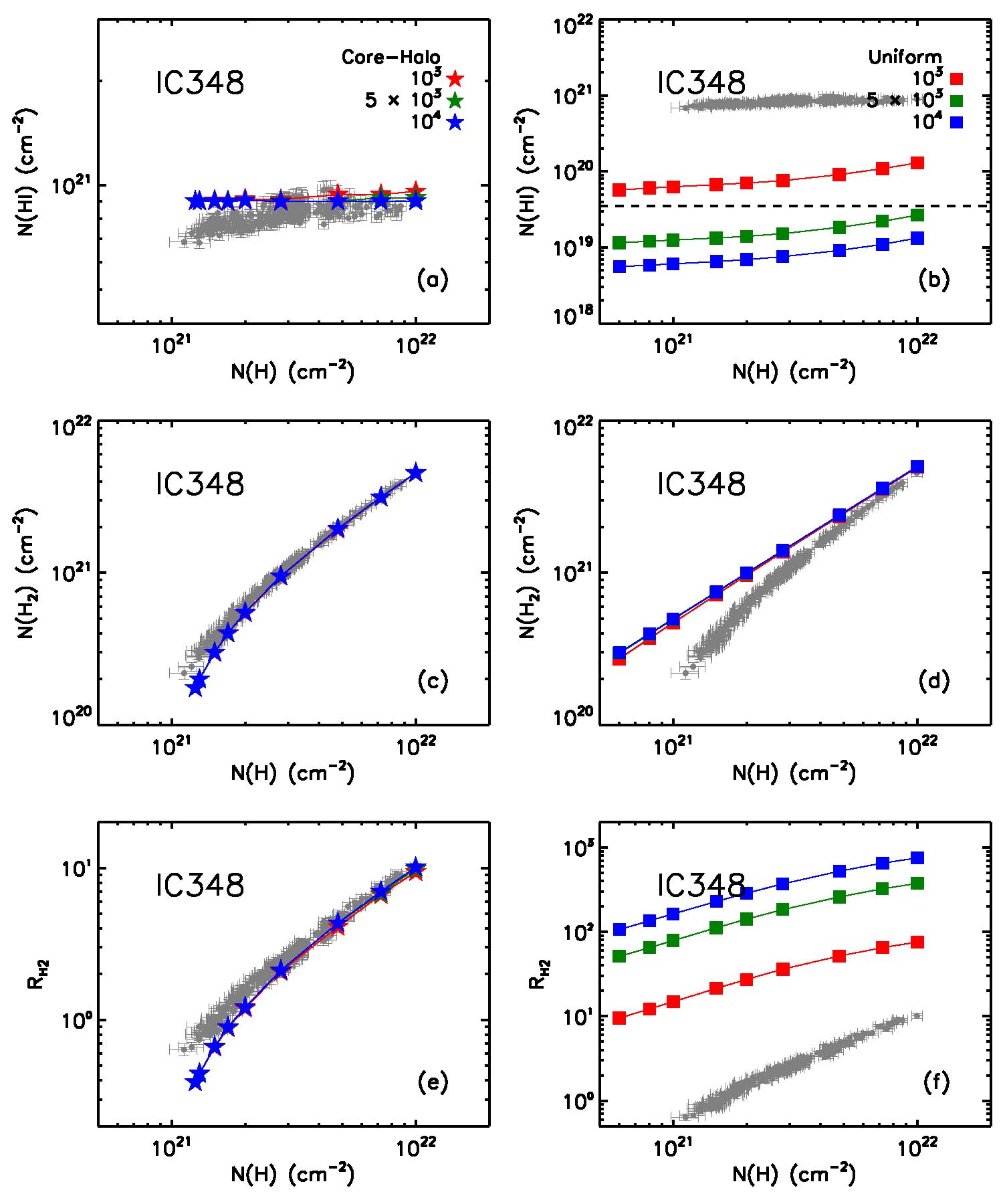}
\caption{\label{f:W_compare} Comparison with the modified W10 model for IC348.
The mean 1$\sigma$ uncertainties of $N$(HI) ($\sim$3.5 $\times$ 10$^{19}$ cm$^{-2}$) and $I_{\rm CO}$ ($\sim$0.09 K km s$^{-1}$) are shown as dashed lines, 
while those of $N$(H) ($\sim$1.6 $\times$ 10$^{20}$ cm$^{-2}$), $N$(H$_{2}$) ($\sim$3.3 $\times$ 10$^{19}$ cm$^{-2}$), 
$R_{\rm H2}$ ($\sim$0.09), and $A_{V}$ ($\sim$0.2 mag) are too small to be shown.  
The median 1$\sigma$ uncertainty of $X_{\rm CO}$ ($\sim$1.5 $\times$ 10$^{18}$) is shown as a dashed line as well. 
(Left) The modified W10 model with the ``core-halo'' structure is shown with red, green, and blue stars 
($n_{\rm core}$ = 10$^{3}$, 5 $\times$ 10$^{3}$, and 10$^{4}$ cm$^{-3}$). 
(Right) The modified W10 model with the uniform density distribution is shown with red, green, and blue squares 
($n$ = 10$^{3}$, 5 $\times$ 10$^{3}$, and 10$^{4}$ cm$^{-3}$).}
\end{center}
\end{figure}

We first compare our data with the S11 model by constructing normalized histograms of
$N$(HI), $N$(H$_{2}$), $N$(H), and $I_{\rm CO}$ in Figure \ref{f:histo_Perseus_s11}.
To construct the histograms, we use the data points with 
$N$(H$_{2}$) > 0 cm$^{-2}$ in Figure \ref{f:H2_cden_display} (``All'' histograms in black), 
as well as those shown in Figure \ref{f:COMPLETE_five_regions} (``Subset'' histograms in grey).  
While the grey histograms are limited to the regions where the CO emission is detected, 
the black histograms represent the whole Perseus cloud.
The simulated data from the S11 model 
(smoothed, regridded, scaled for 7 pc, and the thresholds applied)
are shown as green histograms.
Note that the $I_{\rm CO}$ values from the S11 model are not scaled
and therefore the green $I_{\rm CO}$ histogram likely represents the upper limit of actual histogram
for sub-regions with a size of $\sim$7 pc (indicated as an arrow). 
Because the simulated data (except for $I_{\rm CO}$) are scaled to match the properties of the five regions 
and the thresholds applied to the S11 model are comparable to the minimum $N$(H$_{2}$) and $I_{\rm CO}$ values of the five regions, 
the green histograms can be directly compared with the grey histograms.  
In comparison between our data and the S11 model, 
we find the following.  
 
First, the black and grey $N$(HI) histograms are nearly identical.
This results from the small variation in $N$(HI) across the whole Perseus cloud, as discussed in Section 2.1. 
The green histogram, on the other hand, has a peak at a factor of $\sim$2 smaller $N$(HI)
and even more importantly a factor of $\sim$6 broader distribution than the observed data (the black and grey histograms). 

Second, the grey and green $N$(H$_{2}$) histograms agree very well: 
both peak at a similar $N$(H$_{2}$), have a similar width, and show a lognormal-like distribution.
The black histogram, on the other hand, is broader and has a tail toward small $N$(H$_{2}$).
The difference between the black and grey histograms results from the existence of H$_{2}$
beyond the CO spatial coverage (e.g., ``CO-dark'' H$_{2}$ discussed in Lee et al. 2012).

Third, the green $N$(H) histogram peaks at a similar $N$(H) compared to the grey histogram, 
while showing a broader (a factor of $\sim$2) and lognormal-like distribution.
The simulated distribution is broader mainly because the simulated $N$(HI) has a greater range than what is observed.    
The black and grey histograms, on the other hand, have a tail toward 
$N$(H) $\gtrsim$ 10$^{21.4}$ $\sim$ 2.5 $\times$ 10$^{21}$ cm$^{-2}$.
This tail is consistent with Kainulainen et al. (2009), 
who found a deviation from the lognormal distribution at $A_{V}$ $\gtrsim$ 3 mag for Perseus 
(corresponding to $N$(H) $\sim$ 2.7 $\times$ 10$^{21}$ cm$^{-2}$ with DGR = 1.1 $\times$ 10$^{-21}$ mag cm$^{2}$) 
and interpreted it as a result of self-gravity. 

Lastly, because the simulated $I_{\rm CO}$ is not scaled for the LoS depth of 7 pc,  
we do not compare the exact shapes of the green and grey histograms 
but emphasize that the simulated $I_{\rm CO}$ becomes comparable to the observed $I_{\rm CO}$
only if we use the whole simulation box of 20 pc. 

In summary, we find that the scaled S11 model reproduces the observed range of $N$(H$_{2}$) very well. 
While the predicted $N$(HI) has a relatively similar mean value compared to the observed $N$(HI), 
it has a broader distribution and this leads to a broader range of $N$(H) in the simulation.
The $I_{\rm CO}$ values from the S11 model, on the other hand, cannot be properly compared with our observations 
because of the nontrivial scaling of $I_{\rm CO}$ with the LoS depth. 
However, we find that the simulated $I_{\rm CO}$ is similar with the observed $I_{\rm CO}$ 
only when the CO emission is integrated for the full simulation box of 20 pc. 
 
\subsubsection{Comparison with Observations: $R_{\rm H2}$ and $X_{\rm CO}$}


We plot $N$(HI) against $N$(H) for each dark and star-forming region 
and show predictions from the S11 model 
(smoothed, regridded, scaled for 7 pc, and the thresholds applied)
in Figure \ref{f:HI_tot_S11}(a). 
While the observed $N$(HI) $\sim$ 9 $\times$ 10$^{20}$ cm$^{-2}$ is in the range of the predicted $N$(HI), 
the relation between $N$(HI) and $N$(H) in the S11 model is different from what we find in Perseus: 
not only does the simulated $N$(HI) have a broader distribution, 
but the S11 model predicts a factor of $\sim$7 increase of $N$(HI) for the range of $N$(H) in Perseus,
where we observe less than a factor of 2 variation in $N$(HI).
This suggests that $N$(HI) linearly correlates with $N$(H) in the S11 model
and we indeed estimate Pearson's linear correlation coefficient $r_{\rm p}$ $\sim$ 0.8. 

In addition, the S11 model predicts a factor of $\sim$2 smaller $N$(HI) for a given $N$(H) on average.   
As a result, $R_{\rm H2}$ is slightly larger in the S11 model for a given $N$(H)  
and increases with $N$(H) with a slope smaller than what we observe (Figure \ref{f:HI_tot_S11}b).
While our observations show $R_{\rm H2}$ < 1 for the outskirts of the five regions, 
the simulation has $R_{\rm H2}$ > 1 everywhere, even for the regions with small $n$ < 10$^{2}$ cm$^{-3}$. 


Next, we plot the observed $I_{\rm CO}$ as a function of $A_{V}$ and show the S11 model in Figure \ref{f:HI_tot_S11}(c). 
As discussed in Section 7.2.1, the simulated $N$(HI) and $N$(H$_{2}$) data can be scaled for the five regions in Perseus, 
while the simulated $I_{\rm CO}$ data cannot. 
To properly examine the relation between $I_{\rm CO}$ and $A_{V}$ in the S11 model, therefore, 
we show the predicted $I_{\rm CO}$ vs $A_{V}$ profile without applying the scaling and the thresholds 
and focus on only the general shape of the profile. 
We find that the S11 model describes the relation between $I_{\rm CO}$ and $A_{V}$ reasonably well: 
a steep increase of $I_{\rm CO}$ at small $A_{V}$ and a hint of the saturation of $I_{\rm CO}$ at large $A_{V}$. 
Interestingly, the S11 model predicts that $I_{\rm CO}$ increases with a large scatter at small $A_{V}$. 

Finally, we show the $X_{\rm CO}$ vs $A_{V}$ profile 
for each dark and star-forming region in Figure \ref{f:HI_tot_S11}(d) with the S11 model. 
As in Figure \ref{f:HI_tot_S11}(c), 
the unscaled $N$(HI), $N$(H$_{2}$), and $I_{\rm CO}$ data are used for this comparison.
We find that the S11 model predicts a sharp decrease of $X_{\rm CO}$ at small $A_{V}$ 
and a gradual increase of $X_{\rm CO}$ at large $A_{V}$. 
While a quantitative comparison is not possible without scaling,
the simulated data show the characteristic relation between $X_{\rm CO}$ and $A_{V}$ 
in broad agreement with the observational data (particulary for IC348 and B1E/B1).
This is consistent with Shetty et al. (2011a), 
who performed a number of MHD simulations ($n$ = 100, 300, 1000 cm$^{-3}$ and $Z$ = 0.1, 0.3, 1 Z$_{\odot}$)
and found a steep decrease of $X_{\rm CO}$ at $A_{V}$ $\lesssim$ 7 mag 
and a steady increase of $X_{\rm CO}$ at $A_{V}$ $\gtrsim$ 7 mag for all simulations probing a large range of interstellar environments. 
Relative to the observations,
we find that the simulated $X_{\rm CO}$ shows a significantly larger scatter at small $A_{V}$, 
while the scatter becomes more comparable to what is found in the observations at the high end of the $A_{V}$ range.

\section{Discussion}
 
\subsection{$\boldsymbol{X_{\rm CO}}$ in Perseus and Comparison with Previous Studies}

In their recent review, Bolatto et al. (2013) showed that 
there is some degree of uniformity among the $X_{\rm CO}$ values in the Milky Way obtained from a variety of observational methods. 
Essentially, the typical value for the Milky Way is $X_{\rm CO}$ $\sim$ 2 $\times$ 10$^{20}$
and is known within a factor of $\sim$2.
We, on the other hand, found that the dark and star-forming regions in Perseus have $\langle X_{\rm CO} \rangle$ 
at least five times smaller than the typical value. 
In Appendix, we provide a detailed comparison with two previous studies, 
Dame et al. (2001) and Pineda et al. (2008), 
to understand the reasons behind such a significant difference.  
We summarize our findings here. 

We find three potential sources responsible for the difference: 
the different resolution of $I_{\rm CO}$ and $N$(H$_{2}$) images used to derive $X_{\rm CO}$,  
the application of different DGR, and the treatment of HI in deriving $N$(H$_{2}$).  
For example, Dame et al. (2001) estimated $\langle X_{\rm CO,Dame} \rangle$ $\sim$ 1.2 $\times$ 10$^{20}$ for Perseus, 
which is a factor of $\sim$4 larger than our $\langle X_{\rm CO} \rangle$ $\sim$ 3 $\times$ 10$^{19}$, 
by combining $I_{\rm CO}$ from the CfA survey with $N$(H$_{2}$) derived using the $E(B-V)$ data from Schlegel et al. (1998) 
and the HI data from the Leiden-Argentine-Bonn (LAB) Survey (Kalberla et al. 2005).
Their study, as well as other large-scale studies of $X_{\rm CO}$ in the Milky Way (e.g., Abdo et al. 2010; Paradis et al. 2012),
is at 36$'$ resolution, mainly limited by the LAB HI data. 
In comparison between our original $X_{\rm CO}$ at 4.3$'$ resolution and our $X_{\rm CO}$ smoothed to 36$'$ resolution, 
we find that spatial smoothing results in a factor of $\sim$1.5 increase in $\langle X_{\rm CO} \rangle$. 
Considering a factor of $\sim$8 decrease in angular resolution, 
the effect of resolution on the estimation of $\langle X_{\rm CO} \rangle$ 
appears to be mild and is within the accepted uncertainties,  
although this would be likely more significant when comparing extragalactic observations on $\sim$kpc scales. 
We then find that the rest of the difference between 
our $\langle X_{\rm CO} \rangle$ and $\langle X_{\rm CO, Dame} \rangle$ can be explained by the difference in DGR. 
While both studies measured DGR, 
Dame et al. (2001) calculated $N$(HI) along a whole line of sight (while we focused on the velocity range for Perseus only) 
and estimated DGR using the images smoothed to 10$^{\circ}$ resolution (while we had 4.3$'$ resolution).  
The DGR effect is slightly larger than the resolution effect (a factor of $\sim$1.8) 
and these two factors together account for most of the difference between our study and Dame et al. (2001). 

In the case of Pineda et al. (2008), angular resolution is not an issue 
because essentially the same $A_{V}$ and $I_{\rm CO}$ data were used.  
However, they estimated $X_{\rm CO, Pineda}$ $\sim$ 1.4 $\times$ 10$^{20}$ for Perseus.
Their methodology for deriving $X_{\rm CO}$ is different from our study mainly in two ways. 
First, they assumed that the $N$(HI) contribution to $A_{V}$ is insignificant and therefore did not consider it. 
Second, they adopted the typical DGR for the Milky Way = 5.3 $\times$ 10$^{-22}$ mag cm$^{2}$ (Bohlin et al. 1978). 
In contrast, we accounted for the $N$(HI) contribution and estimated DGR = 1.1 $\times$ 10$^{-21}$ mag cm$^{2}$ (Lee et al. 2012). 
In Appendix, we show that we estimate $\langle X_{\rm CO} \rangle$ $\sim$ 1 $\times$ 10$^{20}$, which is comparable to $X_{\rm CO, Pineda}$, 
when we follow the methodology of Pineda et al. (2008). 
In addition, we find that the application of each of the two assumptions made by Pineda et al. (2008) 
results in a factor of $\sim$2 difference in $\langle X_{\rm CO} \rangle$, 
altogether explaining the difference between our $\langle X_{\rm CO} \rangle$ and $X_{\rm CO, Pineda}$.  

Our detailed comparison with Dame et al. (2001) and Pineda et al. (2008) shows that 
different resolutions and methodologies for deriving $X_{\rm CO}$ can result in a difference in $X_{\rm CO}$ by up to a factor of $\sim$4,  
even for the same method of $X_{\rm CO}$ determination ($X_{\rm CO}$ based on dust emission/absorption in this case).
Other methods of $X_{\rm CO}$ determination, e.g., $X_{\rm CO}$ based on the Virial technique and $\gamma$-ray observations, have their own assumptions. 
This clearly suggests the difficulty in comparing $X_{\rm CO}$ between molecular clouds and/or galaxies when different observational methods are used,  
as pointed out by Bolatto et al. (2013) as well.
The relatively uniform value of $X_{\rm CO}$ for the Milky Way found from many studies with various resolutions and methodologies, 
therefore, appears puzzling. 

\subsection{$\boldsymbol{X_{\rm CO}}$ in Molecular Clouds}

In Section 6.3, we focused on the individual dark and star-forming regions in Perseus 
and found significant spatial variations in $X_{\rm CO}$.
Specifically, $X_{\rm CO}$ varies by up to a factor of $\sim$100 within a single region with a size of $\sim$6--7 pc. 
Our investigation of the large-scale trends in $G$ and $\sigma_{\rm CO}$ (Section 6.1)
and our comparison with the modified W10 model (Section 7.1) suggest that 
changes in physical parameters are responsible for the variations in $X_{\rm CO}$ 
observed both within the individual regions and between the different regions.

While $X_{\rm CO}$ shows significant variations across the cloud, 
we found that there is a characteristic dependence of $X_{\rm CO}$ on $A_{V}$ (particularly evident for IC348 and B1E/B1): 
a steep decrease of $X_{\rm CO}$ at $A_{V}$ $\lesssim$ 3 mag and a moderate increase of $X_{\rm CO}$ at $A_{V}$ $\gtrsim$ 3 mag. 
This relation between $X_{\rm CO}$ and $A_{V}$ appears to result from the strong dependence of $I_{\rm CO}$ on $A_{V}$. 
The location at which most carbon is locked in CO primarily depends on dust shielding (e.g., W10; Glover \& Mac Low 2011). 
Once dust shielding becomes sufficiently strong to prevent photodissociation ($A_{V}$ $\gtrsim$ 1 mag in Perseus),  
the CO abundance and emission strength sharply rise 
and this could result in decreasing $X_{\rm CO}$ with $A_{V}$. 
$I_{\rm CO}$ then saturates to a certain value 
because the CO emission becomes optically thick with increasing depths ($A_{V}$ $\gtrsim$ 3 mag in Perseus) 
and this could result in increasing $X_{\rm CO}$ with $A_{V}$.
These results suggest that CO is a poor tracer of H$_{2}$ 
for those regions where dust shielding is not strong enough to prevent photodissociation, 
e.g., low-metallicity environments (e.g., Leroy et al. 2007, 2009, 2011; Cormier et al. 2014).
In addition, CO is unreliable for those regions where the CO emission is optically thick,   
because it provides only a lower limit on $N$(H$_{2}$).

Overall, our study suggests that one cannot adopt a single $X_{\rm CO}$ to derive the $N$(H$_{2}$) distribution across a resolved molecular cloud. 
The limited dynamic range of CO as a tracer of H$_{2}$ and the complex dependence of $X_{\rm CO}$ on various physical parameters
hamper the derivation of the accurate $N$(H$_{2}$) distribution.
On the other hand, calculation of the H$_{2}$ mass over the CO-observed area, $M$(H$_{2}$)$_{\rm CO}$, 
appears to be less affected by variations in physical parameters. 
For example, we estimate $M$(H$_{2}$)$_{\rm CO}$ = (1799.8 $\pm$ 3.2) M$_{\odot}$ over the COMPLETE CO spatial coverage.  
If we derive $M$(H$_{2}$)$_{\rm CO}$ using our $\langle X_{\rm CO} \rangle$,
we find $M$(H$_{2}$,$X_{\rm CO}$)$_{\rm CO}$ = (1814.1 $\pm$ 0.2) M$_{\odot}$.
These two estimates are comparable for Perseus, 
mainly because a large fraction of the data points ($\sim$60\%) has $X_{\rm CO}$ different from our $\langle X_{\rm CO} \rangle$ within a factor of $\sim$2.  
The agreement between the observed $X_{\rm CO}$ in Perseus and the model predictions (in particular for the PDR model)
suggests that a theory-based $X_{\rm CO}$ could be used to estimate $M$(H$_{2}$)$_{\rm CO}$ for a molecular cloud. 
Once theoretical models, e.g., PDR and MHD models, 
are thoroughly tested against observations of molecular clouds in diverse environments,  
they will be able to provide predictions over a wide range of physical conditions. 
One then can search a large parameter space to select the most appropriate $X_{\rm CO}$
for a target molecular cloud based on reasonable constraints on physical parameters. 
Note, however, that the total H$_{2}$ mass of the cloud would be still uncertain 
if there is significant ``CO-dark'' H$_{2}$ outside the CO-observed area.  

\subsection{Insights from the Microturbulent Time-independent Model}

The good agreement between our data and the modified W10 model with the ``core-halo'' structure (Section 7.1.2)
suggests that the main assumptions of the model, e.g., H$_{2}$/CO formation in chemical equilibrium,  
the microturbulent approximation for CO spectral line formation, 
and the ``core-halo'' density distribution, are valid for Perseus on $\sim$0.4 pc scales. 
This result is consistent with Lee et al. (2012), who found that $N$(HI) and $N$(H$_{2}$) in Perseus 
conform to the time-independent H$_{2}$ formation model by K09.
We now turn to a couple of interesting aspects of the modified W10 model and discuss their implications.

\subsubsection{The Importance of Diffuse HI Halo for H$_{2}$ and CO Formation}

The modified W10 model that is comparable to the observations of Perseus uses  
the ``core-halo'' structure motivated by previous studies of molecular clouds (Section 7.1.1).
We showed that the model with a uniform density distribution predicts $N$(HI) 
much smaller than the uniform $N$(HI) $\sim$ 9 $\times$ 10$^{20}$ cm$^{-2}$ measured across Perseus.  
The uniform density model with the largest density $n$ = 10$^{4}$ cm$^{-3}$ predicts the smallest $N$(HI),
up to a factor of $\sim$160 smaller than what is observed.
The main reason is that in the uniform density model
H$_{2}$ self-shielding alone counteracts H$_{2}$ photodissociation by LW photons.
Traditionally, it has been known that $G$/$n$ determines 
whether H$_{2}$ self-shielding or dust shielding is more important for H$_{2}$ formation 
and controls the location of the transition from HI to H$_{2}$ in a PDR (e.g., Hollenbach \& Tielens 1997).
With $G$ = 0.5 $G_{0}'$ and $n$ = 10$^{3}$ cm$^{-3}$ in the uniform density model, 
$G$/$n$ is 5 $\times$ 10$^{-4}$ cm$^{3}$, small enough that dust shielding is negligible.
In this case, most of the HI is converted into H$_{2}$ because of the strong H$_{2}$ self-shielding.
On the other hand, the ``core-halo'' model with $n_{\rm core}$ = 10$^{3}$ cm$^{-3}$ and $n_{\rm halo}$ = 40 cm$^{-3}$ 
has $G$/$n_{\rm halo}$ $\sim$ 0.01 cm$^{3}$ in the cloud outskirts. 
This increased $G$/$n$ makes H$_{2}$ self-shielding less important for H$_{2}$ formation 
and as a result, the gas remains atomic with $N$(HI) $\sim$ 9 $\times$ 10$^{20}$ cm$^{-2}$.

The fact that the modified W10 model needs a diffuse HI halo to reproduce the observed $N$(HI) 
suggests that dust shielding is important for H$_{2}$ formation in Perseus.
This importance of dust shielding is consistent with what Lee et al. (2012) found from their comparison with the K09 model. 
The K09 model investigates the structure of a PDR in a spherical cloud based on H$_{2}$ formation in chemical equilibrium 
and predicts the following variable as one of the key parameters that determine the location of the transition from HI to H$_{2}$:

\begin{equation}
\chi = 2.3\frac{(1 + 3.1 Z'^{0.365})}{\phi_{\rm CNM}}, 
\end{equation}

\noindent where $Z'$ is the metallicity normalized to the solar neighborhood value 
and $\phi_{\rm CNM}$ is the ratio of the actual CNM density to the minimum CNM density
at which the CNM exists in pressure balance with the warm neutral medium (WNM).
This $\chi$ is the ratio of the rate at which LW photons are absorbed by dust grains (dust shielding) 
to the rate at which they are absorbed by H$_{2}$ (H$_{2}$ self-shielding) 
and is conceptually similar to $G$/$n$. 
K09 predicts $\chi$ $\sim$ 1 in all galaxies where the pressure balance between the CNM and the WNM is valid,
suggesting that dust shielding and H$_{2}$ self-shielding are equally important for H$_{2}$ formation. 
By fitting the K09 model to the observed $R_{\rm H2}$ vs $\Sigma_{\rm HI}$+$\Sigma_{\rm H2}$ profiles,
Lee et al. (2012) indeed found $\chi$ $\sim$ 1 for Perseus.

In the modified W10 model, a diffuse HI halo is also required to reproduce the observed steep increase of $I_{\rm CO}$ 
at $A_{V}$ $\gtrsim$ 1 mag and sharp decrease of $X_{\rm CO}$ at $A_{V}$ $\lesssim$ 3 mag (Section 7.1.3).  
The uniform density model predicts the shallower increase of $I_{\rm CO}$ at smaller $A_{V}$ $\gtrsim$ 0.6 mag, 
suggesting a less sharp transition from CII/CI to CO located closer to the surface of the gas slab. 
The more extended CO distribution eventually results in the reduced ``CO-free'' H$_{2}$ envelope 
and therefore the uniform density model with $n$ = 10$^{4}$ cm$^{-3}$ would have the smallest amount of ``CO-dark'' H$_{2}$. 
The CO distribution deep inside of the gas slab, on the other hand, 
does not appear to be affected by the presence of the diffuse HI halo because of the saturation of $I_{\rm CO}$. 

Even though the modified W10 model with the ``core-halo'' structure reproduces 
the observed $N$(HI), $N$(H$_{2}$), and $I_{\rm CO}$ distributions,  
the agreement is likely to remain only if the halo density is not significantly larger than 40 cm$^{-3}$. 
The current density $n_{\rm halo}$ = 40 cm$^{-3}$ originates from the theoretical (e.g., Wolfire et al. 2003) 
and observational (e.g., Heiles \& Troland 2003) properties of the CNM. 
While large HI envelopes associated with molecular clouds have been frequently observed 
(e.g., Knapp 1974; Wannier et al. 1983, 1991; Reach et al. 1994; 
Rogers et al. 1995; Williams \& Maddalena 1996; Imara \& Blitz 2011; Lee et al. 2012),
a number of fundamental questions still remain to be answered.  
For example, what are the physical properties of the HI halos, such as density, temperature, and pressure? 
What is the ratio of the CNM to the WNM in the halos? 
Is there any correlation between the ratio and the H$_{2}$ abundance/star formation?  
Are the halos expanding or infalling?
The high-resolution HI data from the GALFA-HI survey will be valuable for future studies of the extended HI halos 
around Galactic molecular clouds in a wide range of interstellar environments. 
Finally, further comparisons between observations and theoretical models will be important to fully constrain
the parameter space and density structure of the HI halos.

\subsubsection{Validity of Steady State and Equilibrium Chemistry}

The timescale of H$_{2}$ formation on dust grains, $t_{\rm H2}$, dominates chemical timescales of PDRs (e.g., Hollenbach \& Tielens 1997). 
For the modified W10 model with the ``core-halo'' structure, dense regions have $n_{\rm core}$ $\gtrsim$ 10$^{3}$ cm$^{-3}$
where gas is completely molecular ($n_{\rm H2}$ $\sim$ 0.5$n$). 
In this case, $t_{\rm H2}$ = 0.5/$\mathcal{R}$$n_{\rm core}$ $\lesssim$ 0.5 Myr,  
where $\mathcal{R}$ = 3 $\times$ 10$^{-17}$ cm$^{3}$ s$^{-1}$ is the rate coefficient for H$_{2}$ formation (Wolfire et al. 2008). 
In diffuse regions with $n_{\rm halo}$ = 40 cm$^{-3}$, on the other hand, 
gas is mostly atomic ($n_{\rm H2}$ $\sim$ 0.1$n$) and therefore 
$t_{\rm H2}$ = 0.1/$\mathcal{R}$$n_{\rm halo}$ $\sim$ 2.6 Myr. 
Because $t_{\rm H2}$ of the model is well within the expected age of Perseus, $t_{\rm age}$ $\sim$ 10 Myr, 
the assumption of chemical equilibrium is valid.
In other words, Perseus is old enough to reach chemical equilibrium 
and therefore it is not surprising that the equilibrium chemistry model (W10) fits our observations very well. 


However, for steady state chemistry to be valid, $t_{\rm H2}$ $\lesssim$ $t_{\rm age}$ is not enough: 
$t_{\rm H2}$ should be short compared to the dynamical timescale of a molecular cloud, $t_{\rm dyn}$. 
For Perseus, this requires $t_{\rm dyn}$ $\gtrsim$ 3 Myr. 
As a rough estimate, we calculate a crossing timescale, $t_{\rm cross}$ = $L$/$\sigma$ $\sim$ 10 pc/1.8 km s$^{-1}$ $\sim$ 6 Myr,
where we choose $L$ as the characteristic size of the individual regions in Perseus and $\sigma$ as the mean CO velocity dispersion.
This $t_{\rm cross}$ $\sim$ 6 Myr satisfies the condition for $t_{\rm dyn}$ $\gtrsim$ 3 Myr. 
However, many dynamical processes are involved with the formation and evolution of molecular clouds 
(e.g., cloud-cloud collisions, spiral shocks, stellar feedback; Mac Low \& Klessen 2004; Mckee \& Ostriker 2007)  
and therefore it is difficult to pin down the exact process that is most relavant for the formation of molecular gas.  
The good agreement between our data and the modified W10 model with the ``core-halo'' structure suggests that 
the characteristic $t_{\rm dyn}$ for the formation of molecular gas in Perseus should be $\gtrsim$ 3 Myr.

\subsection{Insights from the Macroturbulent Time-dependent Model}

\subsubsection{The Choice of the Input Parameters in the MHD Simulation} 

In Section 7.2.2, we found that the scaled S11 model predicts $N$(H$_{2}$) comparable to the estimated $N$(H$_{2}$) in Perseus. 
This excellent agreement will likely hold even if some of the input parameters slightly change. 
For example, the S11 model was run with $G$ = 1 $G_{0}'$ 
and this is a factor of $\sim$2 stronger than what we measure across Perseus. 
Considering that S11 found no considerable difference in $N$(H$_{2}$) for their models with $G$ = 1 $G_{0}'$ and 10 $G_{0}'$ (Section 3.1 of S11), 
however, decreasing $G$ from 1 $G_{0}'$ to 0.5 $G_{0}'$ to match the property of Perseus will not make a significant change in $N$(H$_{2}$). 
In addition, increasing $\zeta$ from 10$^{-17}$ s$^{-1}$ to 10$^{-16}$ s$^{-1}$ to be consistent with the modified W10 model 
will not affect $N$(H$_{2}$) very much, 
based on the fact that Glover \& Mac Low (2007b) found a negligible change in $N$(H$_{2}$) 
when $\zeta$ increased from 10$^{-17}$ s$^{-1}$ to 10$^{-15}$ s$^{-1}$ 
in their MHD simulation with initial $n$ = 100 cm$^{-3}$ (Section 6.3 of Glover \& Mac Low 2007b). 
Increasing DGR from 5.3 $\times$ 10$^{-22}$ mag cm$^{2}$ to 1.1 $\times$ 10$^{-21}$ mag cm$^{2}$ for Perseus will lead to more rapid H$_{2}$ formation, 
but the model with the increased DGR will not be substantially different from the current S11 model 
since the S11 model becomes H$_{2}$-dominated rapidly by $t$ $\sim$ 3 Myr (Figure 7 of Glover \& Mac Low 2011). 
Finally, the extension of the simulation run up to $t$ $\sim$ 10 Myr, comparable to the age of Perseus, 
will not significantly increase $N$(H$_{2}$), 
considering that Glover \& Mac Low (2011) found only a factor of $\sim$1.3 increase of the mass-weighted mean H$_{2}$ abundance 
from $t$ $\sim$ 5 Myr to $t$ $\sim$ 10 Myr for their MHD simulation with initial $n$ = 100 cm$^{-3}$ (Section 3.3 of Glover \& Mac Low 2011). 

Similarly, small changes in the model parameters will likely make no substantial difference in $I_{\rm CO}$.
For example, S11 showed that increasing $G$ from 1 $G_{0}'$ to 10 $G_{0}'$ does not change $I_{\rm CO}$ 
for those regions where CO is well shielded against the radiation field (Section 3.1 of S11). 
Therefore, decreasing $G$ from 1 $G_{0}'$ to 0.5 $G_{0}'$ will make only a minor change in $I_{\rm CO}$ at large column densities. 
Increasing the current DGR of 5.3 $\times$ 10$^{-22}$ mag cm$^{2}$ by a factor of $\sim$2 will cause more rapid CO formation, 
but $I_{\rm CO}$ will not be significantly influenced because CO formation in the S11 model reaches chemical equilibrium rapidly by $t$ $\sim$ 2 Myr. 
Lastly, we do not expect that running the S11 model up to $t$ $\sim$ 10 Myr drastically increases $I_{\rm CO}$, 
considering that the MHD simulation with initial $n$ = 100 cm$^{-2}$ in Glover \& Mac Low (2011) predicts 
only a factor of $\sim$2 increase of the mass-weighted mean CO abundance from $t$ $\sim$ 5 Myr to $t$ $\sim$ 10 Myr (Section 3.3 of Glover \& Mac Low 2011). 
Note that changes in CO abundance at $t$ > 2 Myr are stochastic fluctuations after chemical equilibrium is achieved.  

We therefore conclude that the input parameters used in the S11 model are reasonable for the comparison with the observations of Perseus  
and small (a factor of few) changes in the input parameters will not result in significant 
changes in $N$(HI), $N$(H$_{2}$), and $I_{\rm CO}$.
Considering that Perseus has most likely reached chemical equilibrium, 
it provides a suitable testbed for investigating whether results from the time-depedent MHD simulation 
converge to the time-independent PDR model for molecular clouds that are evolved enough.

\subsubsection{The Role of Turbulence in H$_{2}$ and CO Formation}

As shown in Section 7.2.2, the scaled S11 model produces the $N$(H$_{2}$) distribution in excellent agreement 
with our observations as well as the modified W10 model.
This suggests that the time-dependent H$_{2}$ formation model (S11) 
is consistent with the time-independent H$_{2}$ formation model (W10) 
for a low-mass, old molecular cloud such as Perseus. 
Our result agrees with Krumholz \& Gnedin (2011), 
who found that time-dependent effects on H$_{2}$ formation become important 
only at extremely low metallicities $Z$ $\lesssim$ 10$^{-2}$ Z$_{\odot}$. 
While the median $N$(HI) in the S11 model is also in reasonably good agreement with the observations, 
the simulated $N$(HI) distribution is a factor of $\sim$6 broader than the observed one 
and particularly shows a more extended tail toward small $N$(HI) $\lesssim$ 3 $\times$ 10$^{20}$ cm$^{-2}$. 
This broad $N$(HI) distribution in the MHD simulation likely results from strong compressions and rarefactions by turbulence
and the predicted $N$(HI) is on average a factor of $\sim$2 smaller than the observed $N$(HI) for a given $N$(H).
The discrepancy becomes significant at small $N$(H) $\sim$ 10$^{21}$ cm$^{-2}$, 
where the S11 model underesimates $N$(HI) by up to a factor of $\sim$10.
Finally, the S11 model predicts that $N$(HI) increases with $N$(H), 
suggesting no minimum $N$(HI) beyond which the rest of hydrogen is converted into H$_{2}$.  
In the modified W10 model with the ``core-halo'' structure, on the other hand, 
the diffuse halo remains atomic with $N$(HI) $\sim$ 9 $\times$ 10$^{20}$ cm$^{-2}$ 
and the dense core is fully converted into H$_{2}$.
Clearly, this discrepancy in $N$(HI) between the simulation and the observations is significant and interesting.
One potential avenue in exploring this in the future is by using a mixture of neutral phases for initial conditions, 
mimicing in some way the ``core-halo'' structure in the modified W10 model.

In the case of $I_{\rm CO}$, we could not properly compare the S11 model with our observations 
because of the nontrivial scaling of $I_{\rm CO}$ for different line of sight depths. 
Instead, we found that the simulated $I_{\rm CO}$ becomes comparable to the observed $I_{\rm CO}$ 
only if the CO emission is integrated for the full simulation box of 20 pc.
This suggests that the S11 model likely underestimates $I_{\rm CO}$ for the conditions relevant to Perseus.
Interestingly, we estimate $N$(CO) $\sim$ 1 $\times$ 10$^{17}$ cm$^{-2}$ for B5, IC348, B1, and NGC1333 
by using the $^{13}$CO($J$ = 1 $\rightarrow$ 0) excitation temperatures, optical depths, and integrated intensities provided by Pineda et al. (2008)
and assuming $N$(CO) = 76$N$($^{13}$CO) (Lequeux 2005).  
This value is in reasonably good agreement with the simulated mean $N$(CO) $\sim$ 5 $\times$ 10$^{16}$ cm$^{-2}$ 
(calculated from the smoothed, regridded, scaled, and thresholds applied S11 model)\footnote[10]{As S11 did not perform 
any radiative transfer calculations to produce the CO number density cube, it is appropriate to scale the simulated $N$(CO).}.
This comparison suggests that the potential discrepancy in $I_{\rm CO}$ between the observations and the S11 model 
would result from the radiative transfer calculations and/or the difference in velocity range. 
The velocity range of the CO emission, $\Delta$${v}$, directly affects $I_{\rm CO}$ via $I_{\rm CO}$ = $\int T_{\rm B} dv$ 
and a smaller $\Delta$${v}$ would result in a smaller $I_{\rm CO}$ for the same $T_{\rm B}$. 

While we could not compare specific $I_{\rm CO}$ values predicted by the S11 model with our observations, 
we found that the S11 model reproduces the observed shape of the $I_{\rm CO}$ vs $A_{V}$ profiles reasonably well.
This suggests that penetration of UV photons into the ISM and (dust and self-) shielding against the ISRF
are relatively well captured in the CO formation process by S11.
In addition, we noticed that $I_{\rm CO}$ has a much larger scatter at small $A_{V}$. 
This could result from large density fluctuations in the turbulent medium. 
The gas in the S11 model would be strongly compressed and rarefied by turbulence 
and the gas density at a given $A_{V}$ can vary over several orders of magnitude (e.g., Figure 14 of Glover et al. 2010).
In this case, CO can form in dense clumps even at small $A_{V}$ and $I_{\rm CO}$ therefore shows a large scatter. 
This scatter is reduced at large $A_{V}$ where $I_{\rm CO}$ eventually saturates. 
Finally, turbulent mixing could spread the CO distribution, contributing to the large scatter of $I_{\rm CO}$. 

In general, our study shows that the scaled MHD simulation by S11 is successful in reproducing $N$(H$_{2}$) in Perseus,
which is a low-mass, old molecular cloud most likely in chemical equilibrium. 
On the other hand, future model adjustments are required to better match the observed $N$(HI) and $I_{\rm CO}$. 
We have revealed two important areas of future attention:
(1) the role of diffuse halos in the formation of molecular gas 
and (2) the effect of density fluctuations and turbulent mixing in the spatial distribution of molecular gas. 
To investigate these two issues, we plan to compare observations of several Galactic molecular clouds 
with MHD simulations that explore different fractions of neutral phases and a varying degree of turbulence as initial conditions. 
In particular, our future work will include molecular clouds less evolved 
and/or forming more massive stars (therefore more turbulent) than Perseus, 
where the difference between the MHD and PDR models is likely to be more pronounced.

\section{Summary}

In this paper, we combine high-resolution H$_{2}$ and CO measurements to investigate $X_{\rm CO}$ across the Perseus molecular cloud.  
We derive the $X_{\rm CO}$ image at $\sim$0.4 pc spatial resolution 
by using $N$(H$_{2}$) estimated by Lee et al. (2012) and $I_{\rm CO}$ provided by the COMPLETE survey.
We examine the large-scale spatial variations in $X_{\rm CO}$ across the cloud and their correlations with local ISM conditions. 
In addition, we focus on the characteristic dependence of $X_{\rm CO}$ on $A_{V}$. 

The $N$(HI), $N$(H$_{2}$), $I_{\rm CO}$, and $X_{\rm CO}$ images allow us to test two theoretical models of H$_{2}$ and CO formation:  
the modified W10 model (``microturbulent time-independent model'') and the S11 model (``macroturbulent time-dependent model'').  
For several dark and star-forming regions in Perseus (B5, B1E/B1, L1448, IC348, and NGC1333),  
we investigate $N$(HI) vs $N$(H), $R_{\rm H2}$ vs $N$(H), $I_{\rm CO}$ vs $A_{V}$, and $X_{\rm CO}$ vs $A_{V}$
and compare the results with model predictions.
We summarize our main results as follows.

\begin{enumerate}

\item[(1)] We derive $\langle X_{\rm CO} \rangle\sim$ 3 $\times$ 10$^{19}$ for Perseus. 
This value is a factor of $\sim$4 smaller than the previous estimate of $X_{\rm CO}$ $\sim$ 1 $\times$ 10$^{20}$ 
for the same cloud (Dame et al. 2001; Pineda et al. 2008)
and the discrepancy mainly results from different resolutions, DGRs, and our consideration of $N$(HI) in deriving $N$(H$_{2}$).  


\item[(2)] We find a factor of $\sim$3 region-to-region variations in $X_{\rm CO}$. 
The northeastern part of Perseus (B5 and IC348) has on average larger $X_{\rm CO}$ than the southwestern part (B1E/B1, NGC1333, and L1448). 
This could be explained by a stronger $G$ and/or a smaller $\sigma_{\rm CO}$ in the northeastern part, 
although the correlations between $X_{\rm CO}$ and $G$/$\sigma_{\rm CO}$ are mild. 
Additionally, variations in $n$ and/or $A_{V}$ could contribute to the observed regional variations in $X_{\rm CO}$.
Within the individual dark and star-forming regions with a size of $\sim$6--7 pc, 
$X_{\rm CO}$ varies up to a factor of $\sim$100. 

\item[(3)] The observed $X_{\rm CO}$ vs $A_{V}$ profiles show two characteristic features:  
a steep decrease of $X_{\rm CO}$ at small $A_{V}$ and a gradual increase of $X_{\rm CO}$ at large $A_{V}$.
Among the five dark and star-forming regions, IC348 and B1E/B1 clearly show   
the transition from decreasing to increasing $X_{\rm CO}$ at $A_{V}$ $\sim$ 3 mag. 

\item[(4)] The modified W10 model with the ``core-halo'' density distribution reproduces the observed $X_{\rm CO}$ vs $A_{V}$ profiles, 
particularly well for IC348 and B1E/B1. 
In addition, the model predicts a nearly constant $N$(HI) $\sim$ 9 $\times$ 10$^{20}$ cm$^{-2}$  
and a linear increase of $R_{\rm H2}$ with $N$(H), both consistent with what we find in Perseus. 

\item[(5)] The modified W10 model with the uniform density distribution reproduces the observed $N$(H$_{2}$) reasonably well 
but underestimates $N$(HI) by a factor of $\sim$10--160. 
As a result, the model overestimates $R_{\rm H2}$ for a given $N$(H) by up to a factor of $\sim$300. 
In addition, while matching the observed saturation of $I_{\rm CO}$ at $A_{V}$ $\gtrsim$ 3 mag, 
the model predicts a more gradual increase of $I_{\rm CO}$ at $A_{V}$ $\lesssim$ 3 mag. 
This results in the $X_{\rm CO}$ vs $A_{V}$ profile shallower than the observations at $A_{V}$ $\lesssim$ 3 mag.

\item[(6)] The scaled S11 model predicts $N$(H$_{2}$) in excellent agreement with what we estimate in Perseus. 
However, $N$(HI) increases with $N$(H) by a factor of $\sim$7 in the model 
and this is in contrast with the observed small variation of $N$(HI) with $N$(H) (less than a factor of 2). 
While we do not compare specific $I_{\rm CO}$ values in the S11 model with the observed $I_{\rm CO}$
due to a complex issue of scaling $I_{\rm CO}$ for different line of sight depths, 
we stress that the simulated $I_{\rm CO}$ becomes comparable 
only when the CO emission is integrated along the full simulation box of 20 pc, 
suggesting that the model likely underestimates $I_{\rm CO}$ for the conditions relevant to Perseus.  
In addition, we find that the S11 model reproduces the observed shapes of 
$I_{\rm CO}$ vs $A_{V}$ and $X_{\rm CO}$ vs $A_{V}$ profiles reasonably well 
but with a large scatter particulary at small $A_{V}$. 



\end{enumerate}

Our study shows that $X_{\rm CO}$ can vary by up to a factor of $\sim$100 on $\sim$0.4 pc scales
and depends on local ISM conditions such as $G$, $\sigma_{\rm CO}$, $n$, and $A_{V}$. 
The characteristic relation of $X_{\rm CO}$ with $A_{V}$ is
mainly driven by how $I_{\rm CO}$ varies with $A_{V}$.  
At small $A_{V}$, $X_{\rm CO}$ steeply decreases with $A_{V}$, 
likely because CO becomes sufficiently shielded against photodissociation and $I_{\rm CO}$ sharply increases. 
$X_{\rm CO}$ then gradually increases with $A_{V}$, likely due to the saturation of $I_{\rm CO}$.
Our results observationally confirm previous theoretical predictions of the $X_{\rm CO}$ vs $A_{V}$ profile for the first time.  
However, the precise details of the $X_{\rm CO}$ vs $A_{V}$ profile, 
e.g., the location where the transition from decreasing to increasing $X_{\rm CO}$ occurs, 
the slopes of the decreasing and increasing portions, etc., 
again depend on local environmental parameters (e.g., Taylor et al. 1993; Bell et al. 2006; Shetty et al. 2011a). 
In general, our results suggest that a single $X_{\rm CO}$ cannot be used 
to derive the spatial distribution of $N$(H$_{2}$) across a molecular cloud. 

The detailed comparison between our high-resolution data and theory provides 
important insights into H$_{2}$ and CO formation in molecular clouds. 
For example, the good agreement we found with the modified W10 model suggests that 
the steady state and equilibrium chemistry and the microturbulent approximation for CO spectral line formation and cooling  
work well for Perseus on $\sim$0.4 pc scales.
Perseus appears to be old enough to achieve chemical equilibrium 
and the timescale of the dynamical process(es) most revelant for the formation of molecular gas is likely $\gtrsim$ 3 Myr.  
However, the good agreement with the model is achieved only if the density distribution has a diffuse halo component. 
In the modified W10 model, the halo provides dust shielding against H$_{2}$ and CO photodissociation
and is essential to reproduce the observed $N$(HI), $R_{\rm H2}$, $I_{\rm CO}$, and $X_{\rm CO}$ distributions. 
While our results indicate the importance of the diffuse HI halo for the distributions of two most abundant molecular species, H$_{2}$ and CO, 
the properties of the halo have not been observationally well constrained. 

Despite the lack of fine-tuning to match the characteristics of Perseus, 
the S11 model reproduces the observed $N$(HI), $N$(H$_{2}$), and $I_{\rm CO}$ properties reasonably well. 
In particular, the predicted range of $N$(H$_{2}$) in the scaled S11 model is in excellent agreement with our data. 
These results suggest that the time-dependent chemistry model is generally consistent with the time-independent chemistry model
for a low-mass, old molecular cloud such as Perseus. 
However, there are several interesting discrepancies and they likely result from the nature of turbulence in the S11 model.
The strong compressions and rarefactions by turbulence could result in the wider range of $N$(HI) in the S11 model
and unlike the modified W10 model, there is no minimum $N$(HI) beyond which the rest of hydrogen is fully converted into H$_{2}$. 
In addition, density fluctuations in the S11 model allow the formation of dense clumps even at small $A_{V}$ 
and potentially result in a large scatter of $I_{\rm CO}$. 
Turbulent motions could mix and spread the CO distribution,  
likely contributing to the scatter of $I_{\rm CO}$.
Our future studies of other Galactic molecular clouds, 
in particular those clouds much less evolved and/or forming more massive stars (therefore more turbulent) than Perseus,  
will be important for comprehensive tests of the PDR and MHD models.

We sincerely thank the anonymous referee for suggestions that significantly improved this work. 
We also thank Chris Carilli, Paul Clark, Jay Gallagher, Miller Goss, Paul Goldsmith, Harvey Liszt, 
Adam Leroy, J\"urgen Ott, Josh Peek, and Jaime Pineda for stimulating discussions,  
Tom Dame for graciously providing his Dame et al. (2001) data, 
and the GALFA-HI and COMPLETE survey teams for making their data publicly available.   
M.-Y.L. and S.S. acknowledge support from the NSF grant AST-1056780, NASA through contract 145727 issued by JPL/Caltech,
and the University of Wisconsin Graduate School.  
R.S., S.G., F.M., and R.K. acknowledge support from the Deutsche Forschungsgemeinschaft 
(DFG) via the SFB 881 (B1 and B2) ``The Milky Way System'' and the SPP (priority program) 1573. 
The Arecibo Observatory is operated by SRI International under a cooperative agreement with the National Science Foundation (AST-1100968), 
and in alliance with Ana G. M\'endez-Universidad Metropolitana, and the Universities Space Research Association.
We have made use of the KARMA visualization software (Gooch 1996) and NASA's Astrophysics Data System (ADS). 

\clearpage



\begin{table}
\begin{center}
{\bf TABLE 1} \\
\textsc{Physical Properties of the Dark and Star-forming Regions} \\
\vskip 0.1cm
\begin{tabular}{c c c c c c} \hline \hline
Region & $\alpha$$^{\rm a}$ & $\delta$$^{\rm a}$ & Average Size$^{\rm b}$ & $\Sigma$$N$(H$_{2}$)/$\Sigma$$I_{\rm CO}$ & Median $\sigma_{\rm CO}$ \\
 & (J2000) & (J2000) & (pc) & (cm$^{-2}$ K$^{-1}$ km$^{-1}$ s) & (km s$^{-1}$) \\ \hline 
B5 & 3.71$^{\rm h}$ $-$ 3.84$^{\rm h}$ & 32.57$^{\circ}$ $-$ 33.29$^{\circ}$ & 5 & 5.0 $\times$ 10$^{19}$ & 1.3 \\
IC348 & 3.69$^{\rm h}$ $-$ 3.81$^{\rm h}$ & 31.14$^{\circ}$ $-$ 32.50$^{\circ}$ & 6 & 6.6 $\times$ 10$^{19}$ & 1.4 \\ 
B1E/B1 & 3.52$^{\rm h}$ $-$ 3.65$^{\rm h}$ & 30.57$^{\circ}$ $-$ 32.14$^{\circ}$ & 6 & 2.7 $\times$ 10$^{19}$ & 1.8 \\
NGC1333 & 3.44$^{\rm h}$ $-$ 3.52$^{\rm h}$ & 30.50$^{\circ}$ $-$ 32.21$^{\circ}$ & 7 & 1.9 $\times$ 10$^{19}$ & 2.0 \\
L1448 & 3.36$^{\rm h}$ $-$ 3.44$^{\rm h}$ & 30.21$^{\circ}$ $-$ 31.14$^{\circ}$ & 5 & 1.5 $\times$ 10$^{19}$ & 1.3 \\
\hline 
\end{tabular}
\begin{minipage}{16cm}
$^{\rm a}$ The regional boundaries defined in Section 4. \\
$^{\rm b}$ The characteristic size of each region calculated by (total number of pixels)$^{1/2}$ $\times$ 0.38 pc,
where 0.38 pc is the physical size of one pixel.
\end{minipage}
\end{center}
\end{table}

\begin{table} 
\begin{center} 
{\bf TABLE 2} \\ 
\textsc{Predictions from the Modified W10 Model with a ``Core-Halo'' Density Distribution$^{\rm a}$} \\ 
\vskip 0.1cm 
\begin{tabular}{c c c c} \hline \hline 
 & $n_{\rm core}$ = 10$^{3}$ cm$^{-3}$ & $n_{\rm core}$ = 5 $\times$ 10$^{3}$ cm$^{-3}$ & $n_{\rm core}$ = 10$^{4}$ cm$^{-3}$ \\ \hline 
$N$(HI)$^{\rm b}$ (cm$^{-2}$) & $9.00 \times 10^{20}$ $-$ $9.60 \times 10^{20}$ & $8.98 \times 10^{20}$ $-$ $9.20 \times 10^{20}$ & $8.96 \times 10^{20}$ $-$ $9.04 \times 10^{20}$ \\ 
$N$(H$_{2}$)$^{\rm b}$ (cm$^{-2}$) & $1.75 \times 10^{20}$ $-$ $4.52 \times 10^{21}$ & $1.75 \times 10^{20}$ $-$ $4.54 \times 10^{21}$ & $1.75 \times 10^{20}$ $-$ $4.55 \times 10^{21}$ \\
$R_{\rm H2}$$^{\rm b}$ & $0.39 - 9.42$ & $0.39 - 9.87$ & $0.39 - 10.11$ \\ 
$I_{\rm CO}$$^{\rm b}$ (K km s$^{-1}$) & $0.075 - 36.90$ & $0.33 - 41.40$ & $0.39 - 33.90$ \\ 
$L_{\rm halo}$$^{\rm c}$ (pc) & 7.28 & 7.28 & 7.28 \\
$L_{\rm core}$$^{\rm b, d}$ (pc) & $0.11 - 2.94$ & $0.023 - 0.59$ & $0.011 - 0.29$ \\ 
$L_{\rm halo-core}$$^{\rm b, e}$ (pc) & $7.39 - 10.22$ & $7.30 - 7.87$ & $7.29 - 7.57$ \\ 
$\langle n \rangle$$^{\rm b, f}$ (cm$^{-3}$) & $54.74 - 316.66$ & $55.42 - 411.21$ & $55.49 - 427.51$ \\ 
\hline
\end{tabular}
\begin{minipage}{16cm}
$^{\rm a}$ For all three models, $n_{\rm halo}$ = 40 cm$^{-3}$. \\
$^{\rm b}$ The values are provided for the minimum and maximum column densities ($A_{V}$ = 1.25 mag and 10 mag). \\ 
$^{\rm c}$ The size of the diffuse halo $L_{\rm halo}$ = 9 $\times$ 10$^{20}$ cm$^{-2}$/40 cm$^{-3}$. \\  
$^{\rm d}$ The size of the dense core $L_{\rm core}$ = ($N$(H) $-$ 9 $\times$ 10$^{20}$)/$n_{\rm core}$. \\ 
$^{\rm e}$ The total size of the slab $L_{\rm halo-core}$ = $L_{\rm halo}$ + $L_{\rm core}$. \\
$^{\rm f}$ The average density $\langle n \rangle$ = $N$(H)/$L_{\rm halo-core}$. 
\end{minipage}
\end{center}
\end{table}

\begin{table} 
\begin{center} 
{\bf TABLE 3} \\
\textsc{Predictions from the Modified W10 Model with a Uniform Density Distribution} \\
\vskip 0.1cm
\begin{tabular}{c c c c} \hline \hline 
 & $n$ = 10$^{3}$ cm$^{-3}$ & $n$ = 5 $\times$ 10$^{3}$ cm$^{-3}$ & $n$ = 10$^{4}$ cm$^{-3}$ \\ \hline 
$N$(HI)$^{\rm a}$ (cm$^{-2}$) & $5.71 \times 10^{19} - 1.30 \times 10^{20}$ & $1.15 \times 10^{19} - 2.66 \times 10^{19}$ & $5.59 \times 10^{18} - 1.33 \times 10^{19}$ \\ 
$N$(H$_{2}$)$^{\rm a}$ (cm$^{-2}$) & $2.72 \times 10^{20} - 4.94 \times 10^{21}$ & $2.95 \times 10^{20} - 4.99 \times 10^{21}$ & $2.98 \times 10^{20} - 5.00 \times 10^{21}$ \\ 
$R_{\rm H2}$$^{\rm a}$ & $9.53 - 76.00$ & $51.30 - 375.19$ & $106.62 - 751.88$ \\ 
$I_{\rm CO}$$^{\rm a}$ (K km s$^{-1}$) & $0.84 - 46.90$ & $9.79 - 62.40$ & $18.00 - 60.20$ \\ 
$L_{\rm uniform}$$^{\rm a, b}$ (pc) & $0.19 - 3.24$ & $0.039 - 0.65$ & $0.019 - 0.32$ \\
\hline 
\end{tabular}
\begin{minipage}{17cm}
$^{\rm a}$ The values are provided for the minimum and maximum column densities ($A_{V}$ = 0.6 mag and 10 mag). \\ 
$^{\rm b}$ The total size of the slab $L_{\rm uniform}$ = $N$(H)/$n$. \\
\end{minipage} 
\end{center} 
\end{table}
\clearpage 

\clearpage

\begin{figure}
\begin{center}
\includegraphics[scale=0.45]{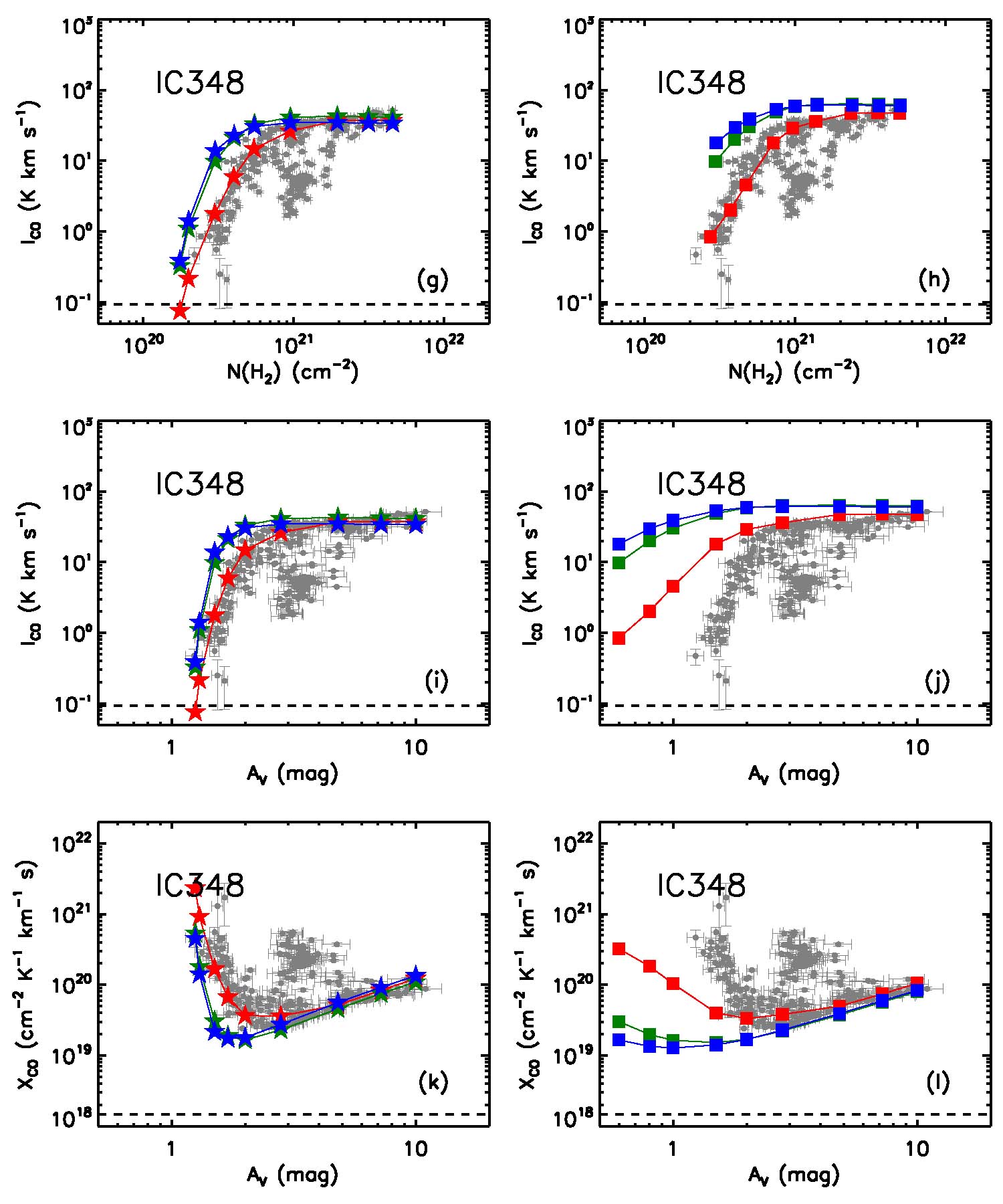}
\vskip 0.3cm
\begin{minipage}{17cm}
Fig. \ref{f:W_compare}.--- (Continued)
\end{minipage}
\end{center}
\end{figure}



\begin{figure} 
\begin{center} 
\includegraphics[scale=0.42]{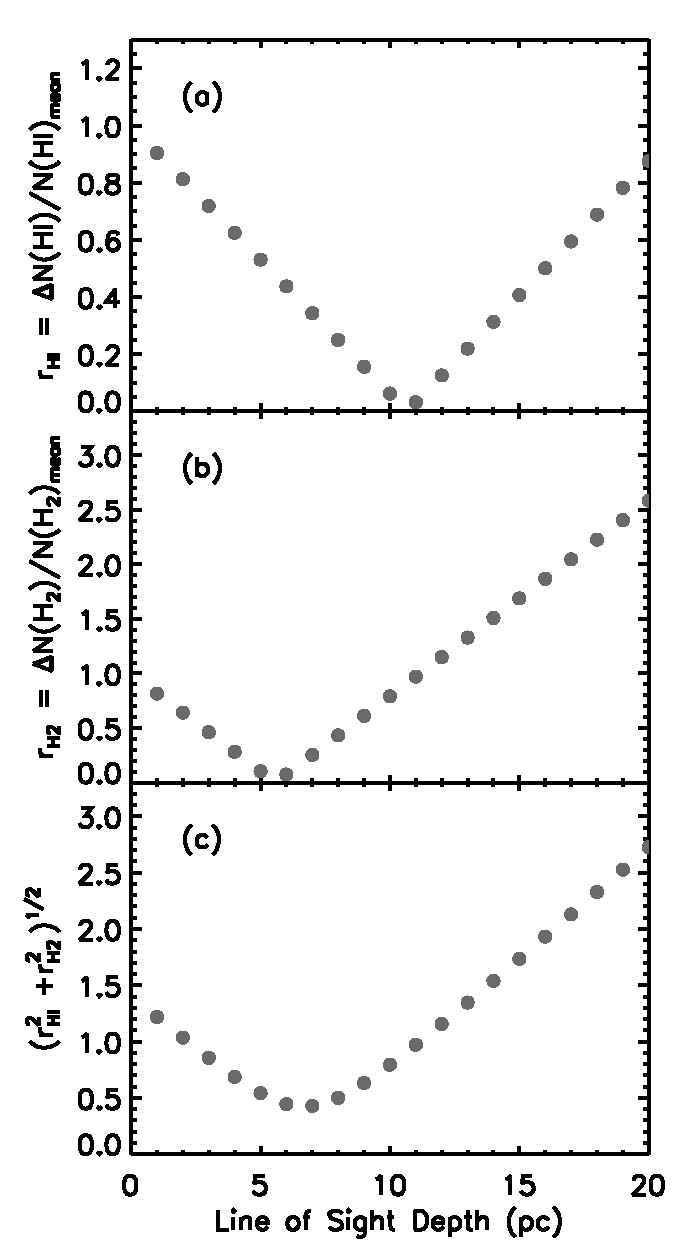}
\caption{\label{f:box_size} 
(a) The difference between the observed mean $N$(HI) and the simulated mean $N$(HI) 
normalized by the observed mean $N$(HI), $r_{\rm HI}$ = $\Delta$$N$(HI)/$N$(HI)$_{\rm mean}$, is plotted as a function of line of sight depth.
(b) Same as (a) but for $N$(H$_{2}$).
(c) The sum of the two normalized differences in quadrature, $\sqrt{r_{\rm HI}^{2} + r_{\rm H2}^{2}}$, is plotted as a function of line of sight depth.
Note that the line of sight depth of 7 pc results in the minimum discrepancy between our and simulated data.}
\end{center}
\end{figure}

\begin{figure}
\begin{center}
\includegraphics[scale=0.4]{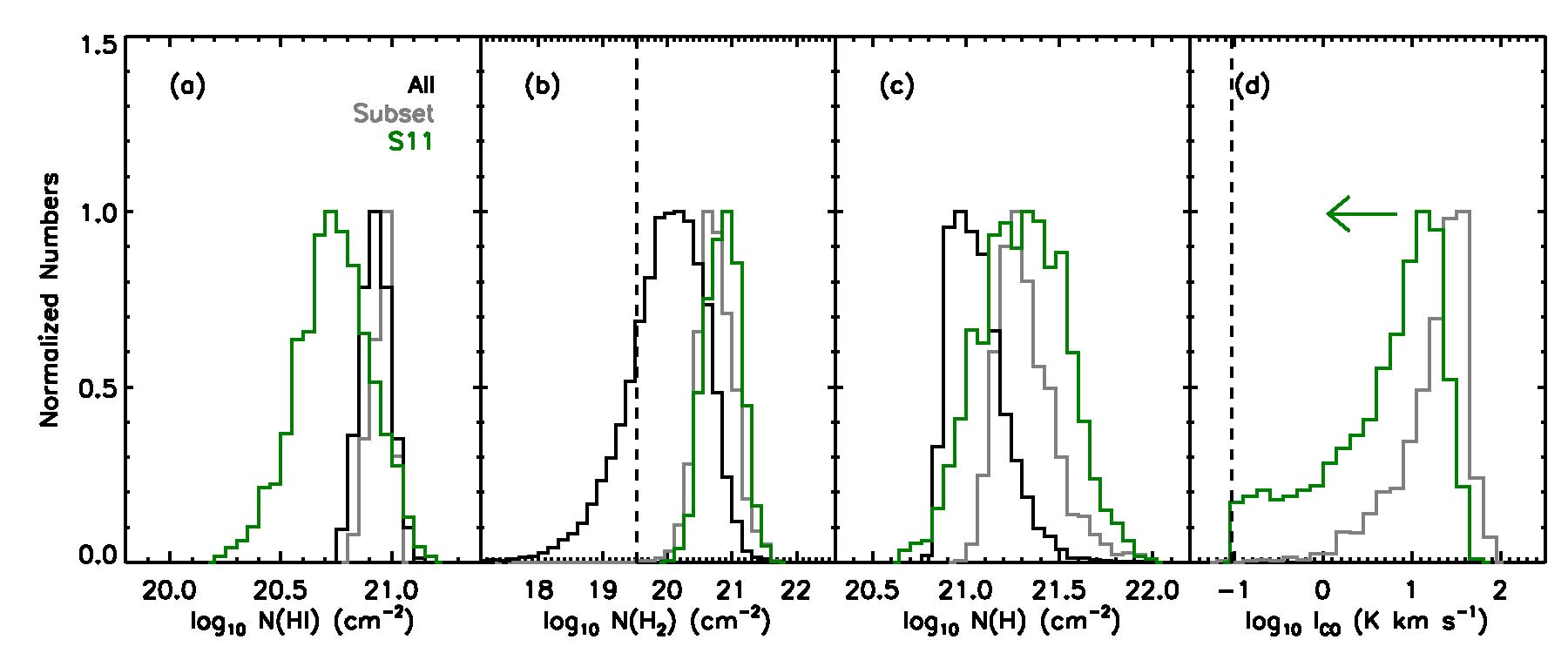}
\caption{\label{f:histo_Perseus_s11} Normalized histograms of $N$(HI), $N$(H$_{2}$), $N$(H), and $I_{\rm CO}$.
The histograms in black, grey, and green are constructed 
using the data points with positive $N$(H$_{2}$), shown in Figure \ref{f:COMPLETE_five_regions}, 
and from the S11 model respectively. 
For the S11 model, the simulated $N$(HI), $N$(H$_{2}$), and $N$(H) data are scaled for 7 pc, 
while the simulated $I_{\rm CO}$ data are not. 
See Section 7.2.2 for details.
The mean 1$\sigma$ uncertainties of $N$(H$_{2}$) ($\sim$3.3 $\times$ 10$^{19}$ cm$^{-2}$) and $I_{\rm CO}$ ($\sim$0.09 K km s$^{-1}$) are shown as dashed lines, 
while those of $N$(HI) ($\sim$3.5 $\times$ 10$^{19}$ cm$^{-2}$) and $N$(H) ($\sim$1.6 $\times$ 10$^{20}$ cm$^{-2}$) are too small to be shown.}
\end{center}
\end{figure}

\begin{sidewaysfigure}
\begin{center}
\includegraphics[scale=0.4]{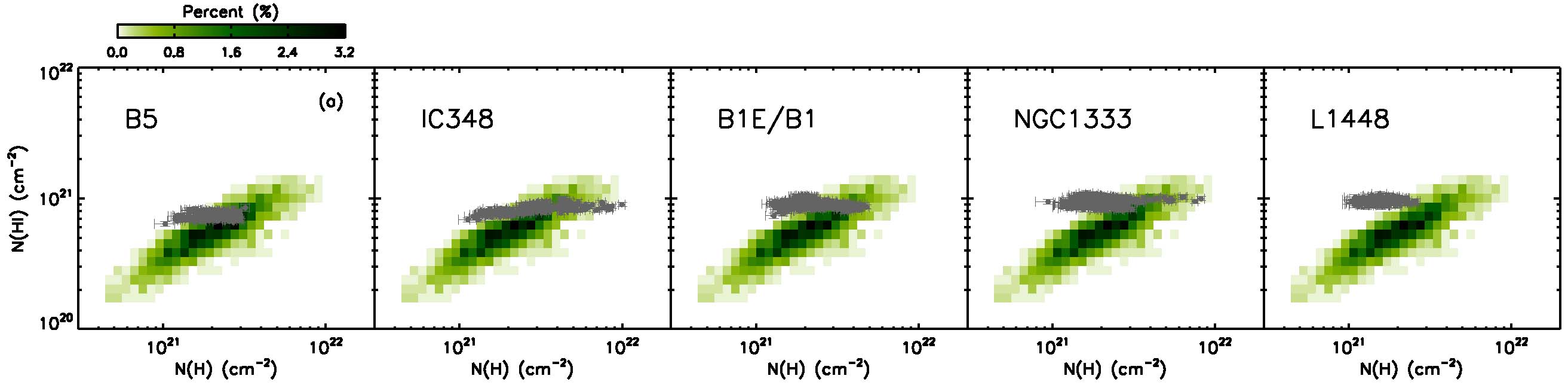}
\includegraphics[scale=0.4]{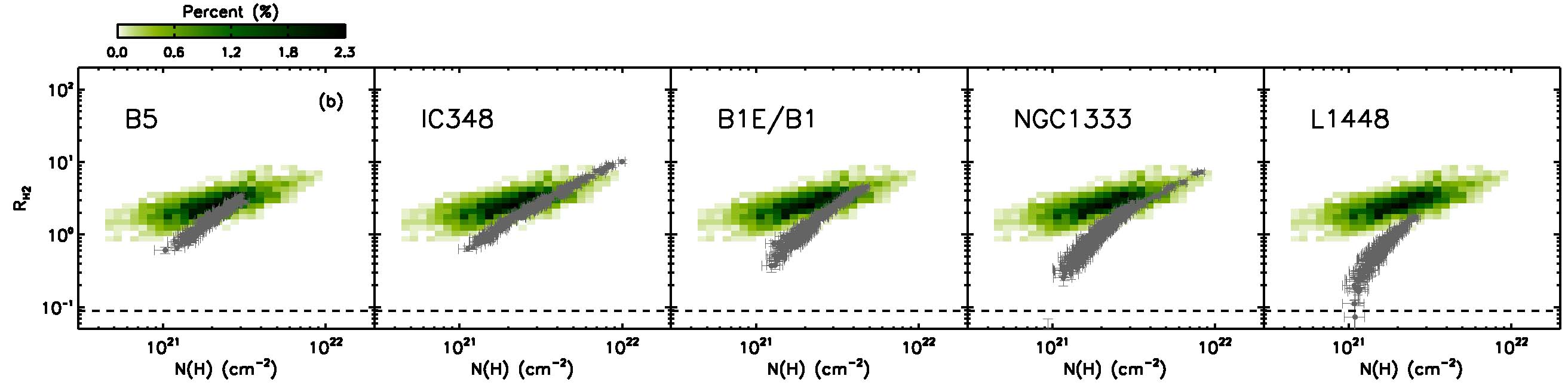}
\includegraphics[scale=0.4]{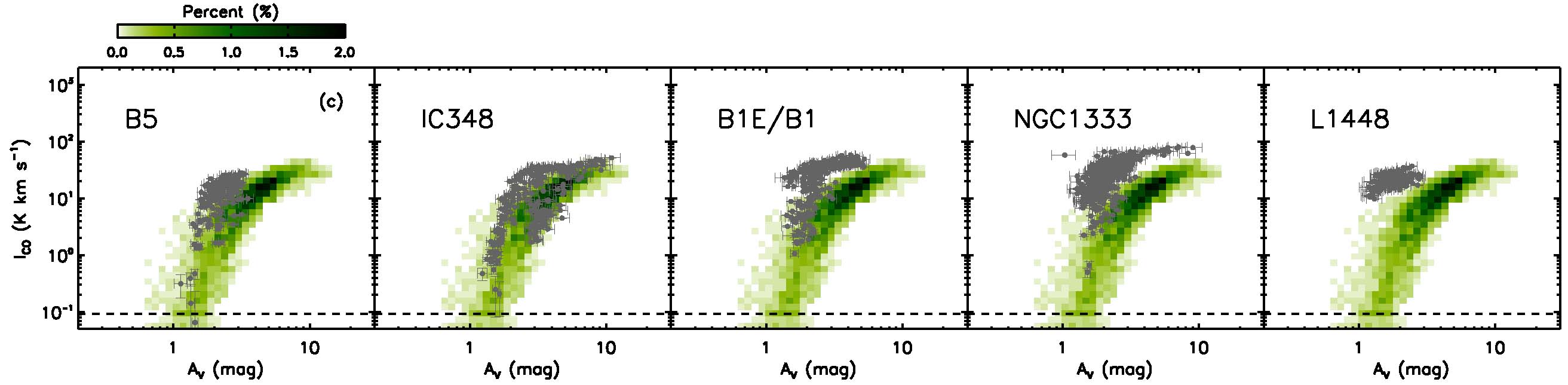}
\includegraphics[scale=0.4]{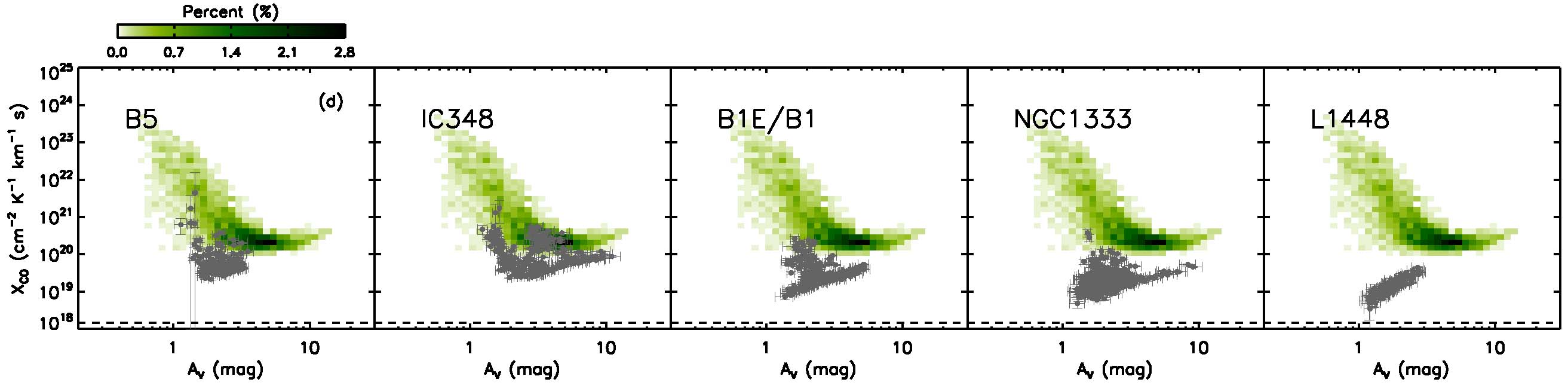}
\caption{\label{f:HI_tot_S11}
Comparison with the S11 model.
The mean 1$\sigma$ uncertainties of $R_{\rm H2}$ ($\sim$0.09) and $I_{\rm CO}$ ($\sim$0.09 K km s$^{-1}$) are shown as dashed lines, 
while those of $N$(HI) ($\sim$3.5 $\times$ 10$^{19}$ cm$^{-2}$), $N$(H) ($\sim$1.6 $\times$ 10$^{20}$ cm$^{-2}$) and $A_{V}$ ($\sim$0.2 mag) are too small to be shown.
The median 1$\sigma$ uncertainty of $X_{\rm CO}$ ($\sim$1.5 $\times$ 10$^{18}$) is shown as a dashed line as well. 
(a) $N$(HI) vs $N$(H).
(b) $R_{\rm H2}$ vs $N$(H). 
(c) $I_{\rm CO}$ vs $A_{V}$. 
(d) $X_{\rm CO}$ vs $A_{V}$.
For (a) and (b), the simulated quantities are smoothed, regridded, and scaled for 7 pc.
In addition, the thresholds for $N$(H$_{2}$) and $I_{\rm CO}$ are applied.
On the other hand, for (c) and (d), the simulated quantities are smoothed and regridded 
but neither the scaling nor the thresholds is applied.
See Section 7.2.3 for details.}
\end{center}
\end{sidewaysfigure}
\clearpage

\appendix

\section{Detailed Comparison With Previous Studies} 

While we found $\langle X_{\rm CO} \rangle$ $\sim$ 3 $\times$ 10$^{19}$ for Perseus, 
Dame et al. (2001) and Pineda et al. (2008) estimated $\sim$1.2 $\times$ 10$^{20}$ and $\sim$1.4 $\times$ 10$^{20}$ respectively. 
These two studies are similar with our study in the sense that they utilized dust as a tracer of total gas column density, 
but applied different methodologies to derive $X_{\rm CO}$. 
We follow their methodologies in order to understand why our result is different.

\subsection{Comparison with Dame et al. (2001)}

Dame et al. (2001) used the $E(B-V)$ data from Schlegel et al. (1998) and the HI data from the LAB survey. 
They estimated DGR on large-scales by smoothing both the $E(B-V)$ and $N$(HI) images to 10$^{\circ}$ resolution 
and calculating the ratio of the smoothed $E(B-V)$ and $N$(HI) images. 
The $E(B-V)$ image was then divided by the large-scale DGR image and $N$(HI) was subtracted to derive $N$(H$_{2}$). 
The derived $N$(H$_{2}$) was finally combined with $I_{\rm CO}$ from the CfA survey to estimate $X_{\rm CO}$. 
The resolution of the HI data was the lowest among all data sets and the estimated $X_{\rm CO}$ values were consequently at 36$'$ resolution. 
We note that most other large-scale studies of $X_{\rm CO}$ in the Milky Way are also at 36$'$ resolution, limited by the LAB HI data 
(e.g., Abdo et al. 2010; Paradis et al. 2012). 

\begin{figure} 
\begin{center} 
\includegraphics[scale=0.4]{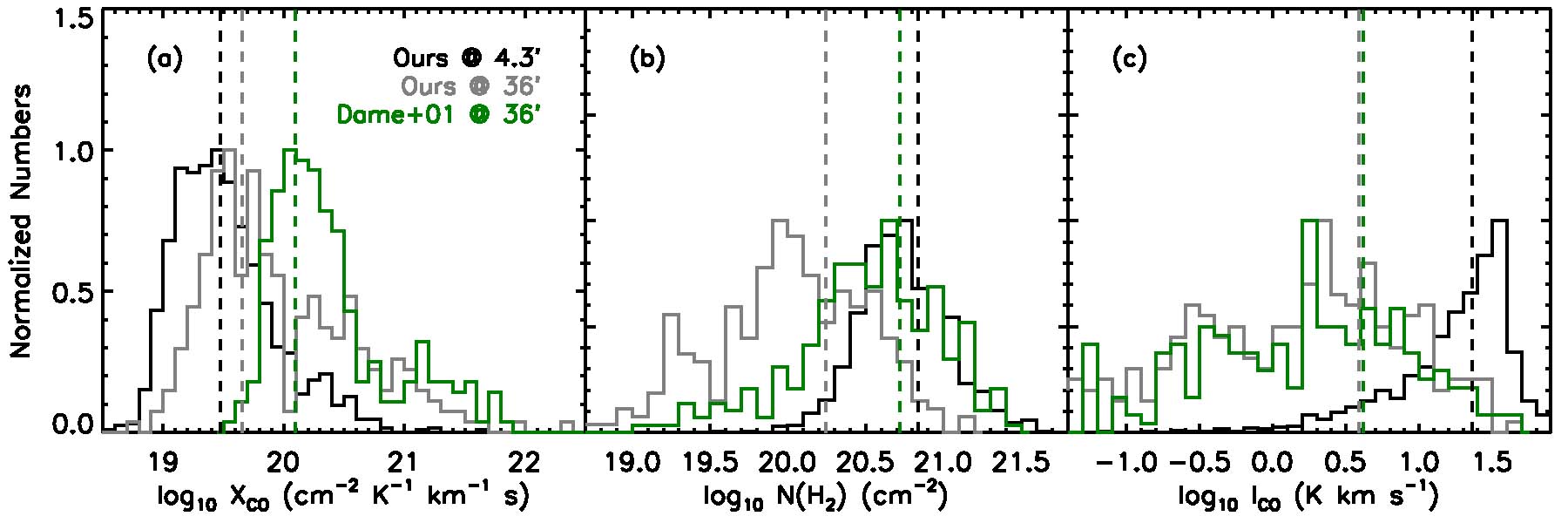}
\includegraphics[scale=0.4]{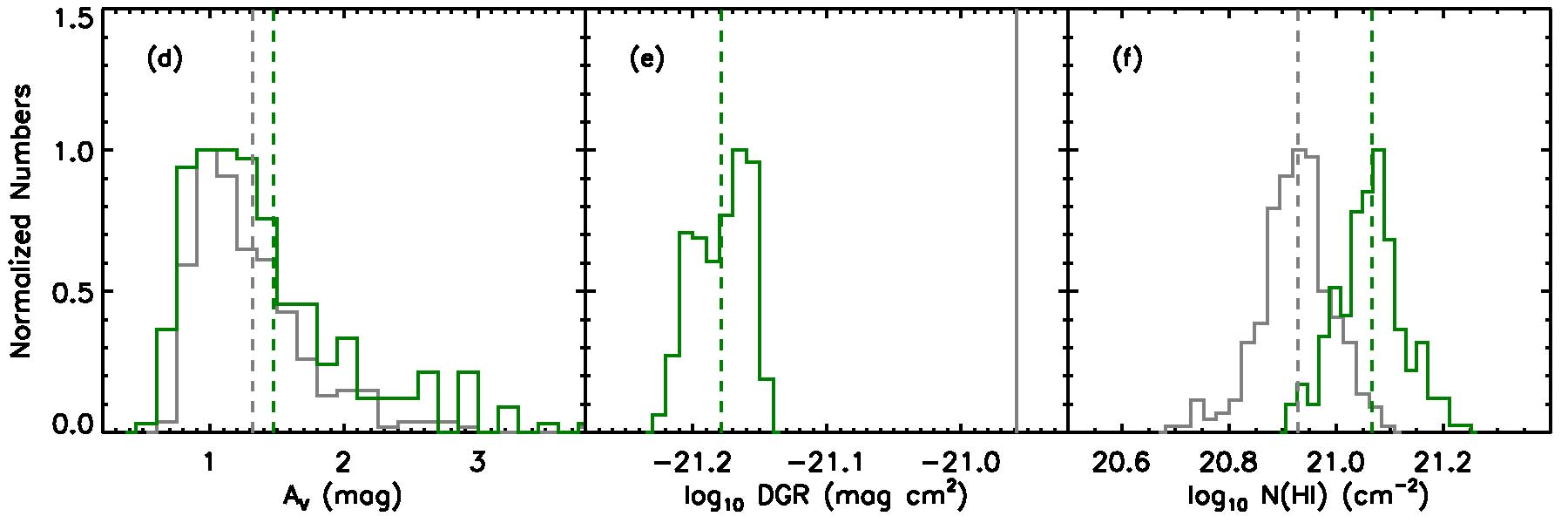}
\caption{\label{f:compare_Dame01} Comparison with Dame et al. (2001). 
The black and grey histograms represent the data from our study at 4.3$'$ and 36$'$ angular resolutions,  
while the green histograms show the data from Dame et al. (2001).
All histograms are normalized for easy comparison.  
Note that the $I_{\rm CO}$ image from the CfA survey was combined with our $N$(H$_{2}$) image 
to derive $X_{\rm CO}$ at 36$'$ angular resolution.
Dashed lines show the mean values of individual quantities, except for those shown in (a),
which represent $\langle X_{\rm CO} \rangle$ = $\rm \Sigma$$N$(H$_{2}$)$/\rm \Sigma$$I_{\rm CO}$.
The local DGR = 1.1 $\times$ 10$^{-21}$ mag cm$^{2}$ Lee et al. (2012) derived for Perseus is shown as a grey solid line.}
\end{center} 
\end{figure}

To show how different resolutions and methodologies affect the estimation of $X_{\rm CO}$, 
we first compare our original data at 4.3$'$ resolution (black histograms; data points for all five regions) 
with (1) our data smoothed to 36$'$ resolution (grey histograms) 
and (2) the data from Dame et al. (2001) (green histograms) in Figure \ref{f:compare_Dame01}. 
Note that we use the CfA CO data here to derive $X_{\rm CO}$ at 36$'$ resolution 
instead of the COMPLETE CO data we used elsewhere in this paper, 
because of their larger spatial coverage 
($\sim$10$^{\circ}$ $\times$ 7$^{\circ}$ for the CfA CO vs $\sim$6$^{\circ}$ $\times$ 3$^{\circ}$ for the COMPLETE CO).
This will not cause any complication with our comparison,  
considering that $\sim$83\% of the data points are consistent within 1$\sigma$ uncertainties 
when the CfA and COMPLETE CO data are compared at the common resolution of 8.4$'$.
For each histogram in Figure \ref{f:compare_Dame01}, we show the mean value of the distribution as a dashed line. 
In the case of $X_{\rm CO}$, $\langle X_{\rm CO} \rangle$ calculated as $\rm \Sigma$$N$(H$_{2}$)/$\rm \Sigma$$I_{\rm CO}$ is shown instead.

In comparison betwen our data at 4.3$'$ and 36$'$ resolutions, 
we find that $\langle X_{\rm CO} \rangle$ increases from $\sim$3 $\times$ 10$^{19}$ (4.3$'$) to $\sim$4.5 $\times$ 10$^{19}$ (36$'$).  
$\langle X_{\rm CO} \rangle$ increases because spatial smoothing affects 
the $I_{\rm CO}$ distribution slightly more than the $N$(H$_{2}$) distribution.
To be precise, $I_{\rm CO}$ decreases by a factor of $\sim$6 from $\sim$23.1 K km s$^{-1}$ to $\sim$3.9 K km s$^{-1}$ on average,
while $N$(H$_{2}$) decreases by a factor of $\sim$4 from $\sim$6.9 $\times$ 10$^{20}$ cm$^{-2}$ to $\sim$1.7 $\times$ 10$^{20}$ cm$^{-2}$ on average. 

While spatial smoothing to 36$'$ resolution results in the slight increase of $X_{\rm CO}$, 
there is still a factor of $\sim$2.7 discrepancy between our $\langle X_{\rm CO} \rangle \sim$ 4.5 $\times$ 10$^{19}$
and the value derived by Dame et al. (2001) for the same area. 
Because the same CfA CO data were used, as shown from the good agreement between the grey and green histograms in Figure \ref{f:compare_Dame01}(c), 
the discrepancy in $X_{\rm CO}$ would come from the difference in $N$(H$_{2}$)
and we indeed find that the mean $N$(H$_{2}$) $\sim$ 5.2 $\times$ 10$^{20}$ cm$^{-2}$ in Dame et al. (2001) is larger than 
our mean $N$(H$_{2}$) $\sim$ 1.7 $\times$ 10$^{20}$ cm$^{-2}$ at 36$'$ resolution by a factor of $\sim$3.
Considering that the equations for deriving $N$(H$_{2}$) in our study and Dame et al. (2001) are essentially the same, 
$N$(H$_{2}$) = ($A_{V}$/DGR $-$ $N$(HI)) $\times$ 0.5, 
we compare our $A_{V}$ and $N$(HI) data smoothed to 36$'$ resolution with the data from Dame et al. (2001) in Figures \ref{f:compare_Dame01}(d,f). 
To convert $E$($B-V$) in Dame et al. (2001) into $A_{V}$, we use the total-to-selective extinction ratio $R_{V}$ $\sim$ 3.1 for the diffuse ISM (Mathis 1990). 
In addition, the local DGR $\sim$ 1.1 $\times$ 10$^{-21}$ mag cm$^{-2}$ Lee et al. (2012) derived for Perseus is compared with 
the DGR data from Dame et al. (2001) in Figure \ref{f:compare_Dame01}(e). 
While we find that our $A_{V}$ at 36$'$ resolution is consistent with $A_{V}$ in Dame et al. (2001), 
our $N$(HI) is slightly smaller than theirs by a factor of $\sim$1.4 on average. 
This difference mainly results from the fact that Dame et al. (2001) integrated the HI emission along a whole line of sight,  
while our $N$(HI) was derived by integrating the HI emission over the velocity range for Perseus, from $-5$ km s$^{-1}$ to $+15$ km s$^{-1}$ (Section 3.1).
The slightly smaller $N$(HI) in our study could affect the estimation of DGR
and we indeed find that the local DGR for Perseus is larger than the mean DGR in Dame et al. (2001) by a factor of $\sim$1.7 on average. 
Another factor that could affect DGR is spatial smoothing to 10$^{\circ}$ resolution done by Dame et al. (2001). 
Specifically, they blanked all pixels whose $I_{\rm CO}$ is larger than 1 K km s$^{-1}$ 
and replaced the blanked pixels with the Gaussian-weighted $E$($B-V$)/$N$(HI) values, the Gaussian having a FWHM of 10$^{\circ}$. 
The angular size of 10$^{\circ}$ is comparable to the size of Perseus 
and in this case spatial smoothing could result in the inclusion of the diffuse ISM with small DGR in the far outskirts of the cloud.

\subsection{Comparison with Pineda et al. (2008)}

\begin{figure} 
\begin{center} 
\includegraphics[scale=0.5]{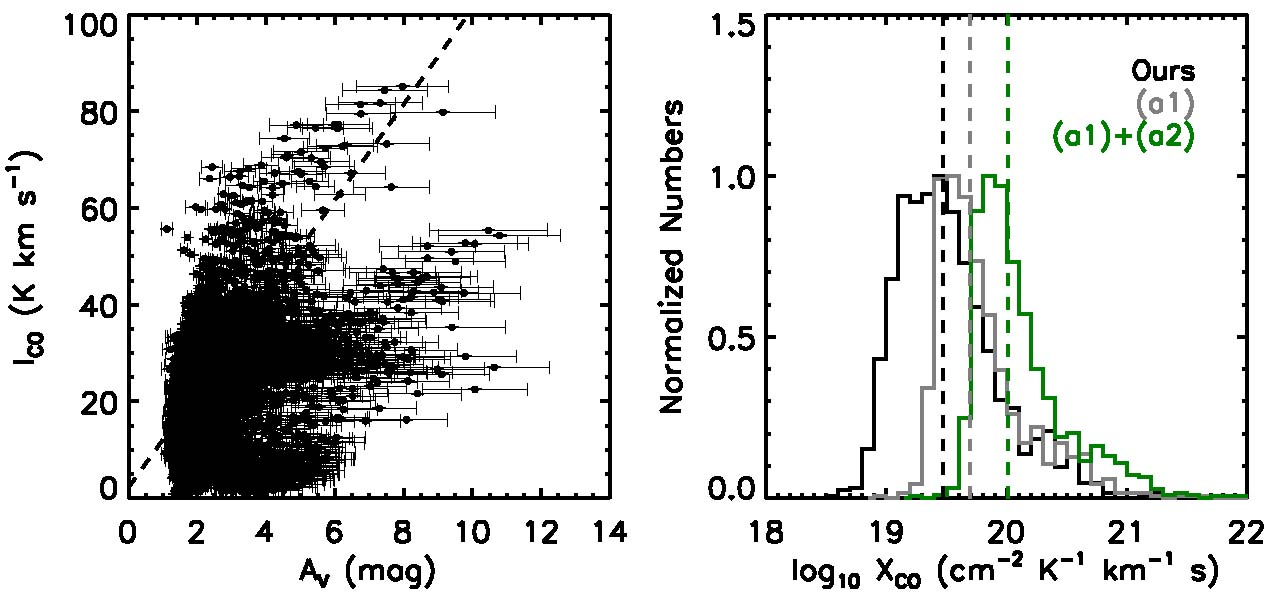}
\caption{\label{f:compare_Pineda08} Comparison with Pineda et al. (2008). 
(Left) $I_{\rm CO}$ vs $A_{V}$. 
The linear fit obtained from the BCES is shown as a dashed line. 
(Right) Normalized histograms of $X_{\rm CO}$.  
The black histogram is our original $X_{\rm CO}$ at 4.3$'$ angular resolution, 
while the grey and green histograms show $X_{\rm CO}$ estimated by assuming (a1) only and (a1) $+$ (a2) respectively.   
Dashed lines show $\langle X_{\rm CO} \rangle=$ $\rm \Sigma$$N$(H$_{2}$)/$\rm \Sigma$$I_{\rm CO}$.  }
\end{center} 
\end{figure}

Pineda et al. (2008) used the $A_{V}$ and $I_{\rm CO}$ images from the COMPLETE survey smoothed to 5$'$ resolution 
and derived $X_{\rm CO}$ for Perseus by fitting a linear function to $I_{\rm CO}$ vs $A_{V}$ under the following two assumptions: 
(a1) all hydrogen traced by $A_{V}$ is in the form of H$_{2}$ and  
(a2) DGR = 5.3 $\times$ 10$^{-22}$ mag cm$^{2}$, the typical Galactic value (Bohlin et al. 1978). 
Considering that our study uses essentially the same data sets, 
the COMPLETE $I_{\rm CO}$ image and the $A_{V}$ image calibrated with the COMPLETE $A_{V}$ data, 
any difference in $X_{\rm CO}$ would come from different methodologies for deriving $X_{\rm CO}$.   
In Figure \ref{f:compare_Pineda08}(a), we plot $I_{\rm CO}$ as a function of $A_{V}$ for Perseus
and fit a linear function to the data as Pineda et al. (2008) did.  
To be consistent with Pineda et al. (2008), we smooth our $I_{\rm CO}$ and $A_{V}$ images 
with Gaussian kernels to obtain a resolution of 5$'$ and regrid the images to a grid of 2.5$'$. 
In addition, we use their primary thresholds, i.e., the CO and $^{13}$CO integrated intensities are positive 
and the CO and $^{13}$CO peak brightness temperatures are at least 10 and 5 times the rms noises of CO and $^{13}$CO, to select data points. 
We do not consider other thresholds adopted in Pineda et al. (2008), 
e.g., exclusion of the data points with a stellar density larger than 10 stars per pixel, 
and expect that they will not make a significant change in the linear fit, 
considering that they account only a small fraction of the total number of data points ($\sim$7\%). 
As Pineda et al. (2008) performed, we fit the linear function 
\begin{equation} 
A_{V} = a + b I_{\rm CO} 
\end{equation}  

\noindent using the bivariate correlated errors and intrinsic scatter estimator (BCES; Akritas \& Bershady 1996) 
and find $a$ = $-0.22 \pm 0.13$ mag and $b$ = $0.10 \pm 0.01$ mag K$^{-1}$ km$^{-1}$ s. 
This result is consistent with Pineda et al. (2008), 
once our fitted parameters $a$ and $b$ are converted into the quantities in Equation (18) of Pineda et al. (2008), 
$A_{V12}$ = $a$ = $-0.22 \pm 0.13$ mag and 
$X_{2}$ = $b$ $\times$ 9.4 $\times$ 10$^{20}$ = ($1.0 \pm 0.1$) $\times$ 10$^{20}$, 
\noindent where $A_{V12}$ is the minimum $A_{V}$ below which there is no CO emission and $X_{2}$ is essentially $X_{\rm CO}$.  
Our $X_{2}$ = ($1.0 \pm 0.1$) $\times$ 10$^{20}$ is very close to $X_{2}$ = ($1.4 \pm 0.1$) $\times$ 10$^{20}$ in Pineda et al. (2008).  
In summary, our result is consistent with Pineda et al. (2008) if we use exactly the same methodology for deriving $N$(H$_{2}$) and $X_{\rm CO}$. 

However, instead of fitting a linear function to $I_{\rm CO}$ vs $A_{V}$,  
we derive $X_{\rm CO}$ on a pixel-by-pixel basis in this paper. 
To investigate whether this could result in a significant difference, we perform additional tests. 
First, we derive $X_{\rm CO}$ by assuming (a1) but with our DGR = 1.1 $\times$ 10$^{-21}$ mag cm$^{2}$. 
The result is shown in Figure \ref{f:compare_Pineda08}(b) (grey histogram),
along with our original $X_{\rm CO}$ distribution (black histogram). 
Second, we assume both (a1) and (a2) and still derive $X_{\rm CO}$ on a pixel-by-pixel basis. 
The result is shown in the same panel as a green histogram. 
In addition, $\langle X_{\rm CO} \rangle$ for each histogram is calculated as $\rm \Sigma$$N$(H$_{2}$)/$\rm \Sigma$$I_{\rm CO}$ 
and is shown as a dashed line. 
We find $\langle X_{\rm CO} \rangle$ $\sim$ 4.9 $\times$ 10$^{19}$ and 1 $\times$ 10$^{20}$ for the first and second test respectively.  
This suggests that our pixel-by-pixel derivation of $X_{\rm CO}$ is consistent with the linear fit method in Pineda et al. (2008) 
and therefore the discrepancy between our $\langle X_{\rm CO} \rangle$ and $X_{\rm CO}$ in Pineda et al. (2008) results from the assumptions (a1) and (a2).  
Specifically, the neglect of $N$(HI) in derivation of $N$(H$_{2}$) (a1) results in a factor of $\sim$1.6 difference in $\langle X_{\rm CO} \rangle$,  
while the use of the Galactic DGR (a2) results in an additional factor of $\sim$2 difference.

\end{document}